\documentclass[fleqn,usenatbib]{mnras}

\usepackage{newtxtext,newtxmath}

\usepackage[T1]{fontenc}
\usepackage{ae,aecompl}
\usepackage{caption}

\usepackage{graphicx}	
\usepackage{amsmath}	
\usepackage{amssymb}	
\usepackage{longtable}
\usepackage{txfonts}
\usepackage{xcolor}
\usepackage{supertabular,booktabs}

\title[Quantitative spectroscopy of extreme helium stars]{Quantitative spectroscopy of extreme helium stars\\\vspace{0.2cm}\LARGE{Model atmospheres and a non-LTE abundance analysis of BD+10$^\circ$2179\thanks{Based on observations obtained at the European Southern Obser\-vatory, proposals 077.D-0458(A) and 077.D-0458(B).}}}

\author[T. Kupfer et al.]{
T.~Kupfer,$^{1}$\thanks{E-mail: tkupfer@caltech.edu}
N.~Przybilla,$^{2}$
U.~Heber,$^{3}$
C.~S.~Jeffery,$^{4}$
N.~T.~Behara,$^{4}$
and K.~Butler$^{5}$
\\
$^{1}$Division of Physics, Mathematics and Astronomy, California Institute of Technology, Pasadena, CA 91125, USA\\
$^{2}$Institut f\"ur Astro- und Teilchenphysik, Universit\"at Innsbruck, Technikerstr. 25/8, 6020 Innsbruck, Austria\\
$^{3}$Dr. Karl Remeis-Observatory \& ECAP, Astronomical Institute,
Friedrich-Alexander University Erlangen-Nuremberg, Sternwart\-str.~7, 96049 Bamberg, Germany\\
$^{4}$Armagh Observatory and Planetarium, College Hill, Armagh, BT61 9DG, N. Ireland, UK\\
$^{5}$University Observatory Munich, Scheinerstr. 1, 81679 Munich, Germany
}

\date{Accepted XXX. Received YYY; in original form ZZZ}

\pubyear{2016}

\begin{document}
\label{firstpage}
\pagerange{\pageref{firstpage}--\pageref{lastpage}}
\maketitle

\begin{abstract}
Extreme helium stars (EHe stars) are hydrogen-deficient supergiants of spectral type A and B. They are believed to result from mergers in double degenerate systems. In this paper we present a detailed quantitative non-LTE spectral analysis for BD+10$^\circ$2179, a prototype of this rare class of stars, using UVES and FEROS spectra covering the range from $\sim$3100 to 10\,000\,{\AA}. Atmosphere model computations were improved in two ways. First, since the UV metal line blanketing has a strong impact on the temperature-density stratification, we used the {\sc Atlas12} code. Additionally, We tested {\sc Atlas12} against the benchmark code {\sc Sterne3}, and found only small differences in the temperature and density stratifications, and good agreement with the spectral energy distributions. Second, 12 chemical species were treated in non-LTE. Pronounced non-LTE effects occur in individual spectral lines but, for the majority, the effects are moderate to small. The spectroscopic parameters give  $T_\mathrm{eff}$\,=17\,300$\pm$300\,K and $\log g$\,=\,2.80$\pm$0.10, and an evolutionary mass of 0.55$\pm$0.05\,$M_\odot$. The star is thus slightly hotter, more compact and less massive than found in previous studies. The kinematic properties imply a thick-disk membership, which is consistent with the metallicity $[$Fe/H$]$\,$\approx$\,$-1$ and $\alpha$-enhancement. The refined light-element abundances are consistent with the white dwarf merger scenario. We further discuss the observed helium spectrum in an appendix, detecting dipole-allowed transitions from about 150 multiplets plus the most comprehensive set of known/predicted isolated forbidden components to date. Moreover, a so far unreported series of pronounced forbidden He\,{\sc i} components is detected in the optical-UV. 
\end{abstract}

\begin{keywords}
line: formation -- stars: atmospheres -- stars: fundamental parameters -- stars:abundances -- stars: individual: BD+10$^\circ$2179
\end{keywords}



\section{Introduction}
Extreme helium stars (EHe stars) are a rare class of hydrogen-deficient supergiant of spectral types A and B. The atmospheres are strongly enriched in helium, carbon, nitrogen and neon. Helium is the most abundant element, carbon is often the second most abundant, while hydrogen is highly depleted by a factor of 100 or more. Only 18 {\em bona-fide} EHe stars are known today\footnote{\citet{jef96} Table 2 lists 23 stars, including two hot RCrB stars, two N-rich C-poor stars, and two He-rich low-gravity sdO stars. Only one new EHe star has been detected since the late 1970's recently \citep{Jeffery17}.}, which means that they have to be produced by an unusual process or they represent a very short-lived stage of evolution, or both \citep{jef08}. No known EHe star shows  evidence of any binary companion \citep{jef87}, widespread radial velocity variations result from small-amplitude pulsations \citep{jef08}.

The chemical composition indicates highly processed material produced by hydrogen and helium burning. The origin and evolution of these stars remained a puzzle over the decades. Two different evolution models emerged. The double degenerate model (DD) calls for a white dwarf merger in a close binary system \citep{web84, iben84}. The other, the final flash model (FF), evokes a late or very late thermal flash in a post-asymptotic giant branch (post-AGB) star which forces the star to expand and start the post-AGB sequence again \citep[e.g.][]{schoenberner77,iben83}.

\cite{web84} and \cite{iben84} introduced the DD model involving the merger of a carbon/oxygen and a helium white dwarf due to the decay of their orbit. The initial system consists of a binary system of two main sequence stars. The more massive star in the binary system evolves first to become a red giant and fill its Roche Lobe. Unstable mass transfer will lead to a common envelope (CE) phase. Friction, loss of orbital energy and therefore a decay of the binary orbit forces the ejection of the CE. After that the less massive star will evolve to become a red giant and another CE will be formed which leads to a further shrinkage of the orbit. Left over is a short period binary system consisting of a carbon/oxygen and a helium white dwarf. Because of gravitational wave radiation the orbit of the binary system decays until the helium white dwarf fills its Roche Lobe. Due to tidal forces the helium white dwarf will be disrupted. A debris disk around the CO white dwarf is created. When a sufficient amount of helium has been accreted, the helium ignites and forces the star to expand and become a yellow supergiant. The star will probably appear as a R Coronae Borealis star. Due to contraction the star will evolve to become an EHe star and ends up as a massive carbon/oxygen white dwarf. A detailed model was developed by \cite{saio02} and \cite{jef11} from which the surface abundances of the resulting EHe star were predicted. The absence of close companions, observed abundances and number densities prefers the merger of a carbon/oxygen and a helium white dwarf as the origin of the observed EHe stars.

The FF model corresponds to a late or very late thermal pulse when the star is already on the post-AGB sequence. Helium burning in an AGB star is unstable and results in so-called thermal pulses. If the pulse happens when the star is already on the post-AGB sequence the remaining envelope mass is small enough ($\lesssim$~10$^{-4}$ M$_\odot$) that the pulse has a significant impact on the outer layers and forces the star to become a cool supergiant \citep{bloe01}. A convection zone mixes the processed material to the surface. The models predict a remaining hydrogen abundance of about $2$\%. The enrichment of carbon, oxygen and helium depends strongly on the position where the late thermal pulse takes place, how effectively the convective zone brings processed material to the surface and whether overshooting does or does not occur \citep{bloe01}. When the star reaches the supergiant phase again a stable helium burning shell is established and the star starts the post-AGB evolution once more. The FF model predicts much higher oxygen and carbon abundances than observed in EHe stars \citep{her99}. 

The best studied EHe star is BD+10$^\circ$2179 (DN Leo, HIP\,52123) which was identified as an extreme helium star by \cite{kle61}. The first abundance analysis was carried out by \citet{hill65}, and the first fine analysis by \cite{hun69}. The latter found the atmosphere to be dominated by carbon ($55\%$ by mass) and helium ($45\%$ by mass), with hydrogen being only a trace element ($0.01\%$). \cite{heb83} re-analysed BD+10$^\circ$2179 from photographic optical spectra and high-resolution UV spectra and found that the carbon abundance is closer to $1\%$ by number whereas the helium abundance is much higher at $98.9\%$ by number, with 
effective temperature $T_{\rm eff}$\,=\,16\,800\,$\pm$\,600\,K and surface gravity $\log g$\,=\,2.55\,$\pm$\,0.2. The abundances of Ti, Cr, Mn, Fe, Co and Ni were found to be about one tenth  solar, only, indicating that BD+10$^\circ$2179 belongs to an old stellar population. A contemporary analysis of the  flux distribution alone yielded  $T_{\rm eff}$\,=\,17\,700\,$\pm$\,600\,K \citep{drilling84}.   
The most recent analysis of BD+10$^\circ$2179 in which the spectrum was calculated entirely under the approximation of local thermodynamic equilibrium (LTE) was by \cite{pan06}. They found 
$T_{\rm eff}$\,=\,16\,900\,K and $\log g$\,=\,2.55$\pm$0.2 from the UV spectra but $T_{\rm eff}$\,=\,16\,400\,K  and $\log g$\,= 2.35\,$\pm$0.2 from the optical. Their helium and carbon abundances are similar to those of \cite{heb83}, but some of the other elements are highly discrepant, up to $0.82$\,dex for Mg. \citet{jef10} employed hydrodynamical non-LTE model atmospheres in spherical geometry as computed with the PoWR code to investigate mass-loss in EHe stars. For BD+10$\degr$2179 they adopted $T_{\rm eff}$\,=\,18\,500\,K -- a significantly higher value than found in all other studies -- and $\log g$\,=\,2.6.
\cite{pan11} carried out a hydrostatic non-LTE analysis of BD+10$^\circ$2179 using the model atmosphere program TLUSTY including H, He, C, N, O and Ne in non-LTE. They found $T_{\rm eff}$\,=\,16\,375\,$\pm$\,250\,K and $\log g$\,=\,2.45\,$\pm$\,0.2 which is similiar to former LTE analyses. 

The aim of this work is to carry out a quantitative spectral analysis for BD+10$^\circ$2179 using a high-resolution spectrum and comprehensive non-LTE techniques. The observations and data reduction are described in Section~\ref{sec:obser}. Before the spectroscopic analysis was carried out using {\sc Atlas12} models a detailed comparison against the benchmark code {\sc Sterne3} was mandatory since {\sc Atlas12} has never before been applied to such unusual conditions. Section~\ref{sec:atmos} describes the model atmospheres.  Section~\ref{sec:comparison} shows the results of the comparison of {\sc Sterne3} and {\sc Atlas12}. The quantitative spectral analysis is described in Section~\ref{sec:analysis}, while the fundamental stellar parameters and the kinematic properties are discussed in Sect.~\ref{kinematics}. Finally the results are discussed in the context of previous analyses (Sect.~\ref{discussion}) and conclusions are drawn in Sect.~\ref{conclusion}. Details on the observed helium spectrum are collected in an Appendix.

\section{Observations and data reduction}\label{sec:obser}
The spectra of BD+10{\degr}2179 used for the present work were observed on April 12, 2006 with {\sc Feros} \citep[Fiberfed Extended Range Optical Spectro\-graph,][]{kau99} on the MPG/ESO 2.2m telescope in La Silla/Chile and on May 16, 2006 with {\sc Uves} \citep[UV-Visual Echelle Spectrograph,][]{dek00} on the ESO VLT at Paranal/Chile. The resolving power $R$\,=\,$\Delta\lambda$/$\lambda$ provided by {\sc Feros} is 48\,000, covering a useful wavelength range from 3800 up to 9200\,{\AA}. The {\sc Uves} observations employed the dichroic mode {\#}2 to cover the optical-UV region in a useful range $\sim$3200--3850\,{\AA} in the blue and $\sim$6650--8540\,{\AA} and $\sim$8650--10\,240\,{\AA} in the red simultaneously. A measured $R$\,$\approx$\,37\,500 was achieved with a 1{\arcsec} slit. Good atmospheric conditions (0.8{\arcsec} seeing) and an exposure time of 2820\,s yielded a signal-to-noise ratio $S$/$N$\,$\approx$\,300 per pixel near 5000\,{\AA} of the {\sc Feros} data. A $S$/$N$\,$\approx$\,200 was achieved in the 1000\,s exposure with {\sc Uves} at 8700\,{\AA} 
(0.9{\arcsec} seeing). Note that we largely discarded the optical-UV and $IzY$-band spectra for the present quantitative analysis, as the spectral region is dominated by the high series members of the \ion{He}{i} lines originating from the 2$p$\ $^{1,3}$P$^\mathrm{o}$ and 3$s$\ $^{1,3}$S, 3$p$\ $^{1,3}$P$^{\mathrm o}$ and 3$d$\ $^{1,3}$D levels, respectively, which we cannot model because of the lack of appropriate line-broadening data.   

Data reduction was accomplished using ESO-MIDAS pipelines and our own recipes. It covered the usual steps of bad pixel and cosmic correction, bias and dark current subtraction, removal of scattered light, optimal order extraction, flat-fielding, wavelength calibration using Th-Ar exposures, and merging of the \'echelle orders. Finally, large scale variations of the spectral response functions were removed by using the featureless spectrum of the DC white dwarf WD1917-07 as continuum tracer \citep[for the {\sc Uves} data; for details of the method see][]{koe01}, and the (well-modelled) subdwarf B star HD188112 in the case of the {\sc Feros} spectrum, yielding a well-normalised spectrum of BD+10{\degr}2179. 

Overall, our observational data are comparable in quality to the optical spectra employed by \citet{pan06} and \citet{pan11} for the analysis of BD+10{\degr}2179, but having a much wider wavelength coverage. The slightly lower spectral resolution of our data has no consequences for the quantitative analysis, as the instrumental width is negligible compared to rotational/macroturbulent broadening in this star.  

For additional \ion{He}{i} line identifications blueward of our spectra, HST STIS spectra were extracted from the MAST archive\footnote{\tt http://archive.stsci.edu/}, a high-resolution spectrum obtained with the E230M grating ($R$\,$\approx$\,30\,000) with wavelength coverage $\sim$1840--2674\,{\AA} \citep[dataset O6MB01020, as STARCAT high-level science product,][]{ayr10} and a low-resolution spectrum obtained with the G230LB grating ($R$\,$\approx$\,700) with wavelength coverage $\sim$1670--3074\,{\AA} (dataset O66V14010). The latter data were originally discussed by \citet{jef10}.

Finally, in order to assess the spectral energy distribution (SED) of BD+10$^\circ$2179 we have extracted flux-calibrated spectra obtained with the International Ultraviolet Explorer (IUE: exposures SWP04825 and LWR04168) from the MAST archive. These low-dispersion/large-aperture data were first described by \citet{heb83}. Wide-band photometry in the Johnson $UBV$ passbands was adopted from \citet{mer97} and in the 2MASS $JHK$ passbands from \citet{cut03}. The magnitudes were converted to fluxes using the zero points described by \citet{heb02}.

\section{Model atmospheres and spectrum synthesis}\label{sec:atmos}   
The unusual chemical composition of EHe stars requires care in the computation of appropriate stellar model atmospheres.
To date, the majority of modern abundance analyses of EHes \citep[for a review see][]{jef08} have been carried
out using the line-blanketed model atmosphere code {\sc Sterne}, together with the line-formation code {\sc Spectrum}, see \citet{jef01} for an overwiew. Classical assumptions are made, i.e.~a chemically homogeneous stratification in plane-parallel geometry and hydrostatic and radiative equilibrium is considered, with the thermodynamic state of the plasma described by LTE. Line opacities are accounted for using an opacity distribution function (ODF) computed for a hydrogen-deficient mixture by \citet{Moeller90} from the \citet{KuPe75} line list. In a more recent version of the code, {\sc Sterne3} \citep{beh06}, improved continuous opacities were introduced\footnote{Data up to the triply-ionized state for all elements (and for \ion{C}{i--vi}) treated by the Opacity Project were implemented. An exception is iron, where data from the IRON Project 
were considered for \ion{Fe}{i--iii}.}, as well as an opacity-sampling (OS) procedure to deal with the line opacities, accounting for atomic transitions from \citet{KuBe95}. 
This resulted in significantly modified temperature structures of hydrogen-deficient OS models in comparison to the older ODF models. Models computed with {\sc Sterne3} are one basis for the investigations here.

A second, widely distributed, code for the computation of classical line-blanketed LTE model atmospheres for chemically peculiar stars is {\sc Atlas12} \citep{kur93,kur96}. {\sc Atlas12} employs the full set of Kurucz linelists, both the observed set of \citet{KuBe95} as well as predicted lines.
but some less state-of-the-art sources of continuous opacities. We use the code with \ion{He}{i} photoionization cross-sections updated as described by \citet{prz05} as the second basis for our investigation. 
A comparison of {\sc Sterne3} and {\sc Atlas12} atmospheres will be made in Sect.~\ref{sec:comparison}.

\begin{table}
\centering
\footnotesize
\caption[]{Model atoms for non-LTE calculations.\\[-6mm] \label{atoms}}
\setlength{\tabcolsep}{.15cm}
\begin{tabular}{llll}
\hline
\hline
\footnotesize
            Ion     &  Levels & Transitions & Reference \\
\hline\\[-3mm] 
     H            &  20        & 190         & [1]\\
\ion{He}{i}       &  29+6      & 162         & [2]\\
\ion{C}{i/ii/iii} &  80/68/70  & 669/425/373 & [3]\\
\ion{N}{i/ii}     &  89/77     & 668/462     & [4]\\
\ion{O}{i/ii}     &  51/52     & 243/134     & [5]\\
\ion{Ne}{i}       &  153       & 952         & [6]\\
\ion{Mg}{ii}      &  37        & 236         & [7]\\
\ion{Al}{ii/iii}  &  54+6/46+1 & 378/272     & [8]\\
\ion{Si}{ii/iii}  &  52+3/68+4 & 357/572     & [9]\\
\ion{S}{ii}       &  78        & 302         & [10]\\
\ion{Ar}{ii}      &  56        & 596         & [11]\\
\ion{Fe}{ii/iii}  & 265/60+46  & 2887/2446   & [12] \\
\hline\\[-5mm]
\end{tabular}
\begin{flushleft}
{\bf References.} [1]~\citet{PrBu04}; [2]~ \citet{Przybilla05}; [3]~\citet{Przybillaetal01b}, \citet{NiPr06,nie08};
[4]~\citet{PrBu01}; [5]~\citet{Przybillaetal00}, \citet{BeBu88}, updated$^a$; [6]~\citet{MoBu08}, updated$^a$;
[7]~\citet{Przybillaetal01a}; [8]~Przybilla (in prep.); [9]~Przybilla \& Butler (in prep.); [10]~ \citet{Vranckenetal96}, updated$^a$;
[11]~Butler (in prep.); [12]~\citet{Becker98}, \citet{Moreletal06}, corrected$^b$.\\
$^a$ See Sect.~\ref{sec:atmos} and Table~\ref{tab:abundances_nlte} for details.; $^b$ See Sect.~\ref{sec:atmos}.
\end{flushleft}
\end{table}

In addition to the model atmospheres, which are held fixed in the following steps, non-LTE line-formation computations are performed with the package {\sc Detail} and {\sc Surface} \citep{gid81,but85}. The former solves the coupled statistical and radiative transfer equations, providing non-LTE level populations, which are used by the latter to calculate the emergent flux from the formal solution, considering detailed line-broadening theories. The codes have
undergone substantial extension and improvements over the years. Most important in our context are {\sc i)} the implementation of opacity sampling in analogy to Kurucz' method in order to facilitate a realistic treatment of line blocking also for chemically peculiar stars, and {\sc ii)} the inclusion of an Accelerated Lambda Iteration scheme \citep{RyHu91}, which allows elaborate non-LTE model atoms to be used while keeping computational expenses moderate.
Line formation calculations in LTE are performed with {\sc Surface} if non-LTE populations are not provided, with all other input data (background opacities, oscillator strengths, broadening data, etc.) being identical.

An overview of the non-LTE model atoms employed in the present work is 
given in Table~\ref{atoms}, which summarises the number of
explicit non-LTE levels in the different ionization stages, the
number of radiative bound-bound transitions and the reference to the model atom.
Typically, the levels are terms, and the transition multiplet transitions.
In some cases additional 'superlevels' were considered (indicated by the
'+' sign), which were packed over many levels. Additional ionization
stages were also considered, typically consisting either of the ground
state of the next higher ionization stage only, or of simple models of
low complexity. In the case of Ne\,{\sc i}, the levels are
fine-structure components. Several of the model atoms
have been updated recently, mostly by consideration of improved oscillator strengths and collisional data. A major source of 
oscillator strengths are \citet{FFT04} and \citet{FFTI06}, based on extensive computations using the multiconfiguration
Hartree-Fock method that included relativistic effects through a Breit-Pauli Hamiltonian, which improve over earlier Opacity Project and IRON project data both in accuracy and precision. Sources of the data used in the quantitative analysis are summarised in Table~A3. 
Note that the original input data for line-formation computations of \ion{Fe}{iii} were corrected according to \citet{NiPr12}.

\begin{figure}
   \centering
   \includegraphics[angle=270, width=0.48\textwidth]{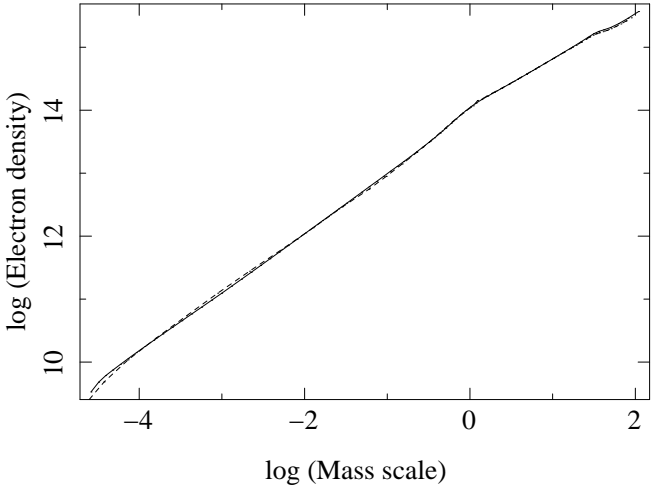}
      \caption{Run of the electron density as a function of mass scale for a helium-rich model with 
      $T_{\rm eff}$\,=\,16\,800\,K and $\log g$\,=\,2.80 as computed with {\sc Sterne3} (dotted) and 
      {\sc Atlas12} (solid).}
         \label{fig:density}
   \end{figure}

Both the hybrid non-LTE approach as well as the model atoms were thoroughly tested and successfully applied in environments similar to those encountered in supergiant B-type extreme helium stars: massive BA-type supergiants \citep[which in particular have similar luminosity-to-mass ratios to supergiant EHes,][]{prz06a} and He-strong B-stars \citep{Przybillaetal16}.
In these cases, significantly better modelling of the observed spectra was achieved than was possible with LTE techniques.
We expect non-LTE effects to be significant here as well.
However, in contrast to these previous cases, hydrogen is almost
completely absent here and therefore does not contribute significantly to
the continuous opacity. Helium also provides little opacity between in
the ultraviolet region long-ward of the He\,{\sc i} ionization threshold.
Instead, the metals take over the r\^ole of main opacity sources, with
carbon being the most important. Consequently, the trace species
approach -- assuming that the individual metals can be treated separately,
because they do not affect the atmospheric structure -- does not hold
here any longer. We therefore solved the rate equations and radiative
transfer for most of the chemical species simultaneously in order to
account for mutual interactions. In order to keep the resulting equation systems and 
run times manageable, three combinations of model atoms were realised:
HHeCNOMgAlSiSAr, HHeCNONeMgSi and HHeCNOMgSiFe. This was driven  
by the high complexity of the current model atoms for neon and iron, and facilitated by 
the trace-species character of Al, S, and Ar because of their lower abundances.

An a-posteriori check of the departure coefficients of the energetically low-lying
states of the main ionization stages of the abundant metals shows that these are close to LTE
at depths relevant for continuum formation, i.e. the continuous
opacities are not driving the atmospheric structure significantly
out of LTE. Concerning the line opacities, non-LTE effects strengthen some and
weaken other transitions. Overall, the LTE opacity sampling used for
the atmospheric structure calculations should therefore give a statistically
meaningful average opacity. In consequence, we expect our
hybrid non-LTE approach to provide a realistic approximation to full
non-LTE model atmosphere calculations, which, however, are beyond the
scope of present capabilities if model atoms as complex as ours should
be used. On the other hand, exactly this complexity is required to
reproduce the minute details of the observed spectrum.

     \begin{figure}
   \centering
   \includegraphics[angle=270,width=0.48\textwidth]{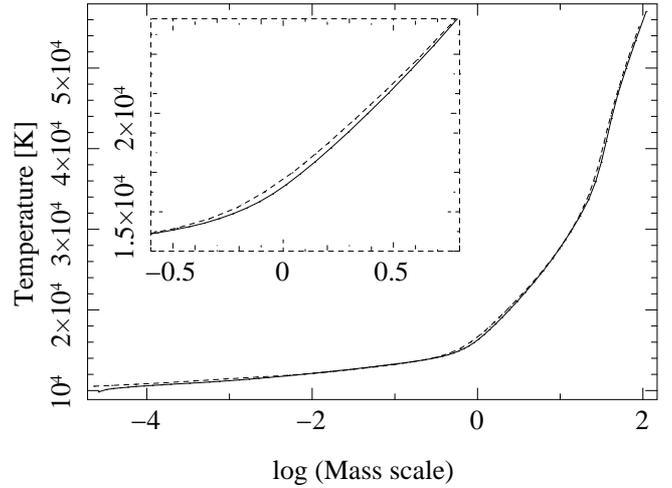}
      \caption{Temperature stratification for {\sc Sterne3} (dotted) and {\sc Atlas12} (solid) using a helium rich atmosphere.}
         \label{fig:temperature}
   \end{figure}
   
A comment has also to be made on line broadening. In a largely ionized plasma like the atmosphere of BD+10{\degr}2179 line broadening 
occurs through the Stark and Doppler effect. Contributors to the Stark effect are electrons and protons in stars of normal composition.
Few protons are present in this case, replaced by single-ionized helium and ionized carbon. It may therefore seem at first glance 
that the standard treatment based on tabulations of broadening coefficients due to electrons and protons are inapplicable. 
However, one has to recall that {\sc i} collisions with protons provide only a minor contribution to the dominating broadening 
by electron collisions, and {\sc ii} the mass-ratio between electrons on the one hand and helium or carbon particles on the other hand 
is larger than for protons, resulting in heavy particle velocities lowered by a factor 2 to $\sim$ 3.5. The contribution of heavy particles 
to the line broadening is therefore neglected in the present case.

A powerful fitting routine was used for a semi-automatic comparison of the observed and theoretical spectra in order to derive atmospheric parameters and elemental abundances. {\sc Spas}\footnote{Spectrum Plotting and Analysing Suite, {\sc Spas} \citep{Hirsch09}.} provides the means to interpolate between model grid points for up to three parameters simultaneously and allows instrumental and rotational broadening functions to be applied to the resulting theoretical profiles. The program uses the downhill simplex algorithm \citep{NeMe65} to minimise $\chi^2$ in order to find a good fit to the observed spectrum. The chemically peculiar nature of EHe stars prohibits the use of large precomputed grids for the analysis as we have employed in previous work \citep{NiPr12,Irrgangetal14}. Instead, microgrids for a small part of parameter space were computed, starting around an initial model based on an LTE analysis of BD+10{\degr}2179, and iteratively refined to bring the various parameter and abundance indicators into agreement, see Sect.~\ref{sec:analysis} for a discussion.

   \begin{figure}
   \centering
    \includegraphics[angle=270,width=0.97\linewidth]{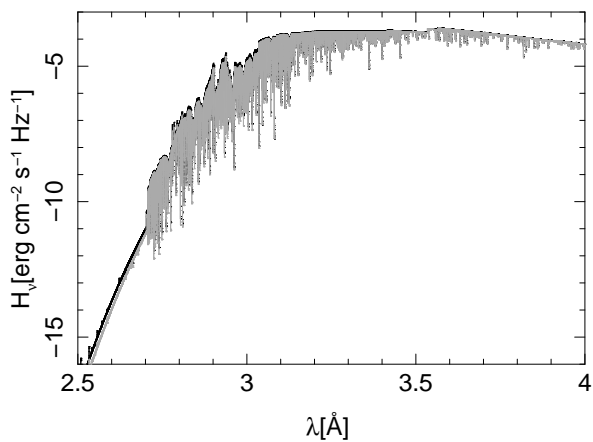}
    \caption{Non-LTE flux distribution from the extreme-UV to the near-IR as based on {\sc Sterne3} (black) and {\sc Atlas12} atmospheres (gray).}
   \label{fig:fluxdistr}
   \end{figure}

   \begin{figure}
    \includegraphics[angle=270,width=0.915\linewidth]{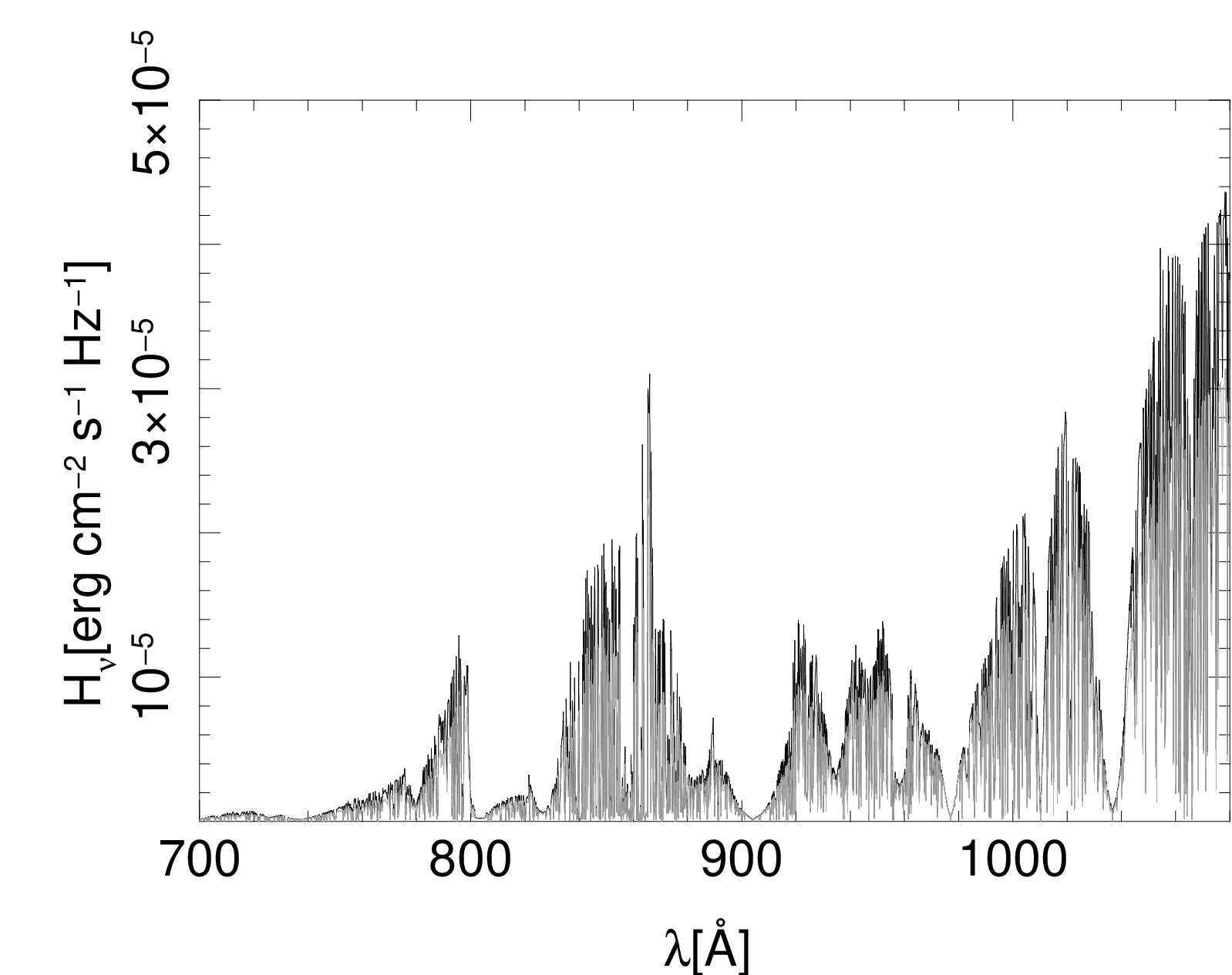}
    \caption{Same as Fig.~\ref{fig:fluxdistr}, concentrating only on the extreme-UV in linear scale to better visualise the differences between the models.}
   \label{fig:fluxdistr2}
    \includegraphics[angle=270,width=0.97\linewidth]{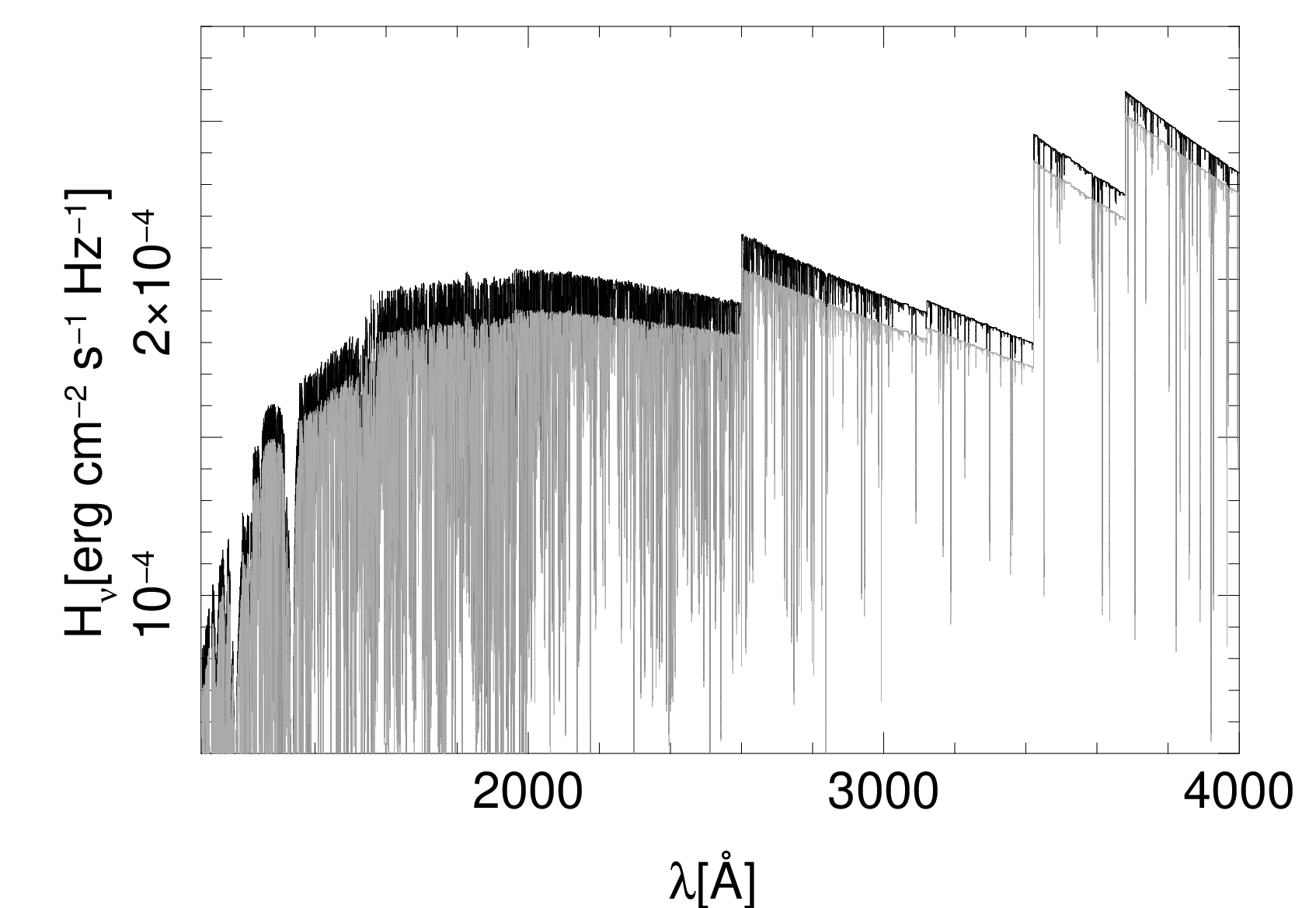}
    \caption{Same as Fig.~\ref{fig:fluxdistr}, concentrating only on the UV/optical in linear scale to better visualise the differences between the models.}
    \label{fig:fluxdistr3}
   \end{figure}


\section{Comparison between {\sc Atlas12} and {\sc Sterne3}}\label{sec:comparison}
Stars with chemical compositions similar to BD+10$^\circ$2179 place high requirements on programs which model the atmospheric stratification. With lower continuous opacities, the observed spectrum samples layers at much higher densities than in the usual H-rich atmospheres. Therefore these stars are good testbeds for a comparison between different model atmosphere codes. Before we describe the spectral analysis of BD+10$^\circ$2179 a comparison between the model structures obtained with {\sc Atlas12} and {\sc Sterne3} and including a non-LTE formal solution with {\sc Detail} was done for a chemical composition typical of that found in EHe stars. \cite{beh06} compared {\sc Sterne3} and {\sc Atlas9} for solar composition and found good agreement between both codes.
Note that because of different normalisations for the abundances a small error in abundance notations could not be avoided\footnote{{\sc Sterne3} uses particle fractions normalised over all elements whereas {\sc Atlas12} normalises over hydrogen and helium with respect to $n_{\rm He}$\,+\,$n_{\rm H}$\,=\,1 where $n_{\rm He}$ is the particle fraction of helium and $n_{\rm H}$ is the particle fraction for hydrogen. For the chosen abundances the error should be in the range of about 1 -- 2 \%.}.
   
   \begin{figure*}
 \centering
    \includegraphics[width=\textwidth]{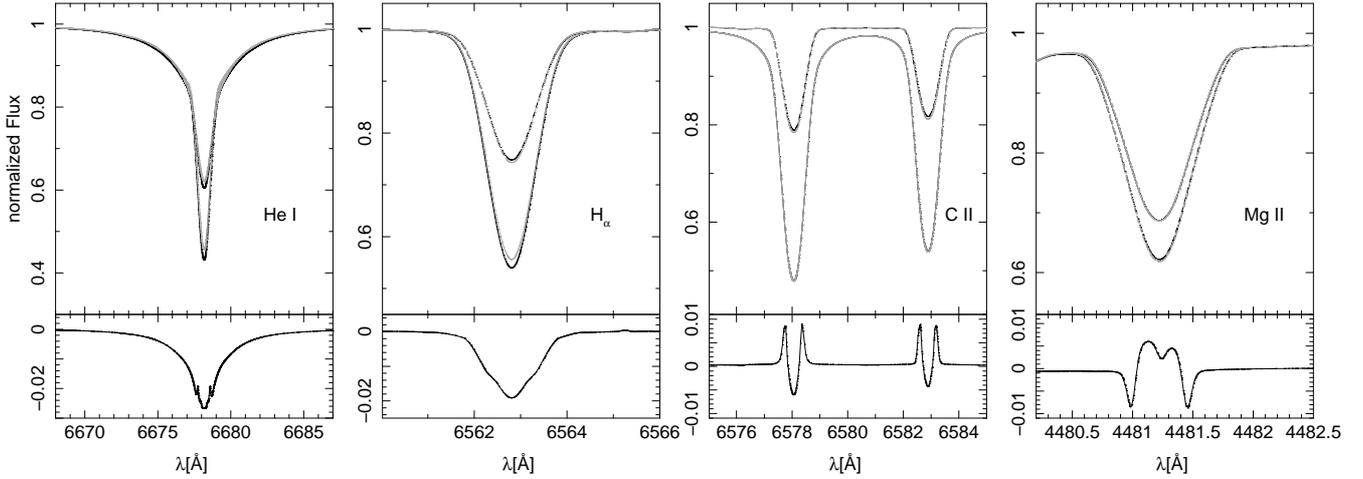}
   \caption{Computed line profiles with strong non-LTE effects using {\sc Atlas12}/{\sc Surface} (black) and {\sc Sterne}/{\sc Surface} (gray) for a helium-rich model with $T_{\rm eff}$\,=\,16\,800\,K and $\log g$\,=\,2.80 in LTE (dashed) and non-LTE (solid line). The lower panel shows the residuals between the non-LTE line profiles.}
  \label{fig:nonLTE_lines}
\end{figure*}

\begin{figure*}
\begin{minipage}[b]{0.48\textwidth}
  \includegraphics[width=0.99\textwidth]{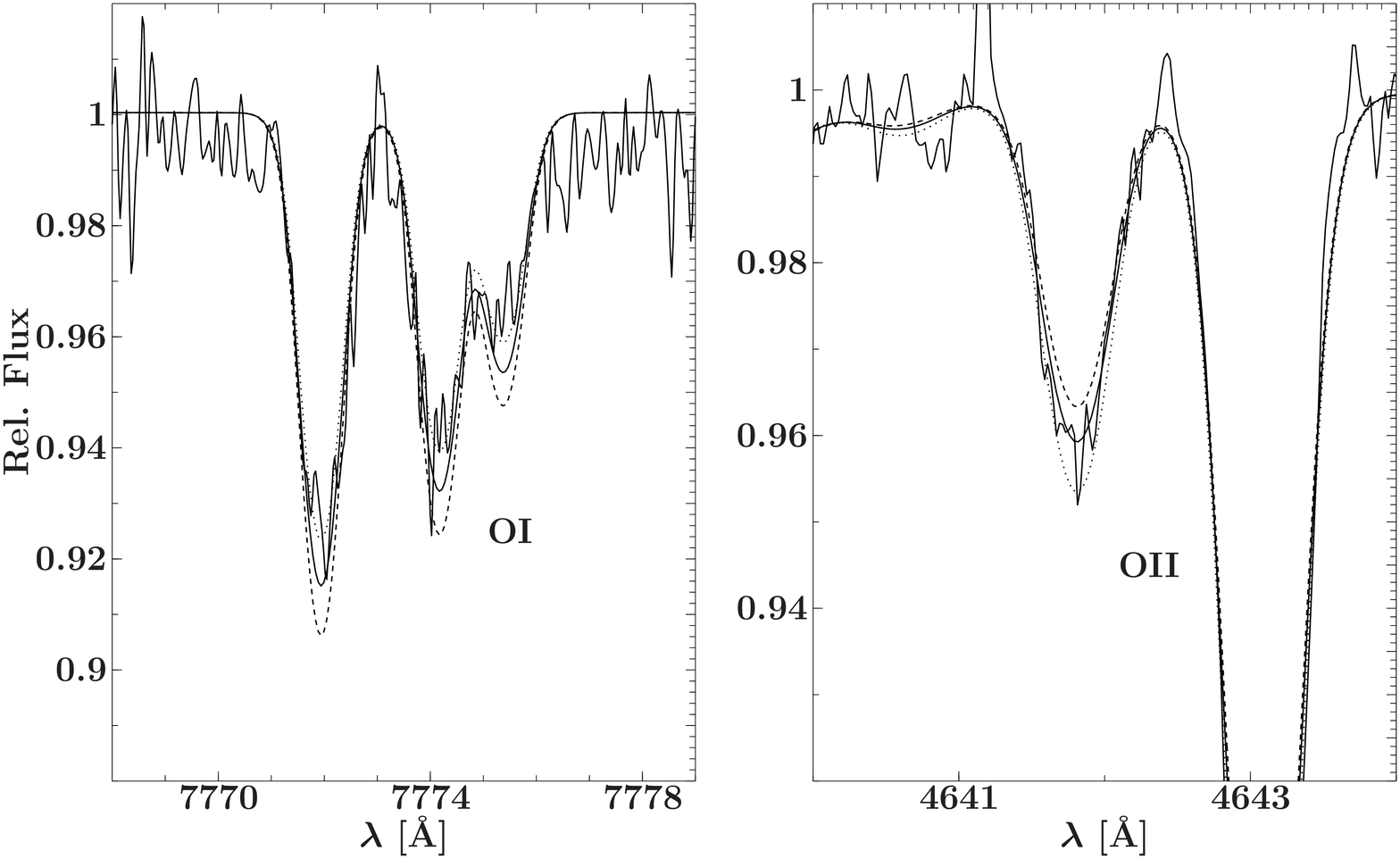}
    \end{minipage}
    \begin{minipage}[b]{0.48\textwidth}
    \includegraphics[width=0.99\textwidth]{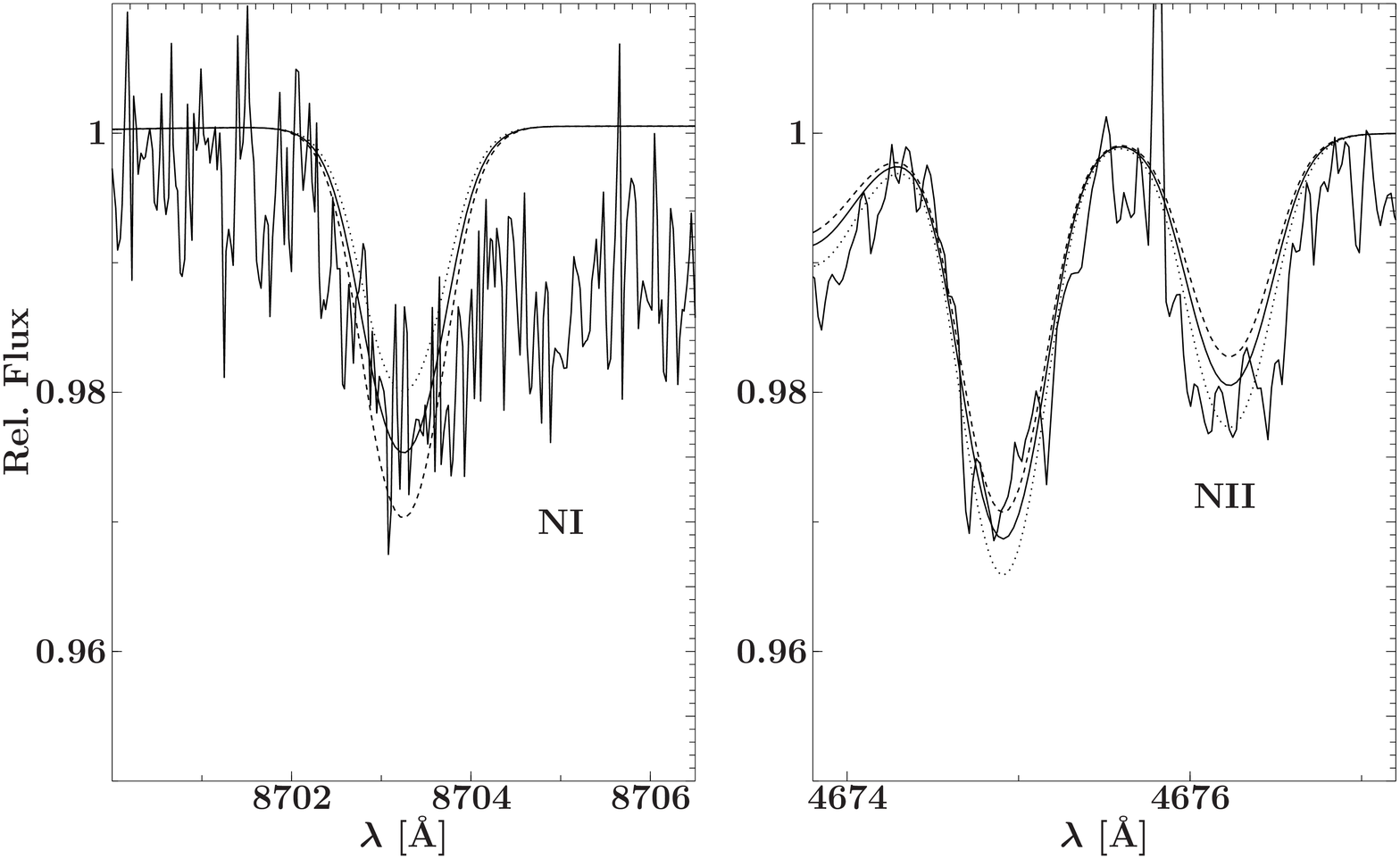}
    \end{minipage}
    \includegraphics[width=.965\textwidth]{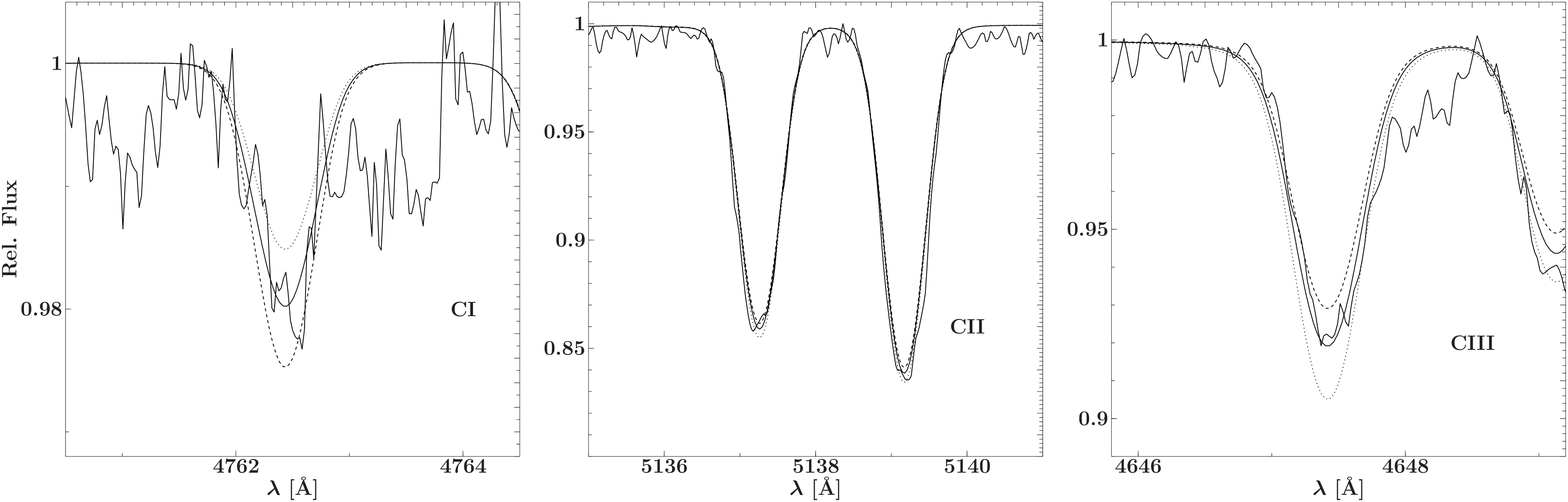}
   \caption{A selection of line profile fits for strategic ions shown to visualise our procedure to establish ionisation balance. Many additional lines of \ion{C}{ii}, \ion{N}{ii} and \ion{O}{ii} were included but are not shown here. Additional lines of \ion{C}{i} and \ion{N}{i} were also used. The dotted/dashed lines are computed line profiles for $T_\mathrm{eff}$ adjusted by +/$-$300\,K with regard to our atmospheric parameter solution.}
  \label{fig:ioni}
\end{figure*}

We compare the model atmosphere structures obtained with {\sc Atlas12} and {\sc Sterne3} by examining the temperature and electron density stratification as a function of mass. Figure~\ref{fig:density} shows the electron density stratification. Both codes match each other well. Small discrepancies can be seen in the temperature stratification (Fig.~\ref{fig:temperature}) especially in the region where line formation takes place, amounting to 2-3\% at maximum. 
The reason is likely due to the different opacities employed and other details in the model assumptions.
From our previous experiences of normal B-type stars \citep{nie07,prz11} one could expect that 
non-LTE effects on the atmospheric structure appear similarly small. Indeed, \citet{pan11} find for the case of BD+10$\degr$2179 that  the introduction of fully consistent non-LTE modelling (model atmosphere plus line analysis) has small effects on the atmospheric parameter determination as well as minor effects on many of the derived abundances when compared to a full LTE approach.  

The structure from the {\sc Atlas12} and {\sc Sterne3} models are the input for the non-LTE calculation which was done using the code {\sc Detail}. The comparison of the non-LTE flux distribution based on the {\sc Sterne3} and {\sc Atlas12} structures overall matches (Figs.~\ref{fig:fluxdistr}). A closer inspection shows that 
the {\sc Sterne3}-based flux is in some regions only slightly higher than the {\sc Atlas12}-based flux (Fig.~\ref{fig:fluxdistr2}), while it is systematically higher in other regions (Fig.~\ref{fig:fluxdistr3}) . The latter differences are explained by the higher local temperatures of the {\sc Sterne3} model at the depths where the 
UV/optical continuum is formed (see Fig.~\ref{fig:temperature}). The differences between the {\sc Detail} non-LTE fluxes on the one hand and the {\sc Atlas12} and {\sc Sterne3} LTE fluxes on the other hand are larger \footnote{In consequence, differences in the total emergent flux occur, e.g. by $\sim$1\% between the {\sc Atlas12} and {\sc Atlas12 + Detail} models, but this is only a minor factor for the error budget, i.e. the hybrid approach is confirmed to be applicable.}. The reasons for these are the non-LTE departures of the \ion{C}{i-iii} levels that determine the continuous opacity and the resolved resonance structure of the photoionisation cross-sections that is accounted for in the {\sc Detail} computations but not in the {\sc Atlas12} calculations. These lead to the jagged appearance of the extreme-UV flux, while the {\sc Atlas12} LTE flux based on parameterised photoionisation cross-sections is much smoother (broad resonance structures are accounted for in the {\sc Sterne3} LTE computations). We want to note that far-/extreme-UV spectroscopy of BD+10$\degr$2179 would offer a unique opportunity to investigate resonance structures in the photoionisation cross-sections of carbon, which dominates the continuous opacity at these wavelengths.

For a comparison of the line profiles, {\sc Surface} was employed to compute synthetic spectra based on the {\sc Sterne3} and {\sc Atlas12} model atmospheres. Including/excluding the {\sc Detail} calculation leads to non-LTE/LTE line profiles for otherwise exactly the same input data. We focused on two issues. First, how large are the non-LTE effects, and second, how well matched are the line-profiles when using either {\sc Atlas12} or {\sc Sterne3} models.
   
BD+10$^\circ$2179 is a blue supergiant and our calculations show that non-LTE effects can be quite strong, at least for some spectral lines (Fig.~\ref{fig:nonLTE_lines}). In non-LTE H$_{\alpha}$ appears to be twice as strong as in LTE. Not only are hydrogen and helium lines affected, lines of heavier elements are also susceptible to non-LTE effects. The strong \ion{C}{ii} doublet close to H$_{\alpha}$ is one example for a significant change in the line profile. Another interesting line is \ion{Mg}{ii} at 4481\,{\AA} (Fig. \ref{fig:nonLTE_lines}). 
Note that these are examples with strong non-LTE effects and that this is not the case for every line. Many features show only moderate to small non-LTE effects. Spectral lines experience both, a non-LTE increase of equivalent width (the majority of cases), as shown in Fig.~\ref{fig:nonLTE_lines}, as well as a reduction of equivalent width in a few cases. Overall, the non-LTE effects become more pronounced towards the red. A comprehensive comparison between LTE and non-LTE profiles for the analysed spectral lines will be made in the next section.

The differences between non-LTE line profiles using {\sc Sterne3} or {\sc Atlas12} are very small (lower panels of Fig. \ref{fig:nonLTE_lines}). Hence, the small discrepancies in the temperature stratification and the flux distribution have a small impact on the line profiles which means that the results from a quantitative spectroscopic analysis should not show significant differences using {\sc Atlas12} instead of {\sc Sterne3} models. In the following section the quantitative non-LTE analysis of BD+10$^\circ$2179 using {\sc Atlas12} models is described in full.
    
\section{Spectral analysis of BD+10$^\circ$2179}\label{sec:analysis}

\begin{table*}
\caption{Stellar parameters and elemental abundances of the non-LTE analysis of BD+10$^\circ$2179. Abundances $\epsilon$ with respect to \mbox{$\log\sum\mu_X\epsilon$($X$)\,=\,12.15}, where $\mu_X$ is the atomic weight of element $X$. } 
\centering
\begin{tabular}{lllllll}
\hline\hline
Stellar parameter& This work  & Pandey$^a$  & Heber$^b$ & Pandey$^c$ & Jeffery$^d$ & Sun$^e$\\
\hline
$T_{\rm eff}$ [K]  & $17\,300 \pm 300$ &  $16\,375 \pm 250$  & $16\,800 \pm 600$ & $16\,400 \pm 500 $ & $18\,500$ \\
$\log(g$[cm\,s$^{-2}$]) & $2.80 \pm 0.10$ &  $2.45 \pm 0.2$  & $2.55 \pm 0.2$ &  $2.35 \pm 0.2$ & $2.6$ (fixed)  \\
$v_{\rm rot}\sin i$ [km\,s$^{-1}$]  & $20 \pm 3$ &  -  &  $20 \pm 20$   & $20 \pm 2$ & -  \\
$\zeta$ [km\,s$^{-1}$] & $20 \pm 3$ & - & - & - & -  \\
$\xi$ [km\,s$^{-1}$]  & $3 \pm 1$ &   $7.5 \pm 1$   &  $7.5 \pm 1.5$  &  $6.5\pm 1$ & - & \\
\hline
Element/Ion  & Abundance & & &    \\
\hline
\vspace{0.05cm}
Hydrogen &  $8.36 \pm 0.02$ & $8.35 \pm 0.13$ & $8.5$ & $8.3$ &  $0$ (fixed) & 12.00\\
\vspace{0.05cm}
Helium  &  $11.53$ & $11.54$ & $11.53$ & $11.54$ &  $11.54$ (fixed)  & $10.93 \pm 0.01$ \\
\ion{C}{i}  &  $9.67 \pm 0.09$  &  $9.35 \pm 0.03$  & - & $9.3$ &  -\\
\ion{C}{ii}  &  $9.78 \pm 0.10$   &   $9.32 \pm 0.14$ & - & $9.3$ & - \\
\ion{C}{iii}  &  $9.72 \pm 0.04$  &  $9.29 \pm 0.00$  & - & $9.4$ & - \\
\vspace{0.05cm}
Carbon  &  $9.75 \pm 0.10$  &  -  & $9.54$ & $9.4$ &  $9.54$ (fixed)   & $8.43 \pm 0.05$\\
\ion{N}{i} & $7.93 \pm 0.06$  & - & - & - & - \\
\ion{N}{ii} & $8.04 \pm 0.10$ &  $8.10 \pm 0.14$ & - & $7.9$  & - \\
\vspace{0.05cm}
Nitrogen&  $8.03 \pm 0.10$ & -  &  $8.11$ & $7.9$ &   $7.9$ (fixed)   &  $7.83 \pm 0.05$\\
\ion{O}{i} & $7.40 \pm 0.00$  & - & - & - &  - \\
\ion{O}{ii} & $7.52 \pm 0.04$ & $7.93 \pm 0.21$ & - & $7.5$ & - \\
\vspace{0.05cm}
Oxygen  &   $7.51 \pm 0.05$  &  - &  $8.1$ & $7.5$ &   - & $8.69 \pm 0.05$\\
\vspace{0.05cm}
Neon   &   $8.00 \pm  0.09$ &  $7.87 \pm 0.10$  & - & - & -  & $7.93 \pm 0.10$\\
\vspace{0.05cm}
Magnesium&  $6.97 \pm  0.05$  &  -  & $8.02$ & $7.2$ & - & $7.60 \pm 0.04$\\
\ion{Al}{ii} & $5.83 \pm 0.03$  & - & - &  -  & - \\
\ion{Al}{iii} & $5.84 \pm 0.05$ & - & - & $5.6$ & - \\
\vspace{0.05cm}
Aluminium &  $5.84 \pm 0.04$  &  -  & $6.25$ & $5.7$ & -  & $6.45 \pm 0.03$\\
\ion{Si}{ii} & $7.12 \pm 0.07$  & - & - &  $6.5$  & - \\
\ion{Si}{iii} & $7.15 \pm 0.05$ & - & - & $6.8$  & - \\
\vspace{0.05cm}
Silicon &  $7.14 \pm 0.06$  &  -  & $7.32$ & $6.8$   & $5.4$ & $7.51 \pm 0.03$\\
\vspace{0.05cm}
Sulphur &  $6.84 \pm 0.05$  &  -  & $7.12$ & $6.5$ &  - & $7.12 \pm 0.03$\\
\vspace{0.05cm}
Argon &  $5.93 \pm 0.07$  &  -  & $6.4$ & $6.1$ & -  &  $6.40 \pm 0.13$\\
\ion{Fe}{ii} & $6.53 \pm 0.11$  & - & - & $6.2$ & - \\
\ion{Fe}{iii} & $6.57 \pm 0.09$ & - & - & $6.3$  & - \\
Iron  &     $6.55 \pm 0.10$  &  - & $6.49$ & $6.2$  & $6.2$ & $7.50 \pm 0.04$\\
\hline
\vspace{-0.3cm}
\end{tabular}
\begin{flushleft}
\textbf{Notes.}~~~$^a$\,\citet{pan11}, $^b$\,\citet{heb83}, $^c$\,\citet{pan06}, $^d$\citet{jef10}, $^e$\,\citet{asp09}, photospheric values\\
\end{flushleft}
\label{tab:abundances_nlte}
\end{table*}

\begin{table}
\caption{Mass fractions of the elements in BD+10$^\circ$2179}
\centering
\begin{tabular}{c c c r}
\hline\hline
X & Solar$^{a}$  & This work & [X/Fe] \\
\hline
H & 73.7$\cdot$10$^{-2}$  & 1.65$\cdot$10$^{-4}$ &$-$2.69\\
He& 24.9$\cdot$10$^{-2}$  & 94.9$\cdot$10$^{-2}$ &  1.55\\
C & 2.36$\cdot$10$^{-3}$  & 4.81$\cdot$10$^{-2}$ &  2.27\\
N & 6.93$\cdot$10$^{-4}$  & 1.07$\cdot$10$^{-3}$ &  1.15\\
O & 5.73$\cdot$10$^{-3}$  & 3.67$\cdot$10$^{-4}$ &$-$0.23\\
Ne& 1.26$\cdot$10$^{-3}$  & 1.42$\cdot$10$^{-3}$ &  1.02\\
Mg& 7.08$\cdot$10$^{-4}$  & 1.84$\cdot$10$^{-4}$ &  0.32\\
Al& 5.56$\cdot$10$^{-5}$  & 1.31$\cdot$10$^{-5}$ &  0.34\\
Si& 6.65$\cdot$10$^{-4}$  & 2.73$\cdot$10$^{-4}$ &  0.58\\
S & 3.09$\cdot$10$^{-4}$  & 1.57$\cdot$10$^{-4}$ &  0.67\\
Ar& 7.34$\cdot$10$^{-5}$  & 2.93$\cdot$10$^{-5}$ &  0.48\\
Fe& 1.29$\cdot$10$^{-3}$  & 1.42$\cdot$10$^{-4}$ &   --  \\
\hline
\end{tabular}
\begin{flushleft}
$^a$ Solar abundances from Table\,1 of \citet{asp09}.
\end{flushleft}
\label{table:abundance_nlte1}
\end{table}

The helium-rich composition of BD+10$^\circ$2179 changes the line-to-continuum opacity ratio and together with the high abundances of some key elements (see below) leads to a very rich line spectrum compared to a supergiant of similar stellar parameters with a normal (hydrogen-rich) composition. 
A large number of indicators is therefore available for the atmospheric parameter determination of the star. The hydrogen Balmer lines H$_\alpha$ to H$_\delta$ were used to constrain the hydrogen abundance, all major helium lines in the optical spectrum and the ionisation equilibria of \ion{C}{i/ii/iii}, \ion{N}{i/ii}, \ion{O}{i/ii}, \ion{Al}{ii/iii}, 
\ion{Si}{ii/iii} and \ion{Fe}{ii/iii} (requiring that the same elemental abundance is derived from each ion within the uncertainties) were employed simultaneously to determine the effective temperature and surface gravity in an iterative way. At all stages, line-profile fits to the observation were used, with each spectral line given the same weight.
A good match of all indicators was found for $T_{\rm eff}$\,=\,17\,300\,$\pm$\,300\,K and $\log g$\,=\,2.80\,$\pm$\,0.10. Figure~\ref{fig:ioni} visualises the process to establish ionization balance. The different lines from the individual ionization stages react differently to modifications of $T_\mathrm{eff}$, in this case to variations of $\pm$300\,K. Similar reactions were found with respect to $\log g$-changes. Other parameters like the microturbulent velocity $\xi$, the (projected) rotational velocity $v_{\rm rot}\sin i$ and the (radial-tangential) macroturbulent velocity $\zeta$ \citep[e.g.][p. 433 ff.]{Gray05} also had to be constrained within the iterative procedure. Standard techniques were used for these,
such as requiring elemental abundances to be independent of the strengths of the lines and minimising residuals from the comparison of synthetic with observed line profiles for varying $v_{\rm rot}\sin i$ and $\zeta$ \citep{FiPr12}.
Finally, abundances for all remaining elements were derived, again using line-profile fits. In total, we derived non-LTE abundances for 12 elements, so far, the most comprehensive non-LTE study of an EHe star. This covers almost all chemical species that show lines in the optical spectrum except for calcium and phosphorus. 

The atmospheric parameters and average abundances for the measured ions, plus values for the total elemental abundance, are summarised in Table~\ref{tab:abundances_nlte}, where results from the present work are also compared with data from the literature \citep{heb83,pan06,jef10,pan11}, and with solar abundances \citep[on the usual abundance scale]{asp09}. The uncertainties stated are 1$\sigma$ standard deviations which, with values of typically 0.04 to 0.10\,dex, are comparable to those obtained for other object classes using the same analysis techniques. Systematic uncertainties of the abundance values due to uncertainties in the atmospheric parameters, continuum normalisation and atomic data (oscillator strengths, photoionisation and collisional cross sections) can be expected to be of the same order of magnitude, based on our experiences for related objects \citep[see e.g.][]{Przybillaetal00,Przybillaetal01a,Przybillaetal01b,PrBu01}.
Note that the standard error of the mean abundance -- often stated in the context of abundance analyses -- would be much lower in many cases, as typically many lines are analysed per ion.
Results from the line-by-line analysis are summarised in Table~A1 which is available in the electronic version of the article. 
There, the transitions are listed by ascending wavelength $\lambda$ per ion. Excitation energies $\chi$ of the lower level of the transition are given, as well as the oscillator strength $\log gf$, an accuracy indicator for the $\log gf$-value and the source of the oscillator strength is identified. The final entry is the non-LTE line abundance $\epsilon$ normalised to $\log\sum\mu_X\epsilon$($X$)\,=\,12.15, where $\mu_X$ is the atomic weight of element $X$. Note that apparently empty entries have been computed as blends, such that only one abundance value is derived from multiple components (only blends with lines from the same ion were considered in this case). Sources of Stark broadening parameters are also summarised in Table~A1.   

\begin{figure*}
\centering
\includegraphics[width=.94\textwidth]{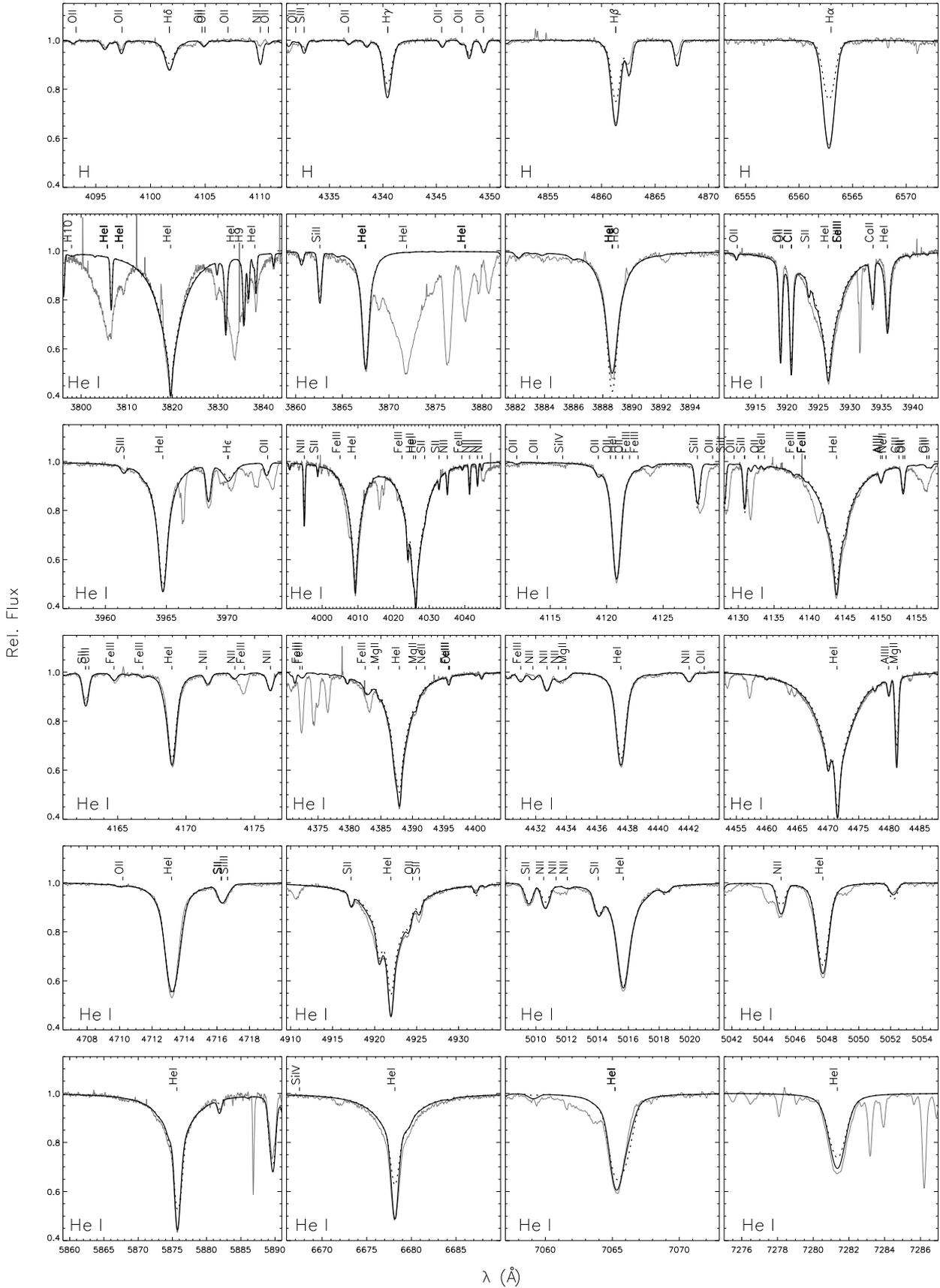}\\[-5mm]
\caption{Fits to the observed optical hydrogen and helium line spectrum. The panels show the {\em global} best-fit non-LTE model (black line), the corresponding LTE model (dotted line) and the FEROS spectrum (grey line). Line identifiers are given on top of the spectra. See the text for a discussion.
}
\label{fig:hhefits}
\end{figure*}

For comparison with the Sun, mass fractions were also calculated, see Table~\ref{table:abundance_nlte1}. 
The chemical composition is dominated by helium and carbon, by $\sim$95\% and $\sim$5\% in mass fraction, followed by nitrogen and neon, both contributing about 0.1\% in mass fraction. All other elements are trace species. Using iron as metallicity substitute, BD+10$^\circ$2179 is confirmed to be in fact a metal-poor star of about 1/10th solar metallicity, showing signatures of $\alpha$-enhancement to a degree characteristic for Population~II stars. An exception is oxygen, which is depleted, probably because of nuclear processing. 

A consequence of the tightly constrained atmospheric parameters and the low scatter of abundance values within the individual elements is a close match of the resulting synthetic and observed spectra over wide wavelength ranges. Examples centered around the hydrogen Balmer and the classical optical \ion{He}{i} lines are shown in Fig.~\ref{fig:hhefits}. One has to stress that the comparison is based on {\em one} model spectrum based on {\em one} set of atmospheric parameters and elemental abundances. Nevertheless the quality of the agreement between model and observation is good to excellent. In particular, the broadening of the \ion{He}{i} lines in the optical is described well by the data provided by \citet{GBKO62}, \citet{Griem64}, \citet{BCS69}, \citet{Sha69} and \citet{DSB90}.

\begin{figure}
   \centering
   \includegraphics[width=0.48\textwidth]{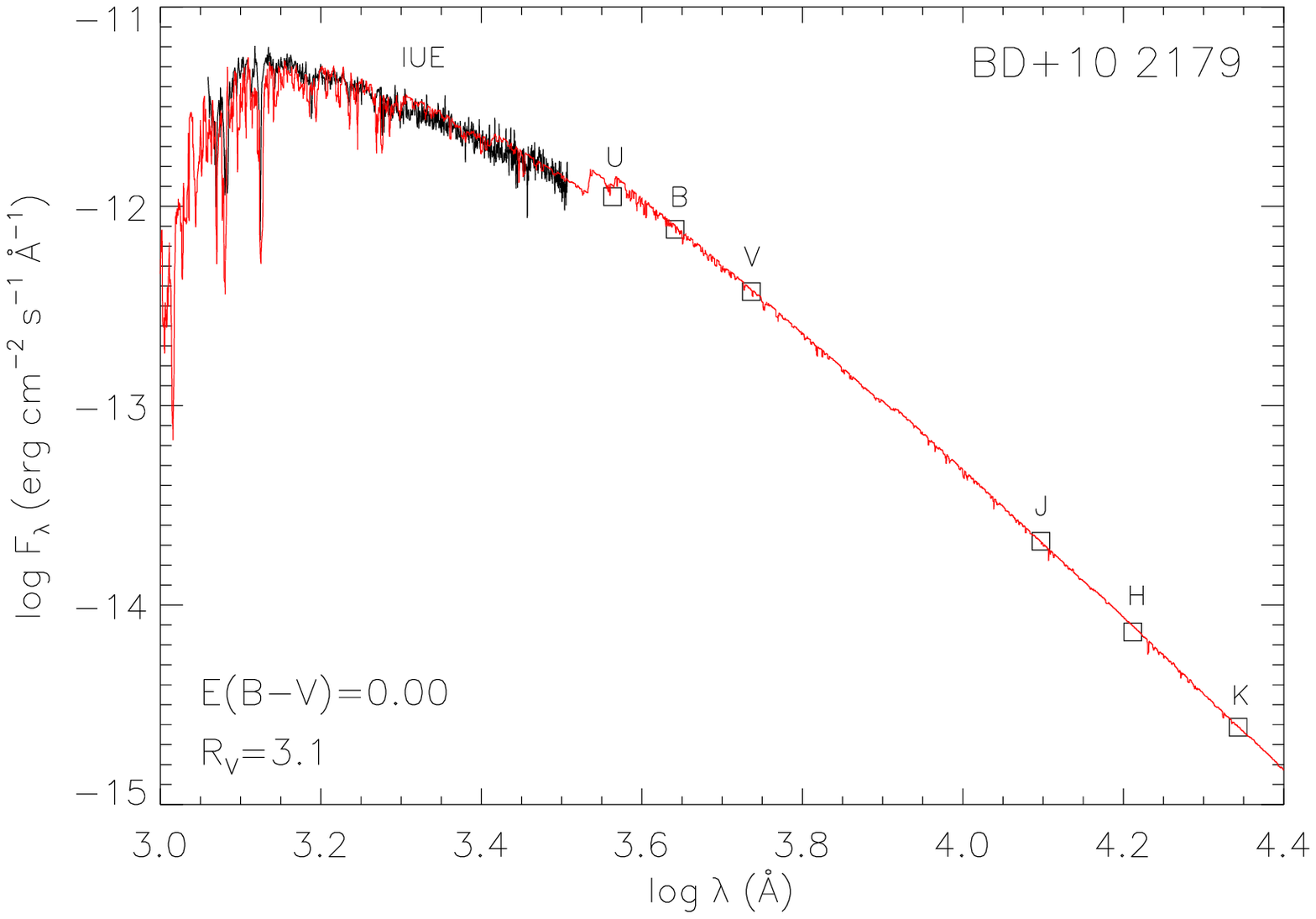}
   \vspace{0.2cm}
      \caption{Spectral energy distribution as a further indicator for $T_{\rm eff}$. 
      Observations (black line and open squares) are over plotted by a synthetic spectrum (red) computed with {\sc Detail} for $T_{\rm eff}$\,=\,17\,300\,K and $\log g$\,=\,2.80.}
         \label{fig:SED}
   \end{figure}
   
For comparison, the corresponding LTE line profiles are also shown in Fig.~\ref{fig:hhefits}. Non-LTE effects strengthen the hydrogen Balmer lines, with the effect diminishing towards the higher series members. Non-LTE effects on the \ion{He}{i} line spectrum are more complex in BD+10$\degr$2179: while the sharp lines in the optical blue are virtually unaffected, the diffuse lines are slightly broader, the red lines are markedly deepened and \ion{He}{i} $\lambda$3889\,{\AA} becomes shallower. 

The broader picture, comprising the metal lines investigated in such detail for the first time, is shown in Figs.~A1--A4 which is available in the electronic version. The \ion{C}{i} lines are systematically weakened in non-LTE (even weak lines), a consequence of the overionisation of this species. Depending on the lower and upper levels involved in the \ion{C}{ii} transitions, one may find non-LTE strengthening or weakening, while some lines are unaffected. The few observed \ion{C}{iii} lines are overall strengthened by non-LTE effects. Lines of \ion{N}{i} and \ion{O}{i}, and about half of the \ion{N}{ii} lines are strengthened by non-LTE effects, while the remainder of the \ion{N}{ii} and the \ion{O}{ii} are close to LTE. The overall largest non-LTE effects are found for \ion{Ne}{i}, with the lines systematically strengthened by large amounts. 
The majority of the (weak) lines of the heavier elements are close to LTE, with the exceptions of several stronger lines, see in particular Figs.~\ref{fig:alphafits} and \ref{fig:sifefits}.
We note that the non-LTE effects found here for BD+10$\degr$2179 cannot be generalised to the whole parameter space spanned by the class of EHe stars.

On the other hand, the observed spectrum contains many lines not included in the model, a situation different to B-type stars of normal composition \citep[see e.g. Figs.~8--11 of][]{NiPr12}. Among the metal lines the reason is the large number of \ion{C}{ii} lines that appear because of the high carbon abundance and because of the unusual line-to-continuum opacity ratio in this star. These are transitions between high-lying energy levels,
which are missing in the available model atom \citep{NiPr06,nie08}. An extension of the model atom is beyond the scope of the present work. This needs to be addressed in a separate study. Another large number of transitions missing in the spectrum synthesis stems from \ion{He}{i}. Again, missing energy levels in the available model atom \citep{prz05} play a role -- lines to upper levels with principal quantum number $n$\,=\,20 are observed, whereas the model atom considers levels up to $n$\,=\,8 only. The absence of reliable line-broadening data for these transitions in the literature is also crucial. Examples in the optical blue can be seen in the first two panels of the second row of Fig.~\ref{fig:hhefits} while the blue wing of \ion{He}{i} $\lambda$7065\,{\AA} is an example of the deficits in the optical red spectrum. Otherwise, only the forbidden \ion{He}{i} components predicted by \citet{BeWe98} are missing in the model of the optical spectrum. Their line-broadening tabulations for the appropriate low densities are not available to us. 
A detailed description of the \ion{He}{i} spectrum of BD+10$^\circ$2179 outside the classical optical range can be found in the Appendix which is available in the electronic version. In total dipole-allowed transitions from almost 150 multiplets are detected within our spectrum, plus 11 more in the HST STIS range. In addition, 23 forbidden components of \ion{He}{i} are identified -- many for the first time (see Appendix).

Other observed lines not accounted for by the modelling are a few narrow interstellar lines like the Na D and Ca H\&K lines, and the telluric absorption spectrum. Narrow features near H$\alpha$ and \ion{He}{i} $\lambda\lambda$\,7065 and 7281\,{\AA} are examples of the latter in Fig.~\ref{fig:hhefits}.

Finally, one can also verify the choice of the correct atmospheric parameters by comparison of the predicted with the observed spectral energy distribution on a global scale. The non-LTE flux as computed with {\sc Detail} is compared to the observed SED in Fig.~\ref{fig:SED}.
Good agreement is found under the assumption of zero reddening, strengthening the conclusions obtained on atmospheric parameters from the analysis of the \ion{He}{i} line wings and metal line ionization equilibria. The very weak interstellar lines -- equivalent widths of the Na D$_1$ and D$_2$ lines are e.g. 104 and 57\,m{\AA}, respectively -- are indeed consistent with a very low reddening. 
\citet{SchFi11} give an $E$($B-V$)\,=\,0.02 towards the direction of BD+10$\degr$2179 obtained from a recalibration of the Milky Way foreground measured by the Cosmic Microwave Background Explorer (COBE), which sets an upper limit.

\begin{table}
\caption{Fundamental parameters of BD+10$^\circ$2179}
\centering
\begin{tabular}{ll}
\hline\hline
Quantity & Value\\
\hline
$\theta$ (10$^{-5}$\,arcsec) & 1.62$\pm$0.05\\
$\log L$/$M$ (solar units)   & 3.55$\pm$0.10\\
$M$/$M_\odot$                & 0.55$\pm$0.05\\
$R$/$R_\odot$                & 4.9$\pm$0.6\\
$\log L$/$L_\odot$           & 3.29$\pm$0.06\\
\hline
\end{tabular}
\vspace{-2mm}
\label{tab:fundamental}
\end{table}

\section{Fundamental parameters and kinematics}\label{kinematics}
Currently, a meaningful parallax measurement, and therefore the distance $d$, of BD+10$^\circ$2179 is unavailable. As a consequence, we cannot infer the fundamental parameters mass $M$, radius $R$ and luminosity $L$ directly on the basis of our atmospheric parameters alone. Only the luminosity-to-mass ratio $L$/$M$ and the angular diameter $\theta$ can be determined. The latter is obtained from the scaling factor required to normalise the theoretical fluxes obtained from a stellar atmosphere calculation to those observed (Fig.~\ref{fig:SED}). Our values for $\log L$/$M$ and $\theta$ (see Table~\ref{tab:fundamental}) are both somewhat smaller than those derived by \citet{heb83}, a consequence of our higher $\log g$ and $T_{\rm eff}$ respectively. 

\begin{figure*}
\includegraphics[width=0.48\textwidth]{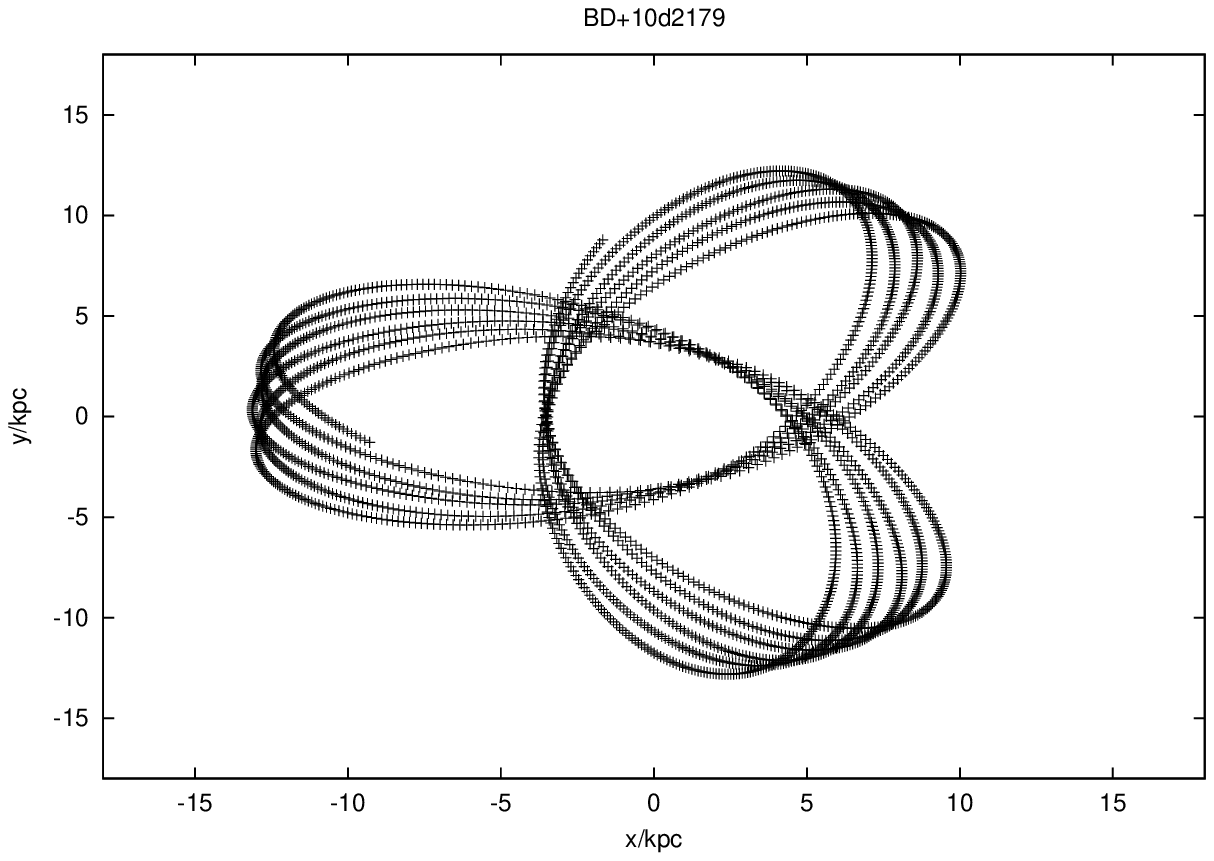}\hfill
\includegraphics[width=0.48\textwidth]{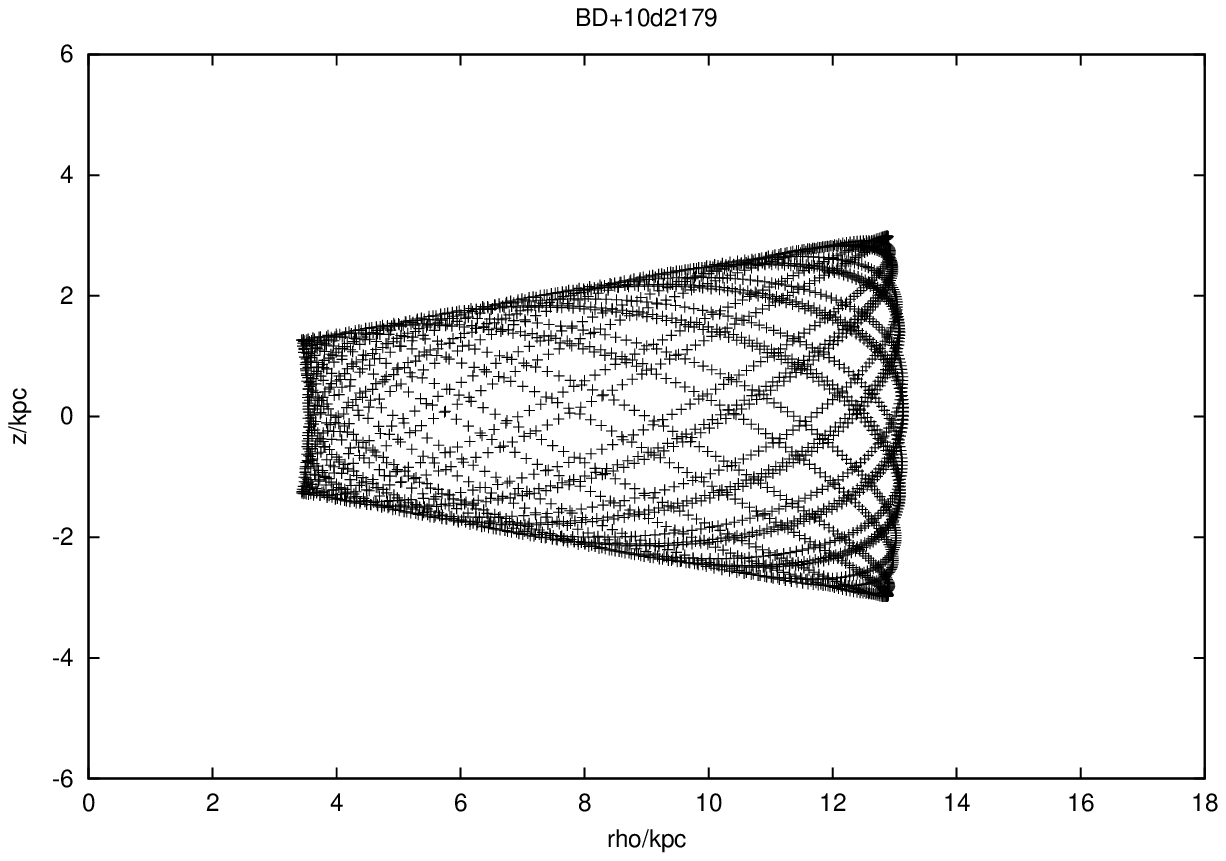}\\
\includegraphics[width=0.48\textwidth]{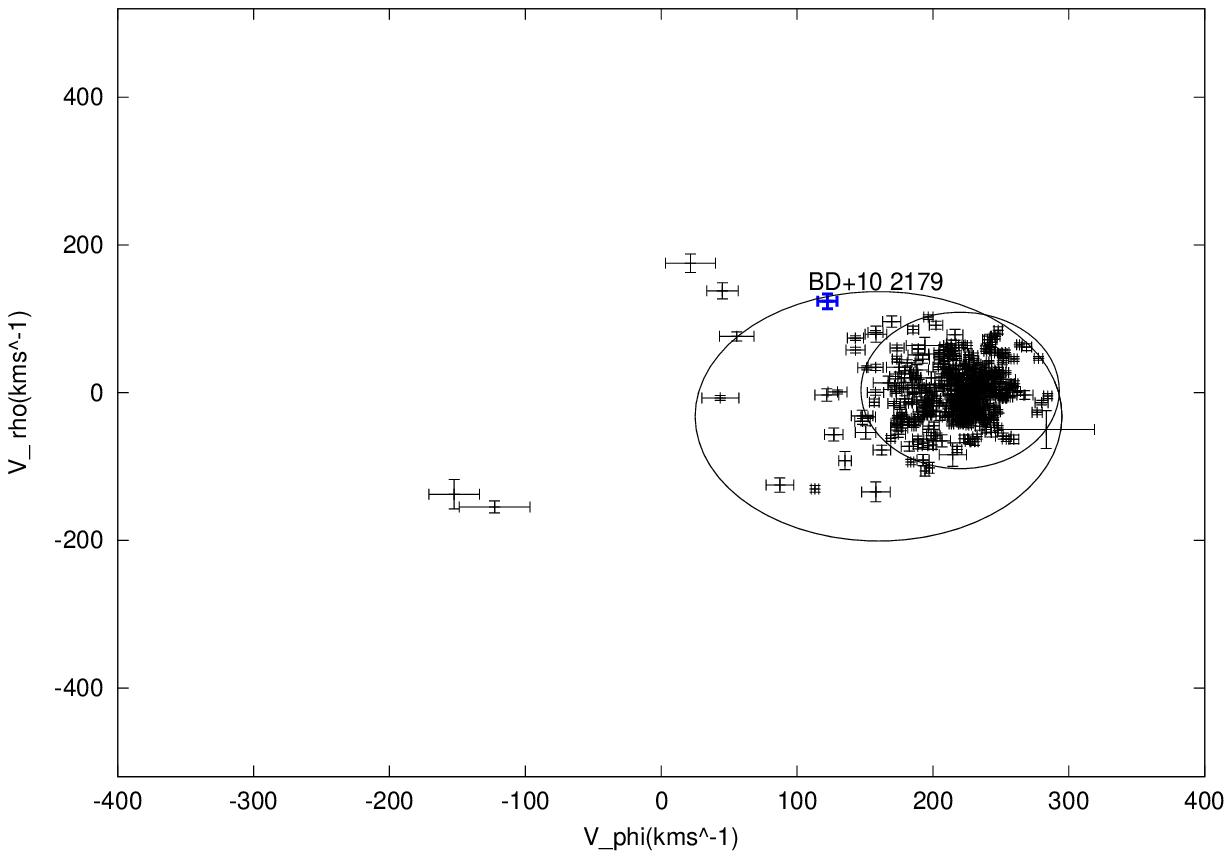}\hfill
\includegraphics[width=0.48\textwidth]{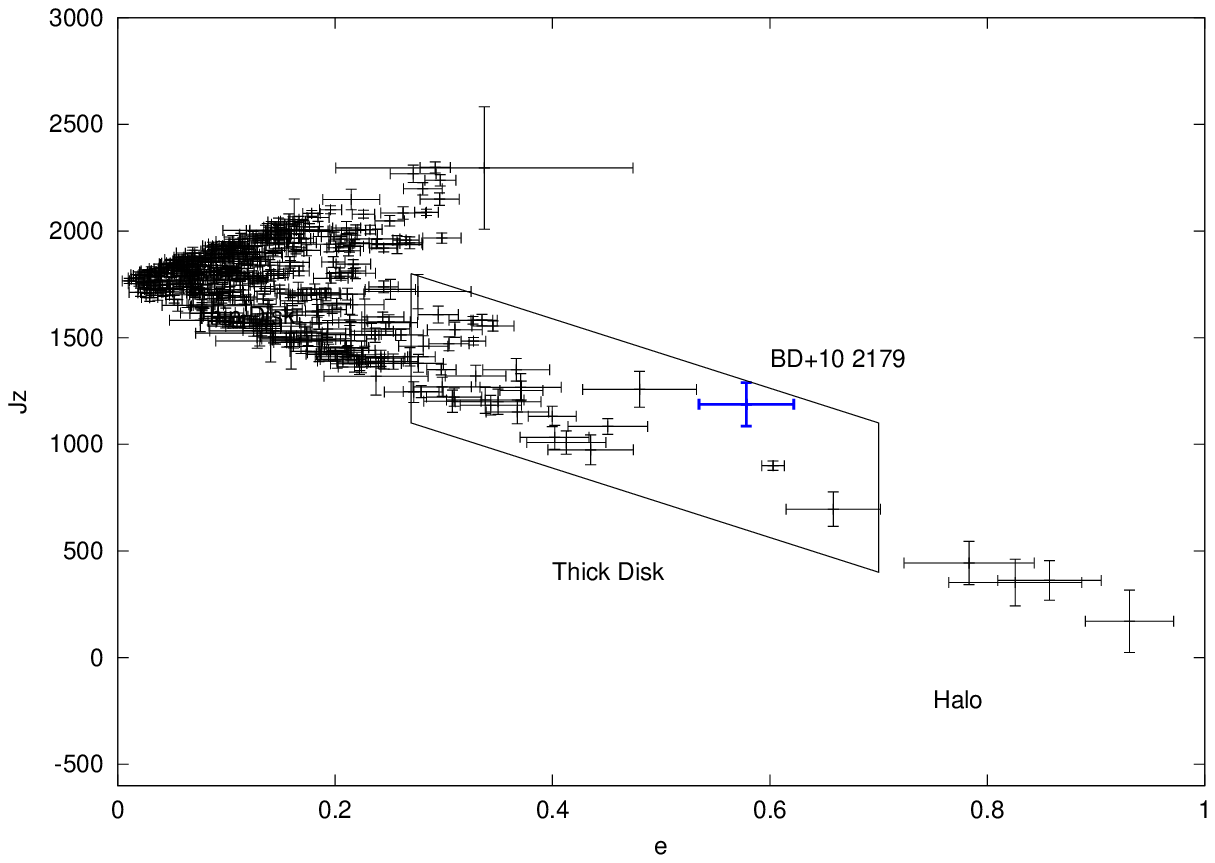}\\
   \vspace{0.2cm}
      \caption{Kinematics of BD+10$^\circ$2179 in comparison with the white dwarf sample of \citet{pauli06}:
      Upper left panel: orbital motion of BD+10$^\circ$2179 in the Galactic plane. Upper right panel: meridional orbit. Integrations in time were done over the interval of 0 to $-$3\,Gyr. Lower panels: 
       $V_\phi$--$V_\rho$ (left) and $e$--$J_{z}$ diagrams (right). The solid ellipses render the 3$\sigma$ thin and thick disk contours in the $V_\phi$--$V_\rho$ diagram, while the solid box in the $e$ -- $J_{z}$ marks the thick disk region as specified by \citet{pauli06}.} 
         \label{fig:kinematics}
   \end{figure*}

\begin{table}
\caption{Kinematical data for BD+10$^\circ$2179}
\centering
\begin{tabular}{lr|lr}
\hline\hline
Quantity & Value & Quantity & Value\\
\hline
$\alpha$\,(J2000)$^1$                & 10:38:55.23523   & $X$ \,(kpc)              & $-$9.3\\
$\delta$\,(J2000)$^1$                & +10:03:48.4975   & $Y$\,(kpc)               & $-$1.2\\
$d$\,(kpc)$^2$                       & 2.6$\pm$0.3      & $Z$\,(kpc)               & 2.1\\     
$v_\mathrm{rad}$\,(km\,s$^{-1}$)$^2$ & 155$\pm$1        & $V_\rho$\,(km\,s$^{-1}$) & 124$\pm$10\\
$\mu_\alpha$\,(mas\,yr$^{-1}$)$^1$   & $-$11.5$\pm$0.6  & $V_\phi$\,(km\,s$^{-1}$) & 122$\pm$7 \\
$\mu_\delta$\,(mas\,yr$^{-1}$)$^1$   & $-$3.9$\pm$0.6   & $W$\,(km\,s$^{-1}$)      & 49$\pm$9\\
\hline
\end{tabular}
\vspace{-2mm}
\begin{flushleft}
$^1$\, Gaia Attitude Star Catalog, \citet{2016A&C....15...29S}; 
$^2$\,this work
\end{flushleft}
\label{tab:kinematics}
\end{table}

In a second step, one can constrain the mass by comparison with predictions from stellar models. The observed surface abundances indicate the presence of both CNO- and 3$\alpha$-processed material. Therefore stellar models based on the accretion of a He white dwarf by a CO white dwarf, such as those of \citet{saio02}, might be considered. The next section discusses this more fully. A comparison in the $\log T_\mathrm{eff}$-$\log g$ diagram \citep[see Fig.~6 of][]{saio02} requires a small amount of extrapolation, as the surface gravity is higher than that of the lowest mass model available (the merger of a 
0.5\,$M_\odot$ CO with a 0.1\,$M_\odot$ He white dwarf), yielding $\sim$0.55\,$M_\odot$. In order to account for possible systematics because of model limitations for this complex evolutionary scenario, we estimate the mass uncertainty to be 0.05\,$M_\odot$. This together with the atmospheric parameters constrains $R$ and $L$, see Table~\ref{tab:fundamental}.

Further constraints on the nature of BD+10$^\circ$2179 can come from the analysis of its kinematics.
The input data for the orbital calculations in the Galactic potential are summarized in the left columns 
of Table~\ref{tab:kinematics}: equatorial coordinates $\alpha$, $\delta$, $d$, radial velocity $v_\mathrm{rad}$, and proper motions $\mu_\alpha$, $\mu_\delta$. The distance was computed following \citet{ram01}. We employed the approach of \citet{pauli06}. The code of \citet{OdBr92} was used for the computation of the orbit and the kinematic parameters; it uses a Galactic potential by \citet{AlSa91} as revised by \citet{Irr13}. The orbit was integrated from the present to 3\,Gyr into the past. 
The right columns of Table~\ref{tab:kinematics} summarize the output, Galactic coordinates $X$, $Y$, $Z$ and the velocity components $V_\rho$, $V_\phi$, $W$. The kinematics of BD+10$^\circ$2179 are visualised in Fig.~\ref{fig:kinematics} where the upper two panels show the orbit projected onto the $X$-$Y$ and the $\rho$\,=\,$\sqrt{X^2+Y^2}$-$Z$ planes.
The lower two panels show the projection of the velocity components of BD+10$^\circ$2179 in the $V_\phi$-$V_\rho$ plane and in the $e$-$J_Z$ plane, where 
$V_\phi$ is the velocity component in the direction of Galactic rotation, $V_\rho$ the component in Galactic radial direction\footnote{$V_\phi$  and $V_\rho$ are often referred to as $V$ and $U$, but may be confused with the cartesian velocities $v_x$  and $v_y$ \citep{1987AJ.....93..864J}, see also \citet{2015A&A...576A..65R}.}.
$J_z$ the component of the angular momentum of the star's Galactic orbit perpendicular 
to the Galactic disk and $e$ the eccentricity of the Galactic orbit.
In both cases a comparison with the kinematic properties of the white dwarf sample of \citet{pauli06} is made in order to further constrain the population membership of BD+10$^\circ$2179.
We conclude that BD+10$^\circ$2179 is a member of the Galactic thick disk population, which is also consistent with our inferred metallicity [Fe/H]\,$\approx$\,$-$1 and $\alpha$-enhancement.

\section{Discussion}\label{discussion}

\subsection{Reassessment of Surface Properties}

Overall, the present reanalysis of BD+10$^\circ$2179 based on improved spectra and models 
is  consistent with the findings of previous work (see Table~\ref{table:abundance_nlte1}), 
Only the present values for $T_\mathrm{eff}$ and $\log g$ are slightly higher than inferred earlier,
as is the carbon abundance. The evolutionary mass  is slightly lower than indicated previously. A significantly lower microturbulence value is found, a consequence of the consistent non-LTE modelling where increasingly stronger non-LTE effects in stronger lines remove the need for large 
microturbulences found in LTE analyses (see Table~\ref{tab:abundances_nlte}).

In the cases where spectral features are included in our line-formation calculations, a good simultaneous reproduction of the observed spectrum is achieved (Figs.~\ref{fig:hhefits}, A1--A4).
This level of agreement between model and observation is certainly the most satisfactory outcome of the present study. However, contrary to stars with normal chemical composition, it became apparent that many observed spectral features are still unaccounted for by our model computations. The very high abundances of helium and carbon produce a plethora of lines that are unobservable under any other circumstances. The particular case of helium is addressed in detail in the Appendix. The quantitative modelling of the complete \ion{He}{i} spectrum present in our observational data has to await an extension of the neutral helium model atom to include all levels up to at least principal quantum number $n$\,=\,20 (and all transitions -- radiative and collisional -- connecting them) and the provision of appropriate line-broadening data. The atomic data required are largely unavailable at present and their calculation will be a considerable challenge to atomic physicists. Also, a level-dissolution formalism in analogy to that of \citet{Hum94} will need to be implemented for an improved treatment of the He\,{\sc i} line overlap near the series limits.

Detailed studies of BD+10$^\circ$2179 covering over half a century open up the possibility to look for secular changes in the spectrum associated with its suspected rapid  evolution. Indeed, comparing the values of  equivalent widths published by \citet{heb83} and based on a photographic spectrum 
taken 1975, with data published  by \citet{pan06}, based on a CCD spectrum taken in 1998, one may suspect atmospheric parameter changes. For example,  
equivalent widths of the Si\,{\sc ii} $\lambda\lambda$ 3853\,{\AA} and 3862\,{\AA} and the Si\,{\sc iii} 
$\lambda\lambda$ 4567\,{\AA} and 4574\,{\AA} lines differ by more than 50\%. However, such pronounced changes 
are not supported by model calculations within reasonable atmospheric parameter variations. Moreover, when we measured equivalent widths from a 1985 CASPEC spectrum \citep[as discussed by][]{jef10} and in our 2006 FEROS spectrum, employing the same criteria for continuum placement and integration limits, consistent 
values resulted within the measurement uncertainties. We conclude that a decisive answer cannot be given here
and suggest that this issue should be addressed based on a consistent reassessment of published high-quality spectra (which are not all available here), plus   additional future observations to cover a timeline of a few decades with CCD spectroscopy.

\begin{figure}
   \centering
   \includegraphics[width=0.5\textwidth]{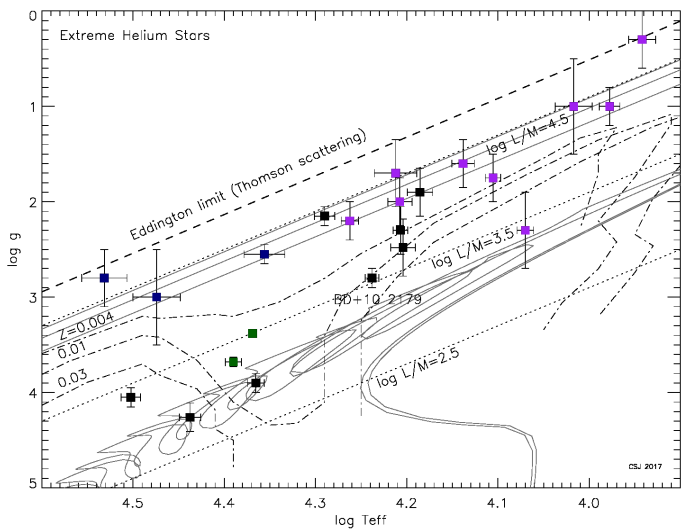}
   \vspace{0.2cm}
      \caption{The $\log T_\mathrm{eff}$-$\log g$ diagram for 
extreme helium stars, including the
position of the Eddington limit (Thomson scattering: dashed), 
luminosity-to-mass contours (solar units: dotted) and lower boundaries for pulsation 
instability (metallicities $Z$\,=\,0.004, 0.01, 0.03: dot-dashed) \citep{jeffery99}. 
In the online version, variable EHes are shown in purple (cool), blue (hot), green (V652\,Her like variables). Non-variables are black. 
Apart from BD+10$^\circ$2179, sources for $T_{\rm eff},g$ are as in \citet{jeffery08.ibvs}.
Post-merger evolution tracks for models of CO+He white dwarf mergers \citep[0.6+0.1, 0.6+0.2 and 
0.6+0.3\,$M_\odot$]{saio02} and He+He white dwarf mergers \citep[0.30+0.25 and 0.30+0.30\,$M_\odot$]{zhang12b} are shown in dark gray.}
         \label{fig:tracks}
   \end{figure}

\subsection{Evolutionary Inferences}

The  problem posed by  BD+10$^\circ$2179 and other EHes is the question of  their evolutionary status and origin.  Analyses by \citet{hun69}, \citet{heb83} and \citet{pan06,pan11} providing effective temperature, surface gravity, and surface abundances established that BD+10$^\circ$2179 is a hydrogen-deficient supergiant. Our current measurements imply $\log L$/$M \approx 3.6$ (solar units). Evidence that there are no pulsations \citep{hill84,grauer84} requires $\log L$/$M < 4$ \citep[solar units,][]{jeffery16a}. 

From its Galactic position, apparent magnitude, effective temperature and observed $L$/$M$ ratio, BD+10$^\circ$2179 must necessarily be a low-mass star, and cannot therefore be in a long-lived core or shell-burning phase of evolution. It must be either expanding towards or contracting away from the giant branch, and its  rate of evolution must be governed primarily by the thermal time scale of the envelope, being that part of the star lying outside any degenerate or inactive core. Absence of evidence for a binary companion \citep{jef87} rules out models comprising a stripped helium star following a mass-transfer episode. 

The expansion of a helium star after shell helium ignition depends on the nature of the progenitor and the mode of ignition. A core helium-burning star will evolve to become a shell-helium burning giant if sufficiently massive \citep{paczynski71,weiss87a}, but only if the mass exceeds about 0.8\,$M_\odot$.  The track for a double white-dwarf merger \citep{web84} depends on the progenitor white dwarfs and on the post-merger accretion rate. The evolution of a star which has a helium-shell flash whilst contracting depends on how late the flash occurs \citep{her99}. In either case, the expansion may be rapid and not in hydrostatic equilibrium \citep[cf. V4334\,Sgr: ][]{jeffery02a}.

The contraction of a post-giant branch helium star \citep{paczynski71,schoenberner77, weiss87a,saio88a} has been studied in the context of a post final helium-shell flash \citep{schoenberner79,iben83,her99} and in the context of a post double white dwarf merger \citep{saio00,saio02,zhang12b,zhang14}. Once a helium-burning shell ceases to capable of supporting a giant envelope, a star contracts at constant luminosity on a roughly thermal timescale determined by the mass (or total thermal energy) of the envelope and the luminosity. In cases where the  precursor has reached hydrostatic and nuclear equilibrium (e.g. as an RCrB star), the luminosity should be linked to the shell luminosity of the helium shell-burning giant as it leaves the giant branch, and hence to the mass of the carbon-oxygen core \citep[cf.][]{jeffery88}. In the case of post-merger models, the luminosity is also governed by the mass of the surviving white-dwarf core \citep{zhang14}. The rate of evolution depends on the residual envelope mass. 

Comparing the observed surface gravity with post-giant branch evolution tracks in order to estimate a mass \citep[e.g. 0.6 $M_\odot$: ][or 0.55 $M_\odot$ above]{heb83} depends on the model adopted. Additional constraints on the evolution are provided by the surface composition, which also depends on history. 

The flash-driven convection which accompanies a post-AGB late thermal pulse mixes helium- and carbon-rich nuclear products part of the way to the surface; opacity-driven convection may complete the process when the star becomes a giant. In a very late thermal pulse, the flash-driven convection provides prompt enrichment of the entire envelope through to the surface. The extent to which surface layers are depleted in hydrogen depends on the relative masses of the residual hydrogen envelope and the helium intershell and the lateness of the thermal pulse \citep{schoenberner77,her99}, but appears to be inversely correlated with carbon enrichment. Most FF models predict too much carbon for the amount of hydrogen remaining on the surface of BD+10$^\circ$2179.  

The merger of a carbon-oxygen plus helium white dwarf binary exhibits multiple nuclear episodes. Temperatures exceeding $10^8$ K during merger produce carbon, oxygen ($^{18}$O), some neon, and p-capture products, amongst other nuclides \citep{clayton05}. Helium-shell ignition produces additional carbon and strong flash-driven convection, enabling fresh carbon to reach the surface when the star becomes a giant \citep{zhang14}. At the same time, any hydrogen remaining from the pre-merger  white dwarfs is diluted throughout the helium-rich envelope. 

\citet{jef11} argued that a double white dwarf merger could account for much of the surface chemistry of EHes, including BD+10$^\circ$2179. Assuming an initial composition scaled to the current iron abundance, the transformation of initial carbon, nitrogen and oxygen, to nitrogen, via CNO burning, and then of helium to carbon and oxygen ($^{16}$O) and of nitrogen to oxygen ($^{18}$O)  and neon ($^{22}$Ne) can account for present day observations. There is too little carbon (or not enough hydrogen) for the late thermal pulse model.  Meanwhile, models for carbon-oxygen plus helium white dwarf mergers remain hard to build \citep{hall16,zhang17}. 

At $0.55\pm0.05 {\rm M_{\odot}}$, the mass inferred for BD+10$^\circ$2179 from the  \citet{saio02} tracks is at the lower limit for a CO+He WD merger. Until recently, the presence of carbon in the photospheres of EHes argued against an origin in the merger of two helium white dwarfs. However \citet{zhang12b} demonstrated that, at the upper limit of the mass range for He+He white dwarf mergers,  carbon produced during the hot phase of the merger could be mixed to the surface in sufficient quantity to explain the carbon-rich surface compositions of the pulsating low-mass EHe star  BX\,Cir and  some helium-rich sdO stars. Indeed, the revised $L$/$M$ ratio for  BD+10$^\circ$2179 and  that measured for BX\,Cir are remarkably similar \citep{drilling98,woolf00,woolf02}. So we must now also consider the more likely possibility that BD+10$^\circ$2179 was produced in a  double helium white dwarf merger. 

\section{Conclusion}\label{conclusion}
We  have carried out a  detailed fine analysis of high-quality optical spectra of the prototype extreme helium star BD+10$^\circ$2179.  For the model atmospheres  we have used  the LTE opacity-sampling code {\sc atlas12}, which we have benchmarked against the equivalent code {\sc sterne3}, finding good agreement between them. We have treated the line spectrum using  a detailed non-LTE approach, which has resulted in greater internal consistency between results from different spectral lines, and a reduction in the required microturbulent velocity.  The resulting global parameters  ($T_{\rm eff}, g$) are slightly  modified over previous analyses;  the star appears to be hotter and smaller.  Chemically, only small differences in the abundances of most elements are found compared with previous analyses, all of which can be attributed to the small change in $T_{\rm eff}$ and improved microphysics.  The exception is carbon which appears to be about 0.3\,dex richer than previously inferred.  

The low metallicity and Galactic position remain consistent with membership of an old stellar population. The origin of BD+10$^\circ$2179 is more difficult to  establish; models favour an origin in a double white dwarf merger, but whether this is a He+He or a CO+He merger is not clear, as the models remain far from mature.

\section*{Acknowledgments}
TK acknowledges support by the Erasmus Internship Program which supported his stay at Armagh observatory where parts of this paper was written.
The authors thank the staff at the ESO Paranal and La Silla observatories for performing the spectroscopic observations in service mode.
We would also like to thank the anonymous referee for constructive criticism that helped to improve the manuscript.
This research has made use of the SIMBAD database, operated at CDS, Strasbourg, France.
Some of the data presented in this paper were obtained from the Mikulski Archive for Space Telescopes
(MAST). STScI is operated by the Association of Universities for Research in Astronomy, Inc., under NASA 
contract NAS5-26555. Support for MAST for non-HST data is provided by the NASA Office of Space
Science via grant NNX09AF08G and by other grants and contracts.
Research at the Armagh Observatory and Planetarium is supported by a grant-in-aid 
from the Northern Ireland Department for Communities.
CSJ acknowledges support from the UK Science and Technology Facilities Council (STFC) Grant No. 
ST/M000834/1.

\bibliographystyle{mnras}
\bibliography{reference,ehe}

\begin{thebibliography}{}
\makeatletter
\relax
\def\mn@urlcharsother{\let\do\@makeother \do\$\do\&\do\#\do\^\do\_\do\%\do\~}
\def\mn@doi{\begingroup\mn@urlcharsother \@ifnextchar [ {\mn@doi@}
  {\mn@doi@[]}}
\def\mn@doi@[#1]#2{\def\@tempa{#1}\ifx\@tempa\@empty \href
  {http://dx.doi.org/#2} {doi:#2}\else \href {http://dx.doi.org/#2} {#1}\fi
  \endgroup}
\def\mn@eprint#1#2{\mn@eprint@#1:#2::\@nil}
\def\mn@eprint@arXiv#1{\href {http://arxiv.org/abs/#1} {{\tt arXiv:#1}}}
\def\mn@eprint@dblp#1{\href {http://dblp.uni-trier.de/rec/bibtex/#1.xml}
  {dblp:#1}}
\def\mn@eprint@#1:#2:#3:#4\@nil{\def\@tempa {#1}\def\@tempb {#2}\def\@tempc
  {#3}\ifx \@tempc \@empty \let \@tempc \@tempb \let \@tempb \@tempa \fi \ifx
  \@tempb \@empty \def\@tempb {arXiv}\fi \@ifundefined
  {mn@eprint@\@tempb}{\@tempb:\@tempc}{\expandafter \expandafter \csname
  mn@eprint@\@tempb\endcsname \expandafter{\@tempc}}}

\bibitem[\protect\citeauthoryear{{Allen} \& {Santillan}}{{Allen} \&
  {Santillan}}{1991}]{AlSa91}
{Allen} C.,  {Santillan} A.,  1991, \rmxaa, \href
  {http://adsabs.harvard.edu/abs/1991RMxAA..22..255A} {22, 255}

\bibitem[\protect\citeauthoryear{{Asplund}, {Grevesse}, {Sauval}  \&
  {Scott}}{{Asplund} et~al.}{2009}]{asp09}
{Asplund} M.,  {Grevesse} N.,  {Sauval} A.~J.,   {Scott} P.,  2009, \mn@doi
  [\araa] {10.1146/annurev.astro.46.060407.145222}, \href
  {http://adsabs.harvard.edu/abs/2009ARA%26A..47..481A} {47, 481}

\bibitem[\protect\citeauthoryear{{Ayres}}{{Ayres}}{2010}]{ayr10}
{Ayres} T.~R.,  2010, \mn@doi [\apjs] {10.1088/0067-0049/187/1/149}, \href
  {http://adsabs.harvard.edu/abs/2010ApJS..187..149A} {187, 149}

\bibitem[\protect\citeauthoryear{{Barnard}, {Cooper}  \& {Shamey}}{{Barnard}
  et~al.}{1969}]{BCS69}
{Barnard} A.~J.,  {Cooper} J.,   {Shamey} L.~J.,  1969, \aap, \href
  {http://adsabs.harvard.edu/abs/1969A%26A.....1...28B} {1, 28}

\bibitem[\protect\citeauthoryear{{Bates} \& {Damgaard}}{{Bates} \&
  {Damgaard}}{1949}]{BaDa49}
{Bates} D.~R.,  {Damgaard} A.,  1949, \mn@doi [Phil. Trans. R. Soc. London,
  Ser. A,] {10.1098/rsta.1949.0006}, \href
  {http://adsabs.harvard.edu/abs/1949RSPTA.242..101B} {242, 101}

\bibitem[\protect\citeauthoryear{{Beauchamp} \& {Wesemael}}{{Beauchamp} \&
  {Wesemael}}{1998}]{BeWe98}
{Beauchamp} A.,  {Wesemael} F.,  1998, \mn@doi [\apj] {10.1086/305357}, \href
  {http://adsabs.harvard.edu/abs/1998ApJ...496..395B} {496, 395}

\bibitem[\protect\citeauthoryear{{Becker}}{{Becker}}{1998}]{Becker98}
{Becker} S.~R.,  1998. ASP Conf. Ser., 131.
p.~137

\bibitem[\protect\citeauthoryear{{Becker} \& {Butler}}{{Becker} \&
  {Butler}}{1988}]{BeBu88}
{Becker} S.~R.,  {Butler} K.,  1988, \aap, \href
  {http://adsabs.harvard.edu/abs/1988A%26A...201..232B} {201, 232}

\bibitem[\protect\citeauthoryear{{Becker} \& {Butler}}{{Becker} \&
  {Butler}}{1989}]{BeBu89}
{Becker} S.~R.,  {Butler} K.,  1989, \aap, \href
  {http://adsabs.harvard.edu/abs/1989A%26A...209..244B} {209, 244}

\bibitem[\protect\citeauthoryear{{Behara} \& {Jeffery}}{{Behara} \&
  {Jeffery}}{2006}]{beh06}
{Behara} N.~T.,  {Jeffery} C.~S.,  2006, \mn@doi [\aap]
  {10.1051/0004-6361:20053978}, \href
  {http://adsabs.harvard.edu/abs/2006A%26A...451..643B} {451, 643}

\bibitem[\protect\citeauthoryear{{Bloecker}}{{Bloecker}}{2001}]{bloe01}
{Bloecker} T.,  2001, \apss, \href
  {http://adsabs.harvard.edu/abs/2001Ap%26SS.275....1B} {275, 1}

\bibitem[\protect\citeauthoryear{{Butler} \& {Giddings}}{{Butler} \&
  {Giddings}}{1985}]{but85}
{Butler} K.,  {Giddings} J.~R.,  1985, in Newsletter of Analysis of
  Astronomical Spectra, No. 9 (Univ. London)

\bibitem[\protect\citeauthoryear{{Callegari} \& {Trigueiros}}{{Callegari} \&
  {Trigueiros}}{1998}]{CaTr98}
{Callegari} F.,  {Trigueiros} A.~G.,  1998, \mn@doi [\apjs] {10.1086/313158},
  \href {http://adsabs.harvard.edu/abs/1998ApJS..119..181C} {119, 181}

\bibitem[\protect\citeauthoryear{{Cann} \& {Thakkar}}{{Cann} \&
  {Thakkar}}{2002}]{CaTh02}
{Cann} N.~M.,  {Thakkar} A.~J.,  2002, \mn@doi [J. Phys. B: At. Mol. Opt.
  Phys.] {10.1088/0953-4075/35/2/317}, \href
  {http://adsabs.harvard.edu/abs/2002JPhB...35..421C} {35, 421}

\bibitem[\protect\citeauthoryear{{Clayton}, {Herwig}, {Geballe}, {Asplund},
  {Tenenbaum}, {Engelbracht}  \& {Gordon}}{{Clayton} et~al.}{2005}]{clayton05}
{Clayton} G.~C.,  {Herwig} F.,  {Geballe} T.~R.,  {Asplund} M.,  {Tenenbaum}
  E.~D.,  {Engelbracht} C.~W.,   {Gordon} K.~D.,  2005, \apjl, \href
  {http://adsabs.harvard.edu/abs/2005ApJ...623L.141C} {623, L141}

\bibitem[\protect\citeauthoryear{{Cowley}}{{Cowley}}{1971}]{Cowley71}
{Cowley} C.~R.,  1971, The Observatory, \href
  {http://adsabs.harvard.edu/abs/1971Obs....91..139C} {91, 139}

\bibitem[\protect\citeauthoryear{{Cutri} et~al.,}{{Cutri} et~al.}{2003}]{cut03}
{Cutri} R.~M.,  et~al., 2003, VizieR Online Data Catalog, \href
  {http://cdsads.u-strasbg.fr/abs/2003yCat.2246....0C} {2246}

\bibitem[\protect\citeauthoryear{{Dekker}, {D'Odorico}, {Kaufer}, {Delabre}  \&
  {Kotzlowski}}{{Dekker} et~al.}{2000}]{dek00}
{Dekker} H.,  {D'Odorico} S.,  {Kaufer} A.,  {Delabre} B.,   {Kotzlowski} H.,
  2000. SPIE Conf. Ser., 4008.
p.~534

\bibitem[\protect\citeauthoryear{{Dimitrijevic} \&
  {Sahal-Brechot}}{{Dimitrijevic} \& {Sahal-Brechot}}{1990}]{DSB90}
{Dimitrijevic} M.~S.,  {Sahal-Brechot} S.,  1990, \aaps, \href
  {http://adsabs.harvard.edu/abs/1990A%26AS...82..519D} {82, 519}

\bibitem[\protect\citeauthoryear{{Drilling}, {Schonberner}, {Heber}  \&
  {Lynas-Gray}}{{Drilling} et~al.}{1984}]{drilling84}
{Drilling} J.~S.,  {Schonberner} D.,  {Heber} U.,   {Lynas-Gray} A.~E.,  1984,
  \mn@doi [\apj] {10.1086/161786}, \href
  {http://adsabs.harvard.edu/abs/1984ApJ...278..224D} {278, 224}

\bibitem[\protect\citeauthoryear{{Drilling}, {Jeffery}  \& {Heber}}{{Drilling}
  et~al.}{1998}]{drilling98}
{Drilling} J.~S.,  {Jeffery} C.~S.,   {Heber} U.,  1998, \aap, \href
  {http://ukads.nottingham.ac.uk/abs/1998A%26A...329.1019D} {329, 1019}

\bibitem[\protect\citeauthoryear{{Firnstein} \& {Przybilla}}{{Firnstein} \&
  {Przybilla}}{2012}]{FiPr12}
{Firnstein} M.,  {Przybilla} N.,  2012, \mn@doi [\aap]
  {10.1051/0004-6361/201219034}, \href
  {http://adsabs.harvard.edu/abs/2012A%26A...543A..80F} {543, A80}

\bibitem[\protect\citeauthoryear{{Froese Fischer} \& {Tachiev}}{{Froese
  Fischer} \& {Tachiev}}{2004}]{FFT04}
{Froese Fischer} C.,  {Tachiev} G.,  2004, At.~Data Nucl.~Data Tables, 87, 1

\bibitem[\protect\citeauthoryear{{Froese Fischer}, {Tachiev}  \&
  {Irimia}}{{Froese Fischer} et~al.}{2006}]{FFTI06}
{Froese Fischer} C.,  {Tachiev} G.,   {Irimia} A.,  2006, At.~Data Nucl.~Data
  Tables, 92, 607

\bibitem[\protect\citeauthoryear{{Fuhr} \& {Wiese}}{{Fuhr} \&
  {Wiese}}{1998}]{FuWi98}
{Fuhr} J.~R.,  {Wiese} W.~L.,  1998, in CRC Handbook of Chemistry and Physics,
  79th edn., ed. D. R. Lide (Boca Raton: CRC Press)

\bibitem[\protect\citeauthoryear{{Fuhr}, {Martin}  \& {Wiese}}{{Fuhr}
  et~al.}{1988}]{fuhr88}
{Fuhr} J.~R.,  {Martin} G.~A.,   {Wiese} W.~L.,  1988, J. Phys. \& Chem. Ref.
  Data., \href {http://adsabs.harvard.edu/abs/1988JPCRD..17S....F} {17}

\bibitem[\protect\citeauthoryear{{Giddings}}{{Giddings}}{1981}]{gid81}
{Giddings} J.~R.,  1981, PhD thesis, Univ. London

\bibitem[\protect\citeauthoryear{{Grauer}, {Drilling}  \&
  {Sch\"{o}nberner}}{{Grauer} et~al.}{1984}]{grauer84}
{Grauer} A.~D.,  {Drilling} J.~S.,   {Sch\"{o}nberner} D.,  1984, \aap, \href
  {http://ukads.nottingham.ac.uk/abs/1984A%26A...133..285G} {133, 285}

\bibitem[\protect\citeauthoryear{{Gray}}{{Gray}}{2005}]{Gray05}
{Gray} D.~F.,  2005, {The Observation and Analysis of Stellar Photospheres, 3rd
  ed. (Cambridge: Cambridge University Press)}

\bibitem[\protect\citeauthoryear{{Griem}}{{Griem}}{1964}]{Griem64}
{Griem} H.~R.,  1964, {Plasma spectroscopy (New York: Mc-Graw-Hill)}

\bibitem[\protect\citeauthoryear{{Griem}}{{Griem}}{1974}]{Griem74}
{Griem} H.~R.,  1974, {Spectral line broadening by plasmas (New York: Academic
  Press)}

\bibitem[\protect\citeauthoryear{{Griem}, {Baranger}, {Kolb}  \&
  {Oertel}}{{Griem} et~al.}{1962}]{GBKO62}
{Griem} H.~R.,  {Baranger} M.,  {Kolb} A.~C.,   {Oertel} G.,  1962, \mn@doi
  [Phys. Rev.] {10.1103/PhysRev.125.177}, \href
  {http://adsabs.harvard.edu/abs/1962PhRv..125..177G} {125, 177}

\bibitem[\protect\citeauthoryear{{Hall} \& {Jeffery}}{{Hall} \&
  {Jeffery}}{2016}]{hall16}
{Hall} P.~D.,  {Jeffery} C.~S.,  2016, \mn@doi [\mnras]
  {10.1093/mnras/stw2188}, \href
  {http://adsabs.harvard.edu/abs/2016MNRAS.463.2756H} {463, 2756}

\bibitem[\protect\citeauthoryear{{Harrison} \& {Jeffery}}{{Harrison} \&
  {Jeffery}}{1997}]{har97}
{Harrison} P.~M.,  {Jeffery} C.~S.,  1997, \aap, \href
  {http://adsabs.harvard.edu/abs/1997A%26A...323..177H} {323, 177}

\bibitem[\protect\citeauthoryear{{Heber}}{{Heber}}{1983}]{heb83}
{Heber} U.,  1983, \aap, \href
  {http://adsabs.harvard.edu/abs/1983A%26A...118...39H} {118, 39}

\bibitem[\protect\citeauthoryear{{Heber}, {Moehler}, {Napiwotzki}, {Thejll}  \&
  {Green}}{{Heber} et~al.}{2002}]{heb02}
{Heber} U.,  {Moehler} S.,  {Napiwotzki} R.,  {Thejll} P.,   {Green} E.~M.,
  2002, \mn@doi [\aap] {10.1051/0004-6361:20020127}, \href
  {http://adsabs.harvard.edu/abs/2002A%26A...383..938H} {383, 938}

\bibitem[\protect\citeauthoryear{{Herwig}, {Bl{\"o}cker}, {Langer}  \&
  {Driebe}}{{Herwig} et~al.}{1999}]{her99}
{Herwig} F.,  {Bl{\"o}cker} T.,  {Langer} N.,   {Driebe} T.,  1999, \aap, \href
  {http://adsabs.harvard.edu/abs/1999A%26A...349L...5H} {349, L5}

\bibitem[\protect\citeauthoryear{{Hill}}{{Hill}}{1965}]{hill65}
{Hill} P.~W.,  1965, \mnras, \href
  {http://ukads.nottingham.ac.uk/abs/1965MNRAS.129..137H} {129, 137}

\bibitem[\protect\citeauthoryear{{Hill}, {Lynas-Gray}  \& {Kilkenny}}{{Hill}
  et~al.}{1984}]{hill84}
{Hill} P.~W.,  {Lynas-Gray} A.~E.,   {Kilkenny} D.,  1984, \mnras, \href
  {http://ukads.nottingham.ac.uk/abs/1984MNRAS.207..823H} {207, 823}

\bibitem[\protect\citeauthoryear{{Hirsch}}{{Hirsch}}{2009}]{Hirsch09}
{Hirsch} H.,  2009, PhD thesis, University of Erlangen-Nuremberg

\bibitem[\protect\citeauthoryear{{Hubeny}, {Hummer}  \& {Lanz}}{{Hubeny}
  et~al.}{1994}]{Hum94}
{Hubeny} I.,  {Hummer} D.~G.,   {Lanz} T.,  1994, \aap, \href
  {http://adsabs.harvard.edu/abs/1994A%26A...282..151H} {282, 151}

\bibitem[\protect\citeauthoryear{{Hunger} \& {Klinglesmith}}{{Hunger} \&
  {Klinglesmith}}{1969}]{hun69}
{Hunger} K.,  {Klinglesmith} D.,  1969, \mn@doi [\apj] {10.1086/150109}, \href
  {http://cdsads.u-strasbg.fr/abs/1969ApJ...157..721H} {157, 721}

\bibitem[\protect\citeauthoryear{{Iben} \& {Tutukov}}{{Iben} \&
  {Tutukov}}{1984}]{iben84}
{Iben} Jr. I.,  {Tutukov} A.~V.,  1984, \mn@doi [\apjs] {10.1086/190932}, \href
  {http://adsabs.harvard.edu/abs/1984ApJS...54..335I} {54, 335}

\bibitem[\protect\citeauthoryear{{Iben}, {Kaler}, {Truran}  \&
  {Renzini}}{{Iben} et~al.}{1983}]{iben83}
{Iben} Jr. I.,  {Kaler} J.~B.,  {Truran} J.~W.,   {Renzini} A.,  1983, \apj,
  \href {http://ukads.nottingham.ac.uk/abs/1983ApJ...264..605I} {264, 605}

\bibitem[\protect\citeauthoryear{{Irrgang}, {Wilcox}, {Tucker}  \&
  {Schiefelbein}}{{Irrgang} et~al.}{2013}]{Irr13}
{Irrgang} A.,  {Wilcox} B.,  {Tucker} E.,   {Schiefelbein} L.,  2013, \mn@doi
  [\aap] {10.1051/0004-6361/201220540}, \href
  {http://adsabs.harvard.edu/abs/2013A%26A...549A.137I} {549, A137}

\bibitem[\protect\citeauthoryear{{Irrgang}, {Przybilla}, {Heber}, {B{\"o}ck},
  {Hanke}, {Nieva}  \& {Butler}}{{Irrgang} et~al.}{2014}]{Irrgangetal14}
{Irrgang} A.,  {Przybilla} N.,  {Heber} U.,  {B{\"o}ck} M.,  {Hanke} M.,
  {Nieva} M.-F.,   {Butler} K.,  2014, \mn@doi [\aap]
  {10.1051/0004-6361/201323167}, \href
  {http://adsabs.harvard.edu/abs/2014A%26A...565A..63I} {565, A63}

\bibitem[\protect\citeauthoryear{{Jeffery}}{{Jeffery}}{1988}]{jeffery88}
{Jeffery} C.~S.,  1988, \mnras, \href
  {http://adsabs.harvard.edu/abs/1988MNRAS.235.1287J} {235, 1287}

\bibitem[\protect\citeauthoryear{{Jeffery}}{{Jeffery}}{2008a}]{jef08}
{Jeffery} C.~S.,  2008a. ASP Conf. Ser., 391.
p.~53

\bibitem[\protect\citeauthoryear{{Jeffery}}{{Jeffery}}{2008b}]{jeffery08.ibvs}
{Jeffery} C.~S.,  2008b, Information Bulletin on Variable Stars, \href
  {http://ukads.nottingham.ac.uk/abs/2008IBVS.5817....1J} {5817, 1}

\bibitem[\protect\citeauthoryear{{Jeffery}}{{Jeffery}}{2017}]{Jeffery17}
{Jeffery} C.~S.,  2017, \mnras, in press [arXiv: 1706.03377]

\bibitem[\protect\citeauthoryear{{Jeffery} \& {Hamann}}{{Jeffery} \&
  {Hamann}}{2010}]{jef10}
{Jeffery} C.~S.,  {Hamann} W.-R.,  2010, \mn@doi [\mnras]
  {10.1111/j.1365-2966.2010.16410.x}, \href
  {http://adsabs.harvard.edu/abs/2010MNRAS.404.1698J} {404, 1698}

\bibitem[\protect\citeauthoryear{{Jeffery} \& {Pollacco}}{{Jeffery} \&
  {Pollacco}}{2002}]{jeffery02a}
{Jeffery} C.~S.,  {Pollacco} D.,  2002, \apss, \href
  {http://adsabs.harvard.edu/abs/2002Ap%26SS.279...15J} {279, 15}

\bibitem[\protect\citeauthoryear{{Jeffery} \& {Saio}}{{Jeffery} \&
  {Saio}}{2016}]{jeffery16a}
{Jeffery} C.~S.,  {Saio} H.,  2016, \mn@doi [\mnras] {10.1093/mnras/stw388},
  \href {http://adsabs.harvard.edu/abs/2016MNRAS.458.1352J} {458, 1352}

\bibitem[\protect\citeauthoryear{{Jeffery}, {Drilling}  \& {Heber}}{{Jeffery}
  et~al.}{1987}]{jef87}
{Jeffery} C.~S.,  {Drilling} J.~S.,   {Heber} U.,  1987, \mn@doi [\mnras]
  {10.1093/mnras/226.2.317}, \href
  {http://adsabs.harvard.edu/abs/1987MNRAS.226..317J} {226, 317}

\bibitem[\protect\citeauthoryear{{Jeffery}, {Heber}, {Hill}, {Dreizler},
  {Drilling}, {Lawson}, {Leuenhagen}  \& {Werner}}{{Jeffery}
  et~al.}{1996}]{jef96}
{Jeffery} C.~S.,  {Heber} U.,  {Hill} P.~W.,  {Dreizler} S.,  {Drilling} J.~S.,
   {Lawson} W.~A.,  {Leuenhagen} U.,   {Werner} K.,  1996. ASP Conf. Ser., 96.
p.~471

\bibitem[\protect\citeauthoryear{{Jeffery}, {Hill}  \& {Heber}}{{Jeffery}
  et~al.}{1999}]{jeffery99}
{Jeffery} C.~S.,  {Hill} P.~W.,   {Heber} U.,  1999, \aap, \href
  {http://ukads.nottingham.ac.uk/abs/1999A%26A...346..491J} {346, 491}

\bibitem[\protect\citeauthoryear{{Jeffery}, {Woolf}  \& {Pollacco}}{{Jeffery}
  et~al.}{2001}]{jef01}
{Jeffery} C.~S.,  {Woolf} V.~M.,   {Pollacco} D.~L.,  2001, \mn@doi [\aap]
  {10.1051/0004-6361:20010954}, \href
  {http://adsabs.harvard.edu/abs/2001A%26A...376..497J} {376, 497}

\bibitem[\protect\citeauthoryear{{Jeffery}, {Karakas}  \& {Saio}}{{Jeffery}
  et~al.}{2011}]{jef11}
{Jeffery} C.~S.,  {Karakas} A.~I.,   {Saio} H.,  2011, \mn@doi [\mnras]
  {10.1111/j.1365-2966.2011.18667.x}, \href
  {http://adsabs.harvard.edu/abs/2011MNRAS.414.3599J} {414, 3599}

\bibitem[\protect\citeauthoryear{{Johnson} \& {Soderblom}}{{Johnson} \&
  {Soderblom}}{1987}]{1987AJ.....93..864J}
{Johnson} D.~R.~H.,  {Soderblom} D.~R.,  1987, \mn@doi [\aj] {10.1086/114370},
  \href {http://cdsads.u-strasbg.fr/abs/1987AJ.....93..864J} {93, 864}

\bibitem[\protect\citeauthoryear{{Kaufer}, {Stahl}, {Tubbesing},
  {N{\o}rregaard}, {Avila}, {Francois}, {Pasquini}  \& {Pizzella}}{{Kaufer}
  et~al.}{1999}]{kau99}
{Kaufer} A.,  {Stahl} O.,  {Tubbesing} S.,  {N{\o}rregaard} P.,  {Avila} G.,
  {Francois} P.,  {Pasquini} L.,   {Pizzella} A.,  1999, The Messenger, \href
  {http://adsabs.harvard.edu/abs/1999Msngr..95....8K} {95, 8}

\bibitem[\protect\citeauthoryear{{Klemola}}{{Klemola}}{1961}]{kle61}
{Klemola} A.~R.,  1961, \mn@doi [\apj] {10.1086/147135}, \href
  {http://cdsads.u-strasbg.fr/abs/1961ApJ...134..130K} {134, 130}

\bibitem[\protect\citeauthoryear{{Koester} et~al.,}{{Koester}
  et~al.}{2001}]{koe01}
{Koester} D.,  et~al., 2001, \mn@doi [\aap] {10.1051/0004-6361:20011235}, \href
  {http://adsabs.harvard.edu/abs/2001A%26A...378..556K} {378, 556}

\bibitem[\protect\citeauthoryear{{Kurucz}}{{Kurucz}}{1993}]{kur93}
{Kurucz} R.,  1993, Kurucz CD-ROM No.~13.~ Cambridge, Mass.: SAO, \href
  {http://adsabs.harvard.edu/abs/1993KurCD..13.....K} {}

\bibitem[\protect\citeauthoryear{{Kurucz}}{{Kurucz}}{1996}]{kur96}
{Kurucz} R.~L.,  1996. ASP Conf. Ser., 108.
p.~160

\bibitem[\protect\citeauthoryear{{Kurucz} \& {Bell}}{{Kurucz} \&
  {Bell}}{1995}]{KuBe95}
{Kurucz} R.,  {Bell} B.,  1995, Kurucz CD-ROM No.~23.~Cambridge, Mass.: SAO,
  \href {http://adsabs.harvard.edu/abs/1995KurCD..23.....K} {}

\bibitem[\protect\citeauthoryear{{Kurucz} \& {Peytremann}}{{Kurucz} \&
  {Peytremann}}{1975}]{KuPe75}
{Kurucz} R.~L.,  {Peytremann} E.,  1975, SAO Special Report, \href
  {http://adsabs.harvard.edu/abs/1975SAOSR.362.....K} {362}

\bibitem[\protect\citeauthoryear{{Lanz}, {Dimitrijevic}  \& {Artru}}{{Lanz}
  et~al.}{1988}]{Lanzetal88}
{Lanz} T.,  {Dimitrijevic} M.~S.,   {Artru} M.-C.,  1988, \aap, \href
  {http://adsabs.harvard.edu/abs/1988A%26A...192..249L} {192, 249}

\bibitem[\protect\citeauthoryear{{Luo} \& {Pradhan}}{{Luo} \&
  {Pradhan}}{1989}]{LuPr89}
{Luo} D.,  {Pradhan} A.~K.,  1989, \mn@doi [J. Phys. B: At. Mol. Opt. Phys.]
  {10.1088/0953-4075/22/21/005}, \href
  {http://adsabs.harvard.edu/abs/1989JPhB...22.3377L} {22, 3377}

\bibitem[\protect\citeauthoryear{{Mar}, {P{\'e}rez}, {Gonz{\'a}lez}, {Gigosos},
  {del Val}, {de la Rosa}  \& {Aparicio}}{{Mar} et~al.}{2000}]{mar00}
{Mar} S.,  {P{\'e}rez} C.,  {Gonz{\'a}lez} V.~R.,  {Gigosos} M.~A.,  {del Val}
  J.~A.,  {de la Rosa} I.,   {Aparicio} J.~A.,  2000, \mn@doi [\aaps]
  {10.1051/aas:2000225}, \href
  {http://adsabs.harvard.edu/abs/2000A%26AS..144..509M} {144, 509}

\bibitem[\protect\citeauthoryear{{Mermilliod}}{{Mermilliod}}{1997}]{mer97}
{Mermilliod} J.~C.,  1997, VizieR Online Data Catalog, \href
  {http://adsabs.harvard.edu/abs/1997yCat.2168....0M} {2168}

\bibitem[\protect\citeauthoryear{{M\"oller}}{{M\"oller}}{1990}]{Moeller90}
{M\"oller} R.~U.,  1990, Master's thesis, University Kiel

\bibitem[\protect\citeauthoryear{{Morel} \& {Butler}}{{Morel} \&
  {Butler}}{2008}]{MoBu08}
{Morel} T.,  {Butler} K.,  2008, \mn@doi [\aap] {10.1051/0004-6361:200809924},
  \href {http://adsabs.harvard.edu/abs/2008A%26A...487..307M} {487, 307}

\bibitem[\protect\citeauthoryear{{Morel}, {Butler}, {Aerts}, {Neiner}  \&
  {Briquet}}{{Morel} et~al.}{2006}]{Moreletal06}
{Morel} T.,  {Butler} K.,  {Aerts} C.,  {Neiner} C.,   {Briquet} M.,  2006,
  \mn@doi [\aap] {10.1051/0004-6361:20065171}, \href
  {http://adsabs.harvard.edu/abs/2006A%26A...457..651M} {457, 651}

\bibitem[\protect\citeauthoryear{{Nahar}}{{Nahar}}{2002}]{Nahar02}
{Nahar} S.~N.,  2002, \mn@doi [At. Data Nucl. Data Tables]
  {10.1006/adnd.2002.0879}, \href
  {http://adsabs.harvard.edu/abs/2002ADNDT..80..205N} {80, 205}

\bibitem[\protect\citeauthoryear{{Nelder} \& {Mead}}{{Nelder} \&
  {Mead}}{1965}]{NeMe65}
{Nelder} J.~A.,  {Mead} R.,  1965, Computer Journal, 7, 308

\bibitem[\protect\citeauthoryear{{Nieva} \& {Przybilla}}{{Nieva} \&
  {Przybilla}}{2006}]{NiPr06}
{Nieva} M.~F.,  {Przybilla} N.,  2006, \mn@doi [\apjl] {10.1086/501124}, \href
  {http://adsabs.harvard.edu/abs/2006ApJ...639L..39N} {639, L39}

\bibitem[\protect\citeauthoryear{{Nieva} \& {Przybilla}}{{Nieva} \&
  {Przybilla}}{2007}]{nie07}
{Nieva} M.~F.,  {Przybilla} N.,  2007, \mn@doi [\aap]
  {10.1051/0004-6361:20065757}, \href
  {http://adsabs.harvard.edu/abs/2007A%26A...467..295N} {467, 295}

\bibitem[\protect\citeauthoryear{{Nieva} \& {Przybilla}}{{Nieva} \&
  {Przybilla}}{2008}]{nie08}
{Nieva} M.~F.,  {Przybilla} N.,  2008, \mn@doi [\aap]
  {10.1051/0004-6361:20078203}, \href
  {http://adsabs.harvard.edu/abs/2008A%26A...481..199N} {481, 199}

\bibitem[\protect\citeauthoryear{{Nieva} \& {Przybilla}}{{Nieva} \&
  {Przybilla}}{2012}]{NiPr12}
{Nieva} M.-F.,  {Przybilla} N.,  2012, \mn@doi [\aap]
  {10.1051/0004-6361/201118158}, \href
  {http://adsabs.harvard.edu/abs/2012A%26A...539A.143N} {539, A143}

\bibitem[\protect\citeauthoryear{{Odenkirchen} \& {Brosche}}{{Odenkirchen} \&
  {Brosche}}{1992}]{OdBr92}
{Odenkirchen} M.,  {Brosche} P.,  1992, \mn@doi [Astronomische Nachrichten]
  {10.1002/asna.2113130204}, \href
  {http://adsabs.harvard.edu/abs/1992AN....313...69O} {313, 69}

\bibitem[\protect\citeauthoryear{{Paczy{\'n}ski}}{{Paczy{\'n}ski}}{1971}]{paczynski71}
{Paczy{\'n}ski} B.,  1971, \actaa, \href
  {http://adsabs.harvard.edu/abs/1971AcA....21....1P} {21, 1}

\bibitem[\protect\citeauthoryear{{Pandey} \& {Lambert}}{{Pandey} \&
  {Lambert}}{2011}]{pan11}
{Pandey} G.,  {Lambert} D.~L.,  2011, \mn@doi [\apj]
  {10.1088/0004-637X/727/2/122}, \href
  {http://adsabs.harvard.edu/abs/2011ApJ...727..122P} {727, 122}

\bibitem[\protect\citeauthoryear{{Pandey}, {Lambert}, {Jeffery}  \&
  {Rao}}{{Pandey} et~al.}{2006}]{pan06}
{Pandey} G.,  {Lambert} D.~L.,  {Jeffery} C.~S.,   {Rao} N.~K.,  2006, \mn@doi
  [\apj] {10.1086/498674}, \href
  {http://adsabs.harvard.edu/abs/2006ApJ...638..454P} {638, 454}

\bibitem[\protect\citeauthoryear{{Pauli}, {Napiwotzki}, {Heber}, {Altmann}  \&
  {Odenkirchen}}{{Pauli} et~al.}{2006}]{pauli06}
{Pauli} E.-M.,  {Napiwotzki} R.,  {Heber} U.,  {Altmann} M.,   {Odenkirchen}
  M.,  2006, \mn@doi [\aap] {10.1051/0004-6361:20052730}, \href
  {http://adsabs.harvard.edu/abs/2006A%26A...447..173P} {447, 173}

\bibitem[\protect\citeauthoryear{{Przybilla}}{{Przybilla}}{2005}]{Przybilla05}
{Przybilla} N.,  2005, \mn@doi [\aap] {10.1051/0004-6361:20053412}, \href
  {http://adsabs.harvard.edu/abs/2005A%26A...443..293P} {443, 293}

\bibitem[\protect\citeauthoryear{{Przybilla} \& {Butler}}{{Przybilla} \&
  {Butler}}{2001}]{PrBu01}
{Przybilla} N.,  {Butler} K.,  2001, \mn@doi [\aap]
  {10.1051/0004-6361:20011393}, \href
  {http://adsabs.harvard.edu/abs/2001A%26A...379..955P} {379, 955}

\bibitem[\protect\citeauthoryear{{Przybilla} \& {Butler}}{{Przybilla} \&
  {Butler}}{2004}]{PrBu04}
{Przybilla} N.,  {Butler} K.,  2004, \mn@doi [\apj] {10.1086/421316}, \href
  {http://adsabs.harvard.edu/abs/2004ApJ...609.1181P} {609, 1181}

\bibitem[\protect\citeauthoryear{{Przybilla}, {Butler}, {Becker}, {Kudritzki}
  \& {Venn}}{{Przybilla} et~al.}{2000}]{Przybillaetal00}
{Przybilla} N.,  {Butler} K.,  {Becker} S.~R.,  {Kudritzki} R.~P.,   {Venn}
  K.~A.,  2000, \aap, \href
  {http://adsabs.harvard.edu/abs/2000A%26A...359.1085P} {359, 1085}

\bibitem[\protect\citeauthoryear{{Przybilla}, {Butler}, {Becker}  \&
  {Kudritzki}}{{Przybilla} et~al.}{2001a}]{Przybillaetal01a}
{Przybilla} N.,  {Butler} K.,  {Becker} S.~R.,   {Kudritzki} R.~P.,  2001a,
  \mn@doi [\aap] {10.1051/0004-6361:20010164}, \href
  {http://adsabs.harvard.edu/abs/2001A%26A...369.1009P} {369, 1009}

\bibitem[\protect\citeauthoryear{{Przybilla}, {Butler}  \&
  {Kudritzki}}{{Przybilla} et~al.}{2001b}]{Przybillaetal01b}
{Przybilla} N.,  {Butler} K.,   {Kudritzki} R.~P.,  2001b, \mn@doi [\aap]
  {10.1051/0004-6361:20011384}, \href
  {http://adsabs.harvard.edu/abs/2001A%26A...379..936P} {379, 936}

\bibitem[\protect\citeauthoryear{{Przybilla}, {Butler}, {Heber}  \&
  {Jeffery}}{{Przybilla} et~al.}{2005}]{prz05}
{Przybilla} N.,  {Butler} K.,  {Heber} U.,   {Jeffery} C.~S.,  2005, \mn@doi
  [\aap] {10.1051/0004-6361:200500195}, \href
  {http://adsabs.harvard.edu/abs/2005A%26A...443L..25P} {443, L25}

\bibitem[\protect\citeauthoryear{{Przybilla}, {Butler}, {Becker}  \&
  {Kudritzki}}{{Przybilla} et~al.}{2006}]{prz06a}
{Przybilla} N.,  {Butler} K.,  {Becker} S.~R.,   {Kudritzki} R.~P.,  2006,
  \mn@doi [\aap] {10.1051/0004-6361:20053832}, \href
  {http://adsabs.harvard.edu/abs/2006A%26A...445.1099P} {445, 1099}

\bibitem[\protect\citeauthoryear{{Przybilla}, {Nieva}  \& {Butler}}{{Przybilla}
  et~al.}{2011}]{prz11}
{Przybilla} N.,  {Nieva} M.-F.,   {Butler} K.,  2011, \mn@doi [Journal of
  Physics Conference Series] {10.1088/1742-6596/328/1/012015}, \href
  {http://adsabs.harvard.edu/abs/2011JPhCS.328a2015P} {328, 012015}

\bibitem[\protect\citeauthoryear{{Przybilla} et~al.,}{{Przybilla}
  et~al.}{2016}]{Przybillaetal16}
{Przybilla} N.,  et~al., 2016, \mn@doi [\aap] {10.1051/0004-6361/201527646},
  \href {http://adsabs.harvard.edu/abs/2016A%26A...587A...7P} {587, A7}

\bibitem[\protect\citeauthoryear{{Ramspeck}, {Heber}  \& {Edelmann}}{{Ramspeck}
  et~al.}{2001}]{ram01}
{Ramspeck} M.,  {Heber} U.,   {Edelmann} H.,  2001, \mn@doi [\aap]
  {10.1051/0004-6361:20011334}, \href
  {http://adsabs.harvard.edu/abs/2001A%26A...379..235R} {379, 235}

\bibitem[\protect\citeauthoryear{{Randall}, {Bagnulo}, {Ziegerer}, {Geier}  \&
  {Fontaine}}{{Randall} et~al.}{2015}]{2015A&A...576A..65R}
{Randall} S.~K.,  {Bagnulo} S.,  {Ziegerer} E.,  {Geier} S.,   {Fontaine} G.,
  2015, \mn@doi [\aap] {10.1051/0004-6361/201425251}, \href
  {http://adsabs.harvard.edu/abs/2015A%26A...576A..65R} {576, A65}

\bibitem[\protect\citeauthoryear{{Rybicki} \& {Hummer}}{{Rybicki} \&
  {Hummer}}{1991}]{RyHu91}
{Rybicki} G.~B.,  {Hummer} D.~G.,  1991, \aap, \href
  {http://adsabs.harvard.edu/abs/1991A%26A...245..171R} {245, 171}

\bibitem[\protect\citeauthoryear{{Saio}}{{Saio}}{1988}]{saio88a}
{Saio} H.,  1988, \mnras, \href
  {http://ukads.nottingham.ac.uk/abs/1988MNRAS.235..203S} {235, 203}

\bibitem[\protect\citeauthoryear{{Saio} \& {Jeffery}}{{Saio} \&
  {Jeffery}}{2000}]{saio00}
{Saio} H.,  {Jeffery} C.~S.,  2000, \mnras, \href
  {http://ukads.nottingham.ac.uk/abs/2000MNRAS.313..671S} {313, 671}

\bibitem[\protect\citeauthoryear{{Saio} \& {Jeffery}}{{Saio} \&
  {Jeffery}}{2002}]{saio02}
{Saio} H.,  {Jeffery} C.~S.,  2002, \mn@doi [\mnras]
  {10.1046/j.1365-8711.2002.05384.x}, \href
  {http://adsabs.harvard.edu/abs/2002MNRAS.333..121S} {333, 121}

\bibitem[\protect\citeauthoryear{{Schlafly} \& {Finkbeiner}}{{Schlafly} \&
  {Finkbeiner}}{2011}]{SchFi11}
{Schlafly} E.~F.,  {Finkbeiner} D.~P.,  2011, \mn@doi [\apj]
  {10.1088/0004-637X/737/2/103}, \href
  {http://adsabs.harvard.edu/abs/2011ApJ...737..103S} {737, 103}

\bibitem[\protect\citeauthoryear{{Sch\"{o}nberner}}{{Sch\"{o}nberner}}{1977}]{schoenberner77}
{Sch\"{o}nberner} D.,  1977, \aap, \href
  {http://ukads.nottingham.ac.uk/abs/1977A%26A....57..437S} {57, 437}

\bibitem[\protect\citeauthoryear{{Sch\"{o}nberner}}{{Sch\"{o}nberner}}{1979}]{schoenberner79}
{Sch\"{o}nberner} D.,  1979, \aap, \href
  {http://adsabs.harvard.edu/abs/1979A%26A....79..108S} {79, 108}

\bibitem[\protect\citeauthoryear{{Shamey}}{{Shamey}}{1969}]{Sha69}
{Shamey} L.~J.,  1969, PhD thesis, University of Colorado at Boulder

\bibitem[\protect\citeauthoryear{{Smart}}{{Smart}}{2016}]{2016A&C....15...29S}
{Smart} R.~L.,  2016, \mn@doi [Astronomy and Computing]
  {10.1016/j.ascom.2016.01.004}, \href
  {http://adsabs.harvard.edu/abs/2016A%26C....15...29S} {15, 29}

\bibitem[\protect\citeauthoryear{{Underhill}}{{Underhill}}{1966}]{Underhill66}
{Underhill} A.~B.,  ed. 1966, {The early type stars (Dordrecht: Reidel)}

\bibitem[\protect\citeauthoryear{{Vidal}, {Cooper}  \& {Smith}}{{Vidal}
  et~al.}{1973}]{Vidaletal73}
{Vidal} C.~R.,  {Cooper} J.,   {Smith} E.~W.,  1973, \mn@doi [\apjs]
  {10.1086/190264}, \href {http://adsabs.harvard.edu/abs/1973ApJS...25...37V}
  {25, 37}

\bibitem[\protect\citeauthoryear{{Vrancken}, {Butler}  \& {Becker}}{{Vrancken}
  et~al.}{1996}]{Vranckenetal96}
{Vrancken} M.,  {Butler} K.,   {Becker} S.~R.,  1996, \aap, \href
  {http://adsabs.harvard.edu/abs/1996A%26A...311..661V} {311, 661}

\bibitem[\protect\citeauthoryear{{Webbink}}{{Webbink}}{1984}]{web84}
{Webbink} R.~F.,  1984, \mn@doi [\apj] {10.1086/161701}, \href
  {http://adsabs.harvard.edu/abs/1984ApJ...277..355W} {277, 355}

\bibitem[\protect\citeauthoryear{{Weiss}}{{Weiss}}{1987}]{weiss87a}
{Weiss} A.,  1987, \aap, \href
  {http://adsabs.harvard.edu/abs/1987A%26A...185..165W} {185, 165}

\bibitem[\protect\citeauthoryear{{Wiese} \& {Fuhr}}{{Wiese} \&
  {Fuhr}}{2009}]{WiFu09}
{Wiese} W.~L.,  {Fuhr} J.~R.,  2009, \mn@doi [J. Phys. \& Chem. Ref. Data.]
  {10.1063/1.3077727}, \href
  {http://adsabs.harvard.edu/abs/2009JPCRD..38..565W} {38, 565}

\bibitem[\protect\citeauthoryear{{Wiese}, {Smith}  \& {Miles}}{{Wiese}
  et~al.}{1969}]{WSM69}
{Wiese} W.~L.,  {Smith} M.~W.,   {Miles} B.~M.,  1969, Nat. Stand. Ref. Data
  Ser., Nat. Bur. Stand. (US), NSRDS-NBS 22, Vol. 2

\bibitem[\protect\citeauthoryear{{Wiese}, {Fuhr}  \& {Deters}}{{Wiese}
  et~al.}{1996}]{wiese96}
{Wiese} W.~L.,  {Fuhr} J.~R.,   {Deters} T.~M.,  1996, J. Phys. \& Chem. Ref.
  Data., Monograph 7

\bibitem[\protect\citeauthoryear{{Woolf} \& {Jeffery}}{{Woolf} \&
  {Jeffery}}{2000}]{woolf00}
{Woolf} V.~M.,  {Jeffery} C.~S.,  2000, \aap, \href
  {http://adsabs.harvard.edu/abs/2000A\%26A...358.1001W} {358, 1001}

\bibitem[\protect\citeauthoryear{{Woolf} \& {Jeffery}}{{Woolf} \&
  {Jeffery}}{2002}]{woolf02}
{Woolf} V.~M.,  {Jeffery} C.~S.,  2002, \mn@doi [\aap]
  {10.1051/0004-6361:20021113}, \href
  {http://cdsads.u-strasbg.fr/abs/2002A%26A...395..535W} {395, 535}

\bibitem[\protect\citeauthoryear{{Zhang} \& {Jeffery}}{{Zhang} \&
  {Jeffery}}{2012}]{zhang12b}
{Zhang} X.,  {Jeffery} C.~S.,  2012, \mn@doi [\mnras]
  {10.1111/j.1745-3933.2012.01330.x}, \href
  {http://adsabs.harvard.edu/abs/2012MNRAS.426L..81Z} {426, L81}

\bibitem[\protect\citeauthoryear{{Zhang}, {Jeffery}, {Chen}  \& {Han}}{{Zhang}
  et~al.}{2014}]{zhang14}
{Zhang} X.,  {Jeffery} C.~S.,  {Chen} X.,   {Han} Z.,  2014, \mn@doi [\mnras]
  {10.1093/mnras/stu1741}, \href
  {http://adsabs.harvard.edu/abs/2014MNRAS.445..660Z} {445, 660}

\bibitem[\protect\citeauthoryear{{Zhang}, {Hall}, {Jeffery}  \& {Bi}}{{Zhang}
  et~al.}{2017}]{zhang17}
{Zhang} X.,  {Hall} P.~D.,  {Jeffery} C.~S.,   {Bi} S.,  2017, \mn@doi [\apj]
  {10.3847/1538-4357/835/2/242}, \href
  {http://adsabs.harvard.edu/abs/2017ApJ...835..242Z} {835, 242}

\makeatother
\end{thebibliography}




\appendix
\section{The metal line spectrum in BD+10$^\circ$2179}


\begin{table}
\caption{Line by line abundances of the non-LTE analysis of BD+10$^\circ$2179. Abundances $\epsilon$ with respect to $\log\sum\mu_X\epsilon(X)$\,=\,12.15, where $\mu_X$ is the atomic weight of element $X$. \label{tab:abundances_lines_nlte}}
\begin{tabular}{llrrlrl}
\hline\hline
Line & $\lambda$\,({\AA}) & $\chi$(eV) &$\log gf$ & Accu. & Source &  Abun. \\
\hline
H$\alpha$ & 6562.819 & 10.20 &    0.71000 & AAA & WF & 8.37 \\
H$\beta$  & 4861.333 & 10.20 & $-$0.01996 & AAA & WF & 8.36 \\
H$\gamma$ & 4340.471 & 10.20 & $-$0.44666 & AAA & WF & 8.36 \\
H$\delta$ & 4101.742 & 10.20 & $-$0.75243 & AAA & WF & 8.36 \\[2mm]
\ion{C}{i}  & 4762.313 &  7.48 & $-$2.463 & C & WFD & 9.77 \\
\ion{C}{i}  & 4762.533 &  7.48 & $-$2.335 & C & WFD &      \\
\ion{C}{i}  & 4771.742 &  7.49 & $-$1.866 & C & WFD & 9.70 \\
\ion{C}{i}  & 4775.898 &  7.49 & $-$2.304 & C & WFD & 9.84  \\
\ion{C}{i}  & 4932.049 &  7.68 & $-$1.703 & B & LP  & 9.57  \\
\ion{C}{i}  & 5052.167 &  7.68 & $-$1.447 & C & LP  & 9.67  \\
\ion{C}{i}  & 5380.337 &  7.68 & $-$1.615 & B & WFD & 9.65   \\
\ion{C}{i}  & 6013.213 &  8.65 & $-$1.673 & D & WFD & 9.56  \\
\ion{C}{i}  & 6587.610 &  8.54 & $-$1.003 & B & WFD & 9.58   \\
\ion{C}{i}  & 7115.172 &  8.64 & $-$0.935 & B$-$ & WFD &  9.72  \\
\ion{C}{i}  & 7115.182 &  8.64 & $-$1.473 & B & WFD &    \\[2mm]
\ion{C}{ii} & 4267.001 & 18.05 &    0.562 & C+ & WFD & 9.78 \\
\ion{C}{ii} & 4267.261 & 18.05 &    0.717 & C+ & WFD &  \\
\ion{C}{ii} & 4267.261 & 18.05 & $-$0.584 & C+ & WFD &  \\
\ion{C}{ii} & 4737.966 & 13.72 & $-$3.444 & B+ & WFD &  9.57  \\
\ion{C}{ii} & 4744.766 & 13.72 & $-$3.111 & B+ & WFD & 9.52   \\
\ion{C}{ii} & 4747.279 & 13.72 & $-$3.820 & B+ & WFD &  9.55  \\
\ion{C}{ii} & 5137.257 & 20.70 & $-$0.911 & B  & WFD & 9.76  \\
\ion{C}{ii} & 5139.174 & 20.70 & $-$0.707 & B  & WFD & 9.82  \\
\ion{C}{ii} & 5143.495 & 20.70 & $-$0.212 & B  & WFD & 9.89  \\
\ion{C}{ii} & 5145.165 & 20.71 &    0.189 & B  & WFD & 9.93  \\
\ion{C}{ii} & 5535.353 & 19.49 & $-$1.493 & B  & WFD & 9.93  \\
\ion{C}{ii} & 5537.609 & 19.49 & $-$1.794 & B  & WFD & 9.90  \\
\ion{C}{ii} & 6578.052 & 14.45 & $-$0.087 & C+ & N02 & 9.65  \\
\ion{C}{ii} & 6582.882 & 14.45 & $-$0.388 & C+ & N02 &   \\
\ion{C}{ii} & 6727.070 & 22.53 & $-$1.065 & B  & WFD & 9.74  \\
\ion{C}{ii} & 6727.260 & 22.53 & $-$0.919 & B  & WFD &   \\
\ion{C}{ii} & 6731.070 & 22.53 & $-$0.862 & B  & WFD & 9.77   \\
\ion{C}{ii} & 6733.581 & 22.53 & $-$1.066 & B  & WFD & 9.75   \\
\ion{C}{ii} & 6734.003 & 22.53 & $-$1.009 & B  & WFD &    \\ 
\ion{C}{ii} & 6738.606 & 22.53 & $-$0.529 & B  & WFD & 9.82  \\
\ion{C}{ii} & 6742.428 & 22.53 & $-$0.920 & B  & WFD & 9.78   \\
\ion{C}{ii} & 6750.536 & 22.54 & $-$0.229 & B  & WFD & 9.79   \\
\ion{C}{ii} & 6755.161 & 22.54 & $-$1.010 & B  & WFD & 9.75   \\
\ion{C}{ii} & 6779.942 & 20.70 &    0.025 & B  & WFD & 9.96   \\
\ion{C}{ii} & 6783.908 & 20.71 &    0.304 & B  & WFD & 9.86   \\
\ion{C}{ii} & 6787.207 & 20.70 & $-$0.377 & B  & WFD & 9.93   \\
\ion{C}{ii} & 6791.466 & 20.70 & $-$0.270 & B  & WFD & 9.92   \\
\ion{C}{ii} & 6798.104 & 20.70 & $-$1.077 & B  & WFD & 9.82   \\
\ion{C}{ii} & 6800.683 & 20.71 & $-$0.343 & B  & WFD & 9.80   \\
\ion{C}{ii} & 6812.280 & 20.71 & $-$1.300 & B  & WFD & 9.54   \\[2mm]
\ion{C}{iii}& 4186.900 & 40.01 &    0.918 & B  & WFD & 9.76   \\
\ion{C}{iii}& 4647.418 & 29.53 &    0.070 & B+ & WFD & 9.67   \\
\ion{C}{iii}& 4650.246 & 29.53 & $-$0.151 & B+ & WFD & 9.73   \\[2mm]
\ion{N}{i} & 7442.298 &  10.33 & $-$0.384 & B+ & WFD  & 7.99 \\
\ion{N}{i} & 7468.312 &  10.34 & $-$0.189 & B+ & WFD  & 7.86 \\
\ion{N}{i} & 8680.282 &  10.34 &    0.346 & B+ & WFD  & 7.88 \\
\ion{N}{i} & 8683.403 &  10.33 &    0.086 & B+ & WFD  & 7.93 \\
\ion{N}{i} & 8686.149 &  10.33 & $-$0.305 & B+ & WFD  & 8.01 \\
\ion{N}{i} & 8703.247 &  10.33 & $-$0.310 & B+ & WFD  & 7.93 \\
\hline
\end{tabular}
\end{table}

\begin{table}
\begin{tabular}{llrrlrl}
\multicolumn{2}{}{} {\bf Table A1} continued. \\
\hline\hline
Line & $\lambda$\,({\AA}) & $\chi$(eV) &$\log gf$ & Accu. & Source &  Abun. \\
\hline
\ion{N}{ii}& 3955.851 &  21.15 & $-$0.813 & B  & WFD  & 7.99 \\
\ion{N}{ii}& 3994.997 &  18.50 & 0.163    & B  & FFT  & 8.03 \\
\ion{N}{ii}& 4035.081 &  23.12 & 0.599    & B  & BB89 & 7.89  \\
\ion{N}{ii}& 4041.310 &  23.14 & 0.748    & B  & MAR  & 7.97   \\
\ion{N}{ii}& 4043.532 &  23.13 & 0.440    & C  & MAR  & 8.19   \\
\ion{N}{ii}& 4176.159 &  23.20 & 0.316    & B  & MAR  & 8.04  \\
\ion{N}{ii}& 4227.736 &  21.60 & $-$0.060 & B  & WFD  & 8.03   \\
\ion{N}{ii}& 4236.927 &  23.24 & 0.383    & X  & KB   & 7.84   \\
\ion{N}{ii}& 4241.755 &  23.24 & 0.210    & X  & KB   & 7.88  \\
\ion{N}{ii}& 4447.030 &  20.41 & 0.221    & B  & FFT  & 7.92  \\
\ion{N}{ii}& 4507.560 &  20.67 & $-$0.817 & B  & WFD  & 7.92   \\
\ion{N}{ii}& 4530.410 &  23.47 &    0.604 & C+ & MAR  & 7.99  \\ 
\ion{N}{ii}& 4601.478 &  18.47 & $-$0.452 & B+ & FFT  & 8.10  \\
\ion{N}{ii}& 4607.153 &  18.46 & $-$0.522 & B+ & FFT  & 8.09  \\
\ion{N}{ii}& 4613.868 &  18.47 & $-$0.692 & B+ & FFT  & 7.99  \\
\ion{N}{ii}& 4630.539 &  18.48 &    0.080 & B+ & FFT  & 8.22  \\
\ion{N}{ii}& 4643.086 &  18.48 & $-$0.371 & B+ & FFT  & 8.06  \\
\ion{N}{ii}& 4654.531 &  18.50 & $-$1.404 & C+ & WFD  & 7.84  \\
\ion{N}{ii}& 4667.208 &  18.50 & $-$1.533 & C+ & WFD  & 7.94  \\
\ion{N}{ii}& 4694.642 &  23.57 &    0.100 & X  & KB   & 8.21  \\ 
\ion{N}{ii}& 4788.138 &  20.65 & $-$0.366 & B  & FFT  & 7.99   \\
\ion{N}{ii}& 4793.648 &  20.65 & $-$1.032 & B+ & FFT  & 8.04  \\
\ion{N}{ii}& 4987.376 &  20.94 & $-$0.584 & B  & FFT  & 8.03  \\
\ion{N}{ii}& 5001.474 &  20.65 &    0.435 & B  & FFT  & 8.10  \\
\ion{N}{ii}& 5002.703 &  18.46 & $-$1.022 & B+ & WFD  & 7.99  \\
\ion{N}{ii}& 5005.150 &  20.67 &    0.587 & B  & FFT  & 8.24  \\
\ion{N}{ii}& 5007.328 &  20.94 &    0.145 & B  & FFT  & 8.15   \\
\ion{N}{ii}& 5010.621 &  18.47 & $-$0.607 & B+ & WFD  & 8.07  \\
\ion{N}{ii}& 5025.659 &  20.67 & $-$0.557 & B  & FFT  & 8.11  \\
\ion{N}{ii}& 5452.070 &  21.15 & $-$0.783 & B  & FFT  & 8.13  \\
\ion{N}{ii}& 5462.581 &  21.15 & $-$0.826 & B+ & FFT  & 7.78   \\
\ion{N}{ii}& 5480.050 &  21.16 & $-$0.711 & B+ & FFT  & 8.08 \\
\ion{N}{ii}& 5495.655 &  21.16 & $-$0.220 & B+ & FFT  & 8.01 \\
\ion{N}{ii}& 5666.629 &  18.47 & $-$0.104 & B+ & MAR  & 8.14  \\
\ion{N}{ii}& 5676.017 &  18.46 & $-$0.356 & B+ & MAR  & 8.10 \\
\ion{N}{ii}& 5679.558 &  18.48 & 0.221    & B+ & MAR  & 8.18  \\
\ion{N}{ii}& 5686.213 &  18.47 & $-$0.586 & B+ & FFT  & 8.14  \\
\ion{N}{ii}& 5710.766 &  18.48 & $-$0.466 & B  & MAR  & 8.01  \\
\ion{N}{ii}& 5747.300 &  18.50 & $-$1.092 & B+ & FFT  & 8.09  \\
\ion{N}{ii}& 5767.446 &  18.50 & $-$1.447 & B  & FFT  & 7.98   \\
\ion{N}{ii}& 5931.782 &  21.15 &    0.047 & A  & FFT  & 7.99  \\
\ion{N}{ii}& 5941.654 &  21.16 &    0.307 & A  & FFT  & 8.18  \\
\ion{N}{ii}& 6379.627 &  18.47 & $-$1.188 & C+ & FFT  & 8.05  \\
\ion{N}{ii}& 6482.048 &  18.50 & $-$0.311 & B+ & FFT  & 8.18  \\
\ion{N}{ii}& 6610.562 &  21.60 & 0.440    & B  & FFT  & 8.15  \\[2mm]
\ion{O}{i} &7771.944  &   9.15 & 0.354    & A  & FFT  & 7.40 \\ 
\ion{O}{i} &7774.166  &   9.15 & 0.207    & A  & FFT  & \\
\ion{O}{i} &7775.388  &   9.15 & $-$0.015 & A  & FFT  & \\[2mm]
\ion{O}{ii}& 3911.957 &  25.66 & $-$0.014 & B+ & FFT  &  7.59 \\
\ion{O}{ii}& 3912.107 &  25.66 & $-$0.907 & B+ & FFT  &   \\
\ion{O}{ii}& 4345.560 &  22.98 & $-$0.342 & B+ & FFT  &  7.48 \\
\ion{O}{ii}& 4349.426 &  23.00 &    0.073 & B+ & FFT  &  7.50 \\
\ion{O}{ii}& 4366.892 &  23.00 & $-$0.333 & B+ & FFT  &  7.47 \\
\ion{O}{ii}& 4414.905 &  23.44 &    0.207 & B  & FFT  &  7.54 \\
\ion{O}{ii}& 4596.175 &  25.66 &    0.180 & B+ & FFT  &  7.45 \\
\ion{O}{ii}& 4641.810 &  22.98 &    0.066 & B+ & FFT  &  7.50 \\
\ion{O}{ii}& 4650.839 &  22.97 & $-$0.349 & B+ & FFT  &  7.58 \\
\ion{O}{ii}& 4676.235 &  23.00 & $-$0.410 & B+ & FFT  &  7.53 \\
\ion{O}{ii}& 4705.352 &  26.25 & 0.533    & B+ & FFT  &  7.53 \\
\hline
\end{tabular}
\end{table}

\begin{table}
\begin{tabular}{llrrlrl}
\multicolumn{2}{}{} {\bf Table A1} continued. \\
\hline\hline
Line & $\lambda$\,({\AA}) & $\chi$(eV) &$\log gf$ & Accu. & Source &  Abun. \\
\hline
\ion{Ne}{i}& 5764.419 &  18.56 & $-$0.316 & B &  B08  & 8.14 \\
\ion{Ne}{i}& 5852.488 &  16.85 & $-$0.454 & B &  B08  & 8.07 \\
\ion{Ne}{i}& 5881.895 &  16.62 & $-$0.792 & B &  B08  & 7.91 \\
\ion{Ne}{i}& 5944.834 &  16.62 & $-$0.636 & B &  B08  & 8.05 \\
\ion{Ne}{i}& 6029.997 &  16.67 & $-$1.026 & B  & FFT  & 7.98 \\
\ion{Ne}{i}& 6074.338 &  16.67 & $-$0.473 & B+ & FFT  & 7.92 \\
\ion{Ne}{i}& 6143.063 &  16.62 & $-$0.070 & B+ & FFT  & 7.90 \\
\ion{Ne}{i}& 6163.594 &  16.72 & $-$0.598 & B+ & FFT  & 8.04 \\
\ion{Ne}{i}& 6217.281 &  16.62 & $-$0.943 & B  & FFT  & 7.88 \\
\ion{Ne}{i}& 6266.495 &  16.72 & $-$0.331 & B+ & FFT  & 8.00 \\
\ion{Ne}{i}& 6334.428 &  16.62 & $-$0.277 & B+ & FFT  & 7.92 \\
\ion{Ne}{i}& 6506.528 &  16.67 & $-$0.002 & B+ & FFT  & 7.90 \\
\ion{Ne}{i}& 6532.882 &  16.72 & $-$0.670 & B+ & FFT  & 8.10 \\
\ion{Ne}{i}& 6598.953 &  16.85 & $-$0.360 & B  & B08  & 8.18 \\
\ion{Ne}{i}& 6717.043 &  16.84 & $-$0.346 & B+ & FFT  & 8.01 \\
\ion{Ne}{i}& 7032.413 &  16.62 & $-$0.222 & B+ & FFT  & 7.90 \\[2mm]
\ion{Mg}{ii}& 4481.126&   8.86 &    0.730 & B  & FW   & 6.89 \\
\ion{Mg}{ii}& 4481.150&   8.86 & $-$0.570 & B  & FW   &  \\
\ion{Mg}{ii}& 4481.325&   8.86 &    0.575 & B  & FW   &  \\
\ion{Mg}{ii}& 4851.082&  11.63 & $-$0.424 & C  & CA   & 6.92 \\ 
\ion{Mg}{ii}& 6545.973&  11.63 &    0.408 & C  & CA   & 7.06 \\
\ion{Mg}{ii}& 7877.054&  10.00 &    0.391 & A+ & FFTI & 6.92 \\
\ion{Mg}{ii}& 7896.042&  10.00 & $-$0.308 & A  & FFTI & 7.01 \\
\ion{Mg}{ii}& 7896.366&  10.00 &    0.647 & A+ & FFTI &  \\[2mm]
\ion{Al}{ii}& 4663.046&  10.60 & $-$0.241 & A+ & FFTI & 5.84 \\
\ion{Al}{ii}& 6243.073 & 13.08 & $-$1.250 & X  & KB   & 5.79 \\
\ion{Al}{ii}& 6243.203 & 13.08 & $-$0.080 & X  & KB   & \\
\ion{Al}{ii}& 6243.367 & 13.08 &    0.670 & X  & KB   &  \\
\ion{Al}{ii}& 7042.083 & 11.32 &    0.332 & A+ & FFTI & 5.86 \\[2mm]
\ion{Al}{iii}& 4149.913& 20.55 &    0.626 & A+ & FFTI & 5.88 \\
\ion{Al}{iii}& 4149.968& 20.55 & $-$0.674 & A+ & FFTI &  \\
\ion{Al}{iii}& 4479.885& 20.78 &    0.900 & X  & KB   & 5.78 \\
\ion{Al}{iii}& 4479.971& 20.78 &    1.020 & X  & KB   &  \\
\ion{Al}{iii}& 4480.009& 20.78 & $-$0.530 & X  & KB   &  \\
\ion{Al}{iii}& 4512.565& 17.81 &    0.408 & A+ & FFTI & 5.78 \\
\ion{Al}{iii}& 4528.945& 17.82 & $-$0.291 & A+ & FFTI & 5.80 \\
\ion{Al}{iii}& 4529.189& 17.82 &    0.663 & A+ & FFTI &  \\
\ion{Al}{iii}& 5696.604& 15.64 &    0.232 & A+ & FFTI & 5.89 \\
\ion{Al}{iii}& 5722.730& 15.64 & $-$0.071 & A+ & FFTI & 5.88 \\[2mm]
\ion{Si}{ii}& 4128.054 &  9.84 &    0.31  & C  & WSM  &  7.12  \\
\ion{Si}{ii}& 4130.872 &  9.84 & $-$0.84  & E  & WSM  &  7.03 \\
\ion{Si}{ii}& 4130.894 &  9.84 &    0.462 & C  & WSM  &  \\
\ion{Si}{ii}& 5055.984 & 10.07 &    0.42  & D+ & WSM  &  7.07 \\
\ion{Si}{ii}& 5056.317 & 10.07 & $-$0.53  & E  & WSM  &  \\
\ion{Si}{ii}& 6371.371 &  8.12 & $-$0.126 & C+ & FFTI &  7.24 \\[2mm]
\ion{Si}{iii}&4338.500 & 19.02 & $-$1.673 & C+ & FFTI &  7.08 \\
\ion{Si}{iii}&4567.840 & 19.02 &    0.068 & B+ & FFTI &  7.12 \\
\ion{Si}{iii}&4574.757 & 19.02 & $-$0.409 & B  & FFTI &  7.24  \\
\ion{Si}{iii}&4813.333 & 25.98 &    0.653 & B  & CT   &  7.12 \\
\ion{Si}{iii}&4828.951 & 25.99 &    0.870 & B  & CT   &  7.18 \\
\ion{Si}{iii}&4829.030 & 25.99 & $-$0.661 & B  & CT   &  \\
\ion{Si}{iii}&4829.111 & 25.99 & $-$0.682 & B  & CT   &  \\
\ion{Si}{iii}&4829.214 & 25.99 & $-$2.201 & C  & CT   &  \\
\ion{Si}{iii}&5739.734 & 19.72 & $-$0.078 & B  & FFTI &  7.13 \\
\hline
\end{tabular}
\end{table}

\begin{table}
\begin{tabular}{llrrlrl}
\multicolumn{2}{}{} {\bf Table A1} continued. \\
\hline\hline
Line & $\lambda$\,({\AA}) & $\chi$(eV) &$\log gf$ & Accu. & Source &  Abun. \\
\hline
\ion{S}{ii} & 4153.068 & 15.90 &     0.62  & D   & WSM & 6.84 \\
\ion{S}{ii} & 4162.665 & 15.94 &     0.78  & D   & WSM & 6.76 \\
\ion{S}{ii} & 4189.681 & 15.90 &  $-$0.05  & E   & WSM & 6.90 \\
\ion{S}{ii} & 4282.595 & 16.10 &  $-$0.01  & E   & WSM & 6.76 \\
\ion{S}{ii} & 4815.552 & 13.67 &     0.09  & D   & WSM & 6.80 \\
\ion{S}{ii} & 5428.655 & 13.58 &  $-$0.13  & D   & WSM & 6.83 \\
\ion{S}{ii} & 5432.797 & 13.62 &     0.26  & D   & WSM & 6.86 \\
\ion{S}{ii} & 5473.614 & 13.58 &  $-$0.18  & D   & WSM & 6.89 \\
\ion{S}{ii} & 5509.705 & 13.62 &  $-$0.14  & D   & WSM & 6.89 \\[2mm]
\ion{Ar}{ii}& 4426.001 & 16.75 &     0.195 & B+  & FFTI& 5.89 \\
\ion{Ar}{ii}& 4430.189 & 16.81 &  $-$0.158 & B   & FFTI& 5.90 \\
\ion{Ar}{ii}& 4657.900 & 17.14 &  $-$0.245 & B   & FFTI& 5.86  \\
\ion{Ar}{ii}& 4764.864 & 17.27 &  $-$0.269 & B   & FFTI& 6.06 \\
\ion{Ar}{ii}& 4806.020 & 16.64 &     0.235 & B+  & FFTI& 5.90 \\
\ion{Ar}{ii}& 4847.810 & 16.75 &  $-$0.214 & B   & FFTI& 5.86 \\
\ion{Ar}{ii}& 4879.863 & 17.14 &     0.544 & B+  & FFTI& 6.00 \\[2mm]
\ion{Fe}{ii} &4296.566 &  2.70  & $-$3.01  & D  & FMW  & 6.67 \\
\ion{Fe}{ii} &5018.440 &  2.89  & $-$1.22  & C  & FMW  & 6.51 \\
\ion{Fe}{ii} &5169.033 &  2.89  & $-$0.87  & C  & FMW  & 6.40 \\[2mm]
\ion{Fe}{iii}&4164.731 & 20.63  &    0.92  & X  & KB   & 6.67 \\
\ion{Fe}{iii}&4164.916 & 24.65  &    1.01  & X  & KB   &  \\
\ion{Fe}{iii}&4395.755 &  8.26  & $-$2.60  & X  & KB   & 6.43 \\
\ion{Fe}{iii}&4431.019 &  8.25  & $-$2.572 & X  & KB   & 6.57 \\
\ion{Fe}{iii}&5156.111 &  8.64  & $-$2.02  & X  & KB   & 6.62 \\
\hline
\label{tab:abundances_lines_nlte}
\end{tabular}
\vspace{-4mm}
\begin{flushleft}
accuracy indicators -- uncertainties within: 
AAA: 0.3\%;
AA: 1\%;
A: 3\%; 
B: 10\%; 
C: 25\%;
D: 50\%;
E: >50\%;
X: unknown\\
sources of $gf$-values -- 
B08:   K. Butler, from Breit-Pauli R-matrix calculations as outlined in \citet{MoBu08};
BB89:  \citet{BeBu89};
CA:    Coulomb approximation, \citet{BaDa49};
CT:    \citet{CaTr98}
FFT:   \citet{FFT04}; 
FFTI:  \citet{FFTI06};
FW     \citet{FuWi98};
FMW:   \citet{fuhr88};
KB:    \citet{KuBe95};
LP:    \citet{LuPr89};
MAR:   \citet{mar00};
N02:   \citet{Nahar02};
WF:    \citet{WiFu09};
WFD:   \citet{wiese96};
WSM:   \citet{WSM69}.\\
sources of Stark broadening parameters -- 
\ion{H}{i}:    \citet{Vidaletal73}; 
\ion{C}{i}:    \citet{Griem74}, \citet{Cowley71};
\ion{Si}{ii}:  \citet{Lanzetal88}, \citet{Griem74}, \citet{Cowley71};
\ion{C}{ii}, \ion{N}{i/ii}, \ion{Mg}{ii}:  \citet{Griem64,Griem74}, \citet{Cowley71};
\ion{C}{iii}, \ion{O}{i/ii}, \ion{Ne}{i}, \ion{Al}{ii/iii}, \ion{Si}{iii}, \ion{S}{ii}, \ion{Ar}{ii}, \ion{Fe}{ii/iii}:  \citet{Cowley71}.
\end{flushleft}
\end{table}

\clearpage

\begin{figure*}
\centering
\includegraphics[width=.94\textwidth]{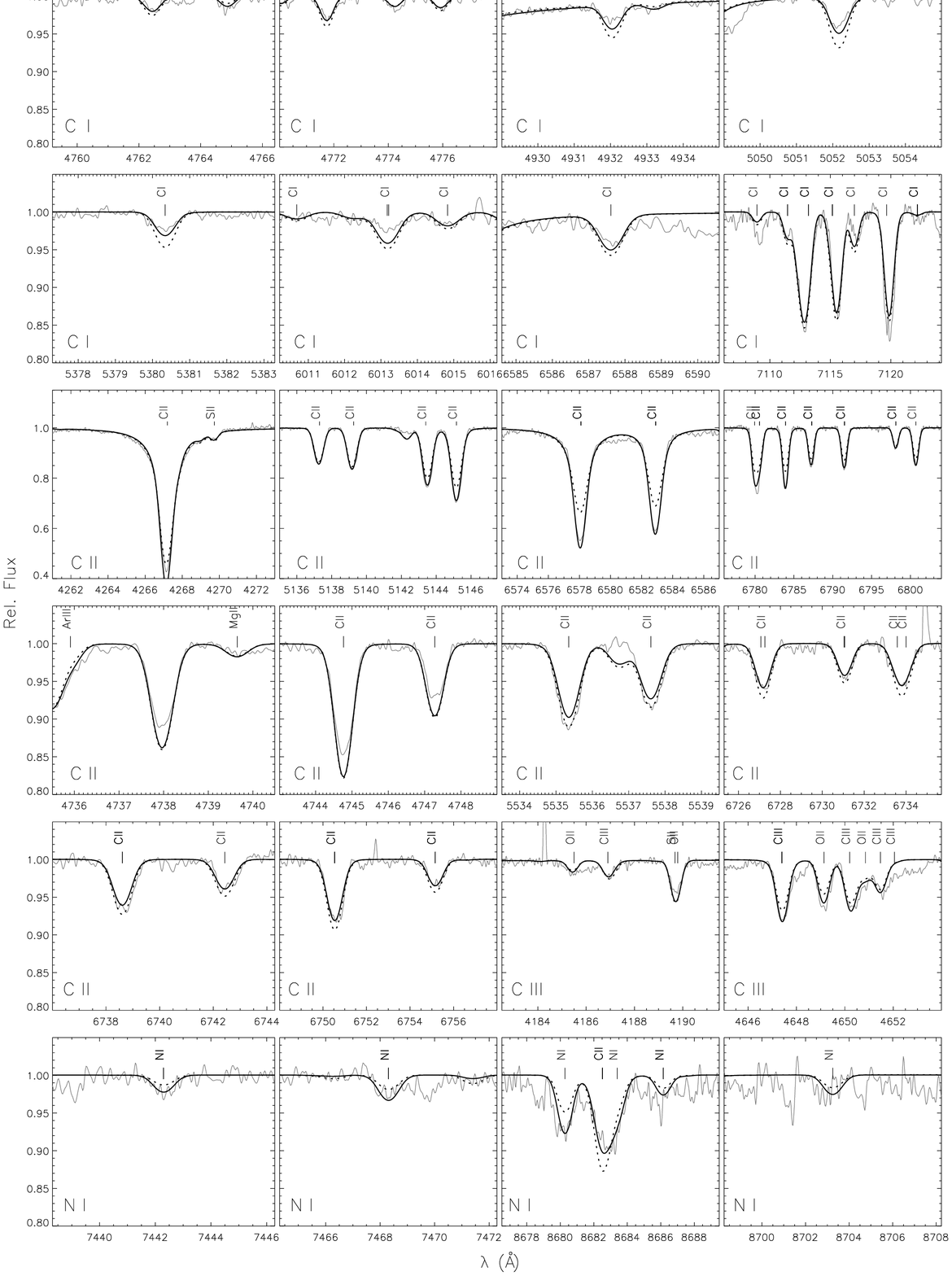}
\caption{Like Fig.~\ref{fig:hhefits}, for analysed \ion{C}{i/ii/iii} and \ion{N}{i} lines.}
\label{fig:cfits}
\end{figure*}

\begin{figure*}
\centering
\includegraphics[width=.94\textwidth]{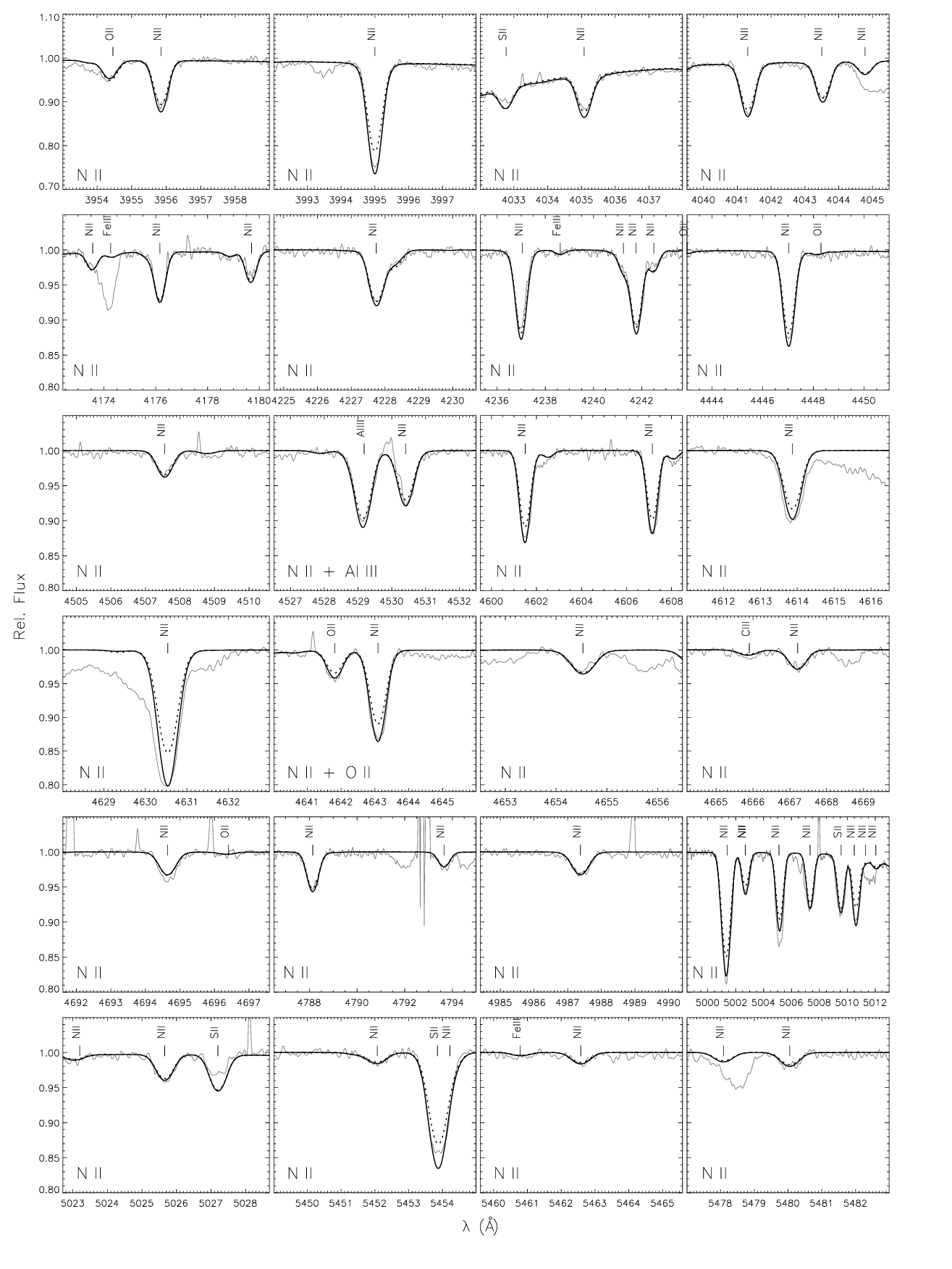}
\caption{Like Fig.~\ref{fig:hhefits}, for analysed \ion{N}{ii} lines.}
\label{fig:nfits}
\end{figure*}

\begin{figure*}
\centering
\includegraphics[width=.94\textwidth]{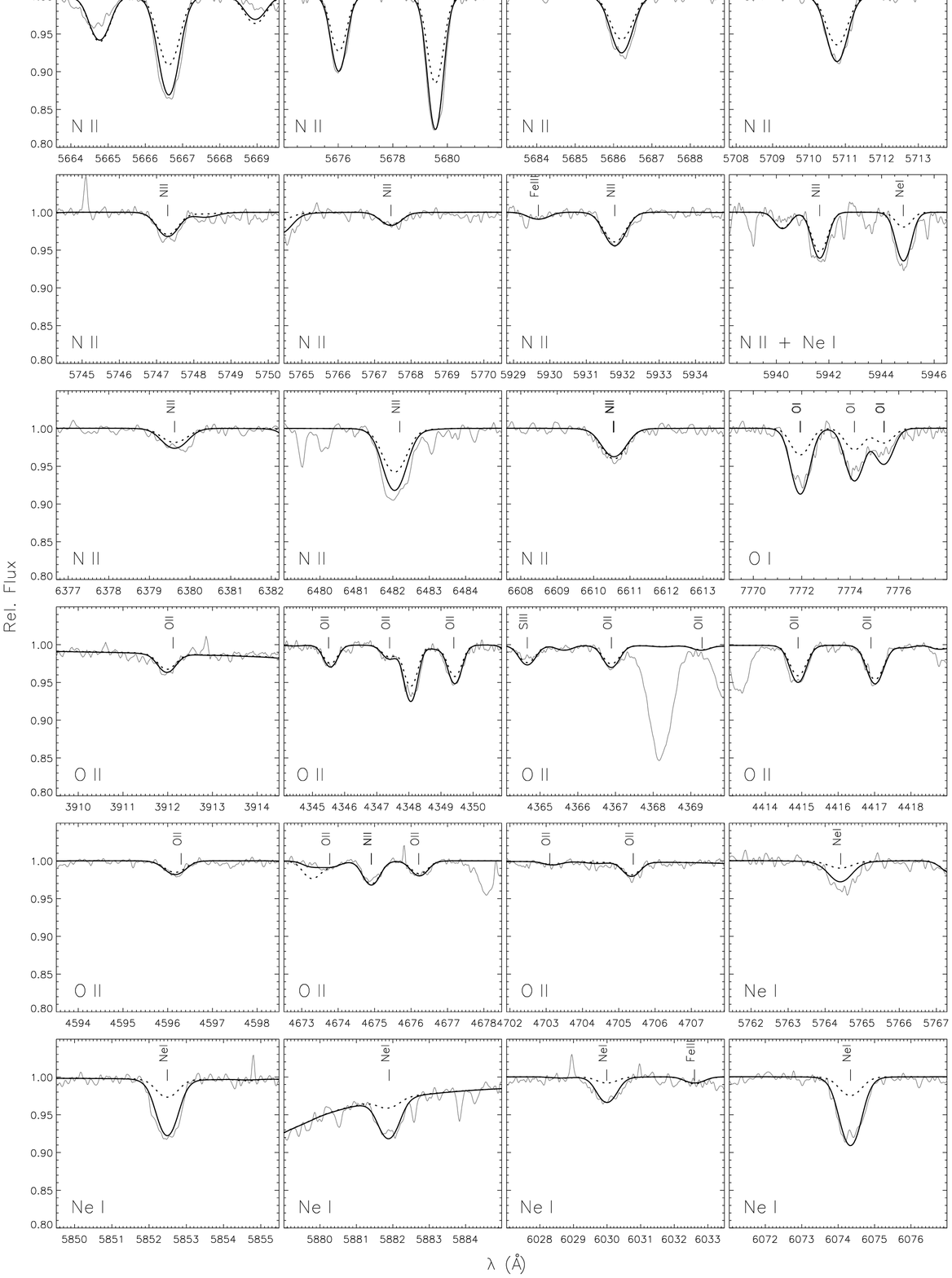}
\caption{Like Fig.~\ref{fig:hhefits}, for analysed \ion{N}{ii}, \ion{O}{i/ii} and \ion{Ne}{i} lines.}
\label{fig:nofits}
\end{figure*}

\begin{figure*}
\centering
\includegraphics[width=.94\textwidth]{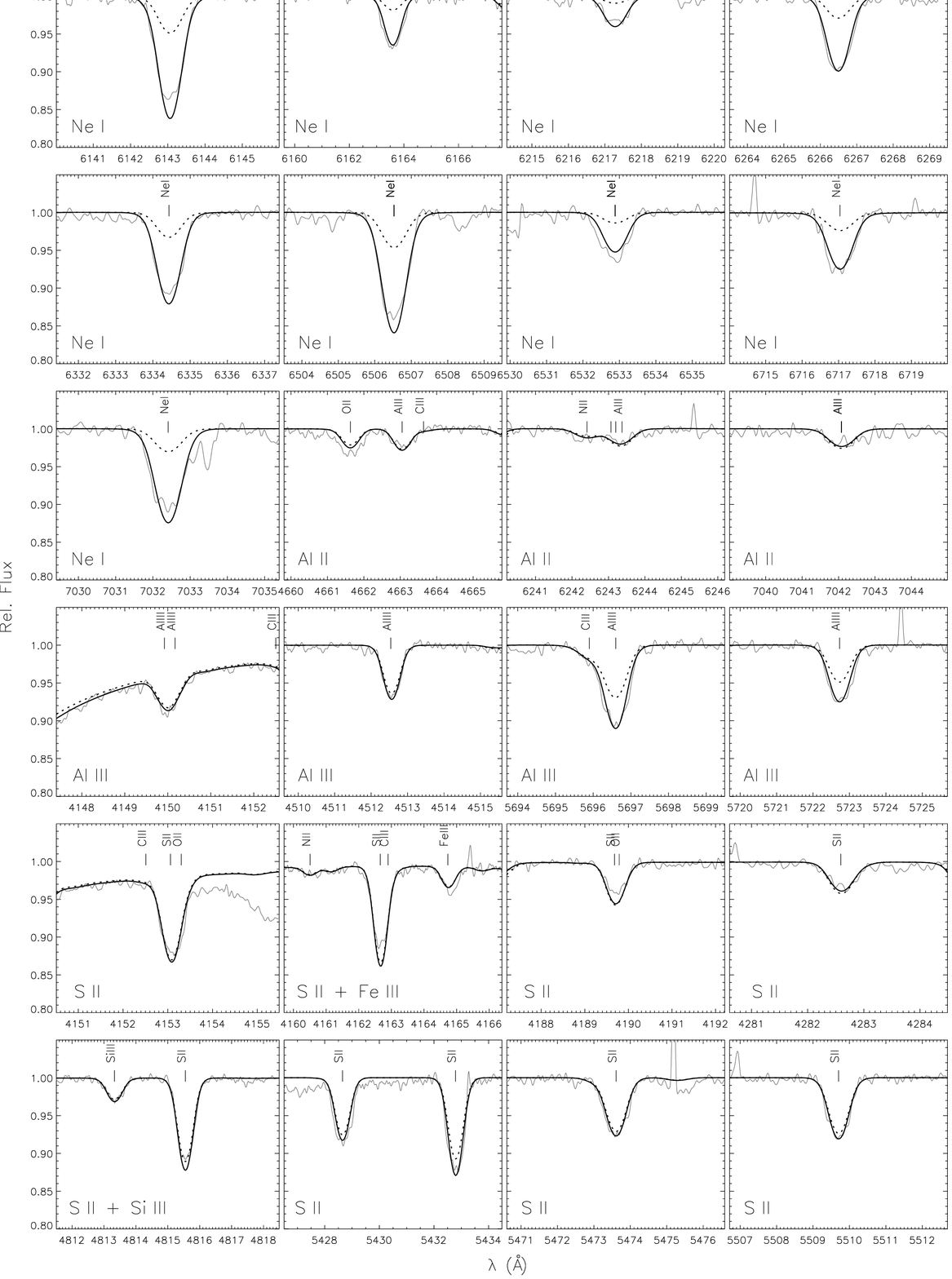}
\caption{Like Fig.~\ref{fig:hhefits}, for analysed \ion{Ne}{i}, \ion{Al}{ii/iii} and \ion{S}{ii} lines.}
\label{fig:alphafits}
\end{figure*}

\begin{figure*}
\centering
\includegraphics[width=.94\textwidth]{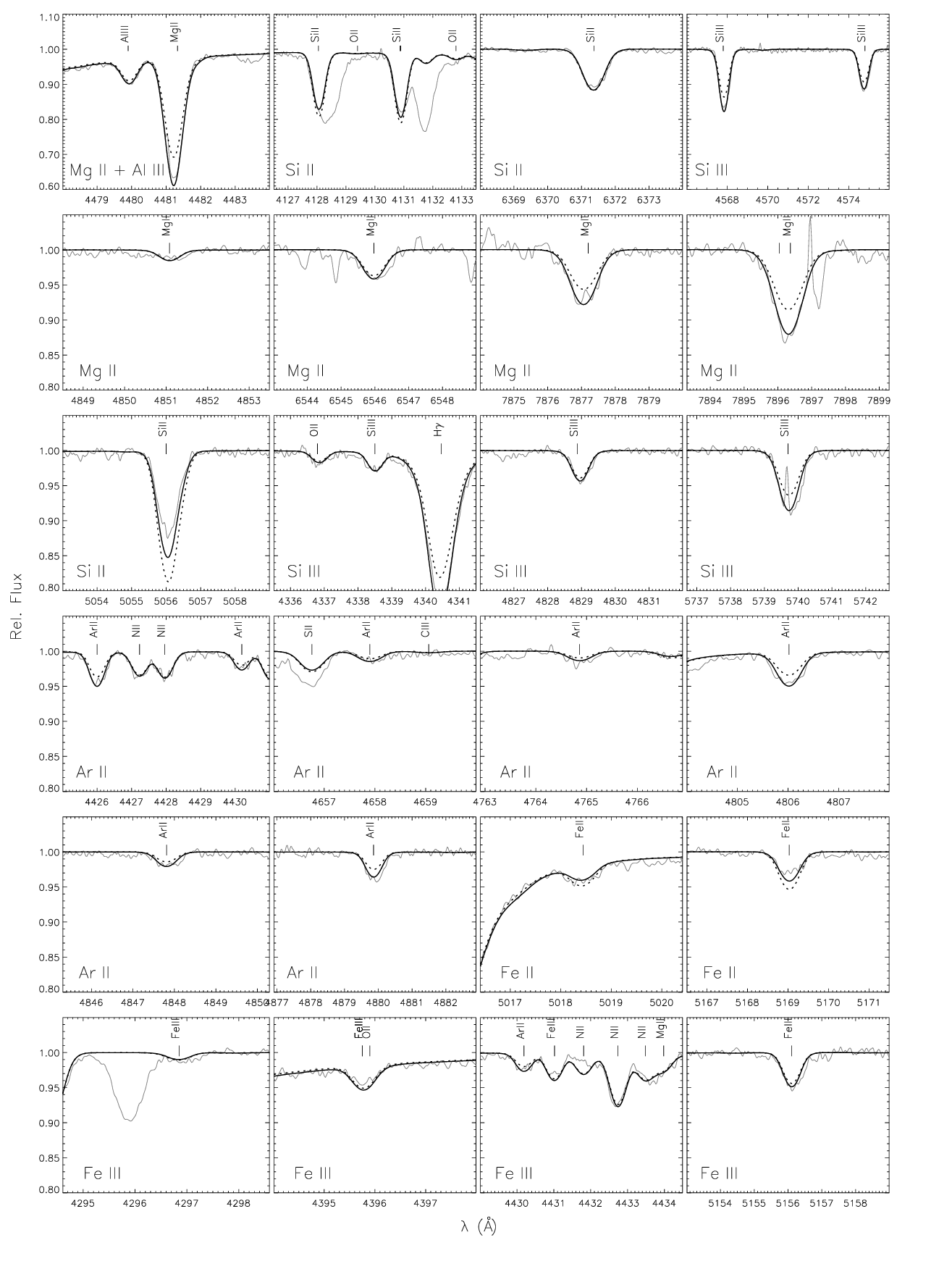}
\caption{Like Fig.~\ref{fig:hhefits}, for analysed \ion{Mg}{ii}, \ion{Si}{ii/iii}, \ion{Ar}{ii} and \ion{Fe}{ii/iii} lines.}
\label{fig:sifefits}
\end{figure*}

\clearpage

\section{The \ion{He}{i} spectrum in BD+10$^\circ$2179}
\normalsize
The peculiar composition of EHe stars in combination with low gravities facilitates a detailed empirical study of the helium line spectrum like in no other class of astronomical objects. While the spectral lines in the optical range can be modelled well with existing line data, see Sect.~\ref{sec:analysis}, this is not the case for neighbouring wavelength ranges to such a degree of completeness. In order to stimulate work to improve, in particular, the availability of line-broadening data, we want to describe the rich \ion{He}{i} line features of BD+10$^\circ$2179 in the optical-UV and the $IzY$-bands covered by our spectra. We therefore also make available the spectrum in electronic form via CDS. Throughout the optical-UV to $IzY$ wavelength range coverd by our spectrum we identify dipole-allowed transitions from about 150 multiplets of \ion{He}{i} (plus 11 in the HST STIS range).
In addition, we want to draw attention to the presence of many forbidden components in the optical spectrum, even several so far unknown ones. The spectral lines were identified based on the compilations of \citet{WiFu09}, \citet{BeWe98}, \citet{CaTh02} and the \ion{He}{i} line list of Kurucz\footnote{\tt http://kurucz.harvard.edu/linelists/gfall/gf0200.all}.

\begin{table}
\caption{\ion{He}{i} line series covered in our spectrum of BD+10$^\circ$2179}
\centering
\begin{tabular}{l@{~~--~~}llr|l@{~~--~~}llr}
\hline\hline
\multicolumn{2}{c}{transition} & $n_\mathrm{min}$ & $n_\mathrm{max}$ &
\multicolumn{2}{c}{transition} & $n_\mathrm{min}$ & $n_\mathrm{max}$\\
\hline
2$s$\,$^3$S         & $np$\,$^3$P$^\circ$ & 3 &  4/15$^*$ & 3$p$\,$^3$P$^\circ$ & $ns$\,$^3$S         & 7 & 14\\
2$s$\,$^1$S         & $np$\,$^1$P$^\circ$ & 3 & 15 & 3$p$\,$^3$P$^\circ$ & $nd$\,$^3$D         & 7 & 14\\
2$p$\,$^3$P$^\circ$ & $ns$\,$^3$S         & 3 & 20 & $3d$\,$^3$D         & $np$\,$^3$P$^\circ$ & 7 & 12\\
2$p$\,$^3$P$^\circ$ & $nd$\,$^3$D         & 3 & 20 & $3d$\,$^3$D         & $nf$\,$^3$F$^\circ$ & 7 & 14\\
2$p$\,$^1$P$^\circ$ & $ns$\,$^1$S         & 3 & 16 & $3d$\,$^1$D         & $nf$\,$^1$F$^\circ$ & 7 & 14\\
2$p$\,$^1$P$^\circ$ & $ns$\,$^1$S         & 3 & 16 & $3d$\,$^1$D         & $np$\,$^1$P$^\circ$ & 7 & 14\\
$3s$\,$^3$S         & $np$\,$^3$P$^\circ$ & 5 & 12 & 3$p$\,$^1$P$^\circ$ & $nd$\,$^1$D         & 7 & 11\\
$3s$\,$^1$S         & $np$\,$^1$P$^\circ$ & 6 & 10 & 3$p$\,$^1$P$^\circ$ & $ns$\,$^1$S         & 8 &  8\\
\hline
\end{tabular}
\vspace{-4mm}
\begin{flushleft}
$^*$ using HST STIS spectra 
\end{flushleft}
\label{table:lineseries}
\end{table}

\subsection{Optical-UV}
Figure~\ref{fig:heuv} displays the spectrum of BD+10$^\circ$2179 in the optical-UV range near the atmospheric cut-off to $\sim$3860\,{\AA}. This region is dominated by the series limits originating from the 2$s$\,$^1$S, 2$p$\,$^3$P$^\circ$ and 2$p$\,$^1$P$^\circ$ levels. The series limit of lines originating from the 2$s$\,$^3$S level is located farther in the UV, covered by HST STIS spectra. Several lines in this series are strong enough to be identified in a low-resolution spectrum, while the series limit is covered at high resolution. Details on the detected components are summarised in Table~\ref{table:lineseries}. Note in particular the very broad transitions to the $^{1,3}$D terms.

\subsection{IzY bands}
Figure~\ref{fig:heir} displays the spectrum of BD+10$^\circ$2179 in the $IzY$ bands from $\sim$7000\,{\AA} to
$\sim$10\,200\,{\AA}. No efforts have been made to correct for telluric lines -- the \ion{He}{i} lines in this spectral range are nonetheless identifiable. This spectral range is dominated by the series members originating from the $3s$, $3p$ and $3d$ levels, see Table~\ref{table:lineseries}. The broad features are typically formed by an overlap of several multiplets, facilitated by the close spacing of the involved energy levels.

\begin{table}
\caption{Known/predicted forbidden \ion{He}{i} lines present in BD+10$^\circ$2179}
\centering
\begin{tabular}{ll@{~~--~~}l|ll@{~~--~~}l}
\hline\hline
$\lambda$\,({\AA}) & \multicolumn{2}{c}{transition} & $\lambda$\,({\AA}) & \multicolumn{2}{c}{transition}\\
\hline
3449     & $2s$\,$^1$S         & $6d$\,$^1$D         & 4025$^*$ & $2p$\,$^3$P$^\circ$ & $5f$\,$^3$F$^\circ$\\
3587$^*$ & $2p$\,$^3$P$^\circ$ & $9f$\,$^3$F$^\circ$ & 4045     & $2p$\,$^3$P$^\circ$ & $5p$\,$^3$P$^\circ$\\
3616     & $2s$\,$^1$S         & $5d$\,$^1$D         & 4142     & 2$p$\,$^1$P$^\circ$ & $6p$\,$^1$P$^\circ$\\
3634$^*$ & $2p$\,$^3$P$^\circ$ & $8f$\,$^3$F$^\circ$ & 4144$^*$ & $2p$\,$^1$P$^\circ$ & $6f$\,$^1$F$^\circ$ \\
3705$^*$ & $2p$\,$^3$P$^\circ$ & $7f$\,$^3$F$^\circ$ & 4383     & 2$p$\,$^1$P$^\circ$ & $5p$\,$^1$P$^\circ$\\
3711     & $2p$\,$^3$P$^\circ$ & $7p$\,$^3$P$^\circ$ & 4387$^*$ & $2p$\,$^1$P$^\circ$ & $5f$\,$^1$F$^\circ$\\
3820$^*$ & $2p$\,$^3$P$^\circ$ & $6f$\,$^3$F$^\circ$ & 4470     & $2p$\,$^3$P$^\circ$ & $4f$\,$^3$F$^\circ$\\
3830     & $2p$\,$^3$P$^\circ$ & $6p$\,$^3$P$^\circ$ & 4517     & $2p$\,$^3$P$^\circ$ & $4p$\,$^3$P$^\circ$\\
3971     & $2s$\,$^1$S         & $4f$\,$^1$F$^\circ$ & 4911     & $2p$\,$^1$P$^\circ$ & $4p$\,$^1$P$^\circ$\\
3972     & $2s$\,$^1$S         & $4d$\,$^1$D         & 4921     & $2p$\,$^1$P$^\circ$ & $4f$\,$^1$F$^\circ$\\
4008     & $2p$\,$^1$P$^\circ$ & $7p$\,$^1$P$^\circ$ & 5042     & $2s$\,$^1$S         & $3d$\,$^1$D        \\
4009$^*$ & $2p$\,$^1$P$^\circ$ & $7f$\,$^1$F$^\circ$\\
\hline
\end{tabular}
\vspace{-2mm}
\begin{flushleft}
$^*$ presence inferred because of line asymmetries 
\end{flushleft}
\label{table:forbidden}
\end{table}

\subsection{Forbidden components in the optical}
Extreme helium stars are a testbed for the detection of isolated forbidden components of \ion{He}{i} transitions.
\citet{har97} detected all ten forbidden components listed by \citet{Underhill66} in their spectrum of the EHe star HD\,144941 in the wavelength range 3840--4960\,{\AA}\footnote{However, resolving the $\lambda$3972\,{\AA} component (and $\lambda$3971\,{\AA})
was hampared by the relatively strong H$\varepsilon$ line in HD\,144931, nearly absent here because of the lower hydrogen content of BD+10$\degr$2179.}. We extend this list to 23 known/predicted lines covered by the wavelength range of our spectra. To our knowledge, these additional forbidden components (electronic quadrupole transitions) are observationally confirmed for the first time by the present work, summarised in Table~\ref{table:forbidden} and visualised in Figure~\ref{fig:heforb}. 
Note the good agreement with the asymmetric line shapes predicted by \citet{BeWe98}.
The presence of several forbidden components as marked in Table~\ref{table:forbidden} is inferred because of line asymmetries.
Line separations between the allowed and forbidden components amount in these cases from a maximum of 0.64\,{\AA} for [\ion{He}{i}] $\lambda$4025\,{\AA} to a minimum value of 0.17\,{\AA}  for [\ion{He}{i}] $\lambda$3705\,{\AA}. 
We cannot confirm the presence of the forbidden component $\lambda\lambda$ 6632\,{\AA} (the region is affected by an artifact from the data reduction). Additional  (weak) transitions at $\lambda\lambda$ 3809, 3165, 4054 and 6068\,{\AA} and forbidden components in the HST STIS range at $\lambda\lambda$ 2935 and 2823\,{\AA} with lower levels of $n$\,=\,2 are also not detected, as well as the forbidden components from lower levels with $n$\,=\,3 like $\lambda\lambda$ 8315, 9360, 9616\,{\AA}. Reasons for these are manifold: the transitions being too weak to be detected, blends, insufficient S/N, insufficient resolution of the low-resolution HST STIS data, uncorrectable contamination with nearly opaque telluric lines.

\begin{figure*}
\centering
\includegraphics[width=.83\textwidth]{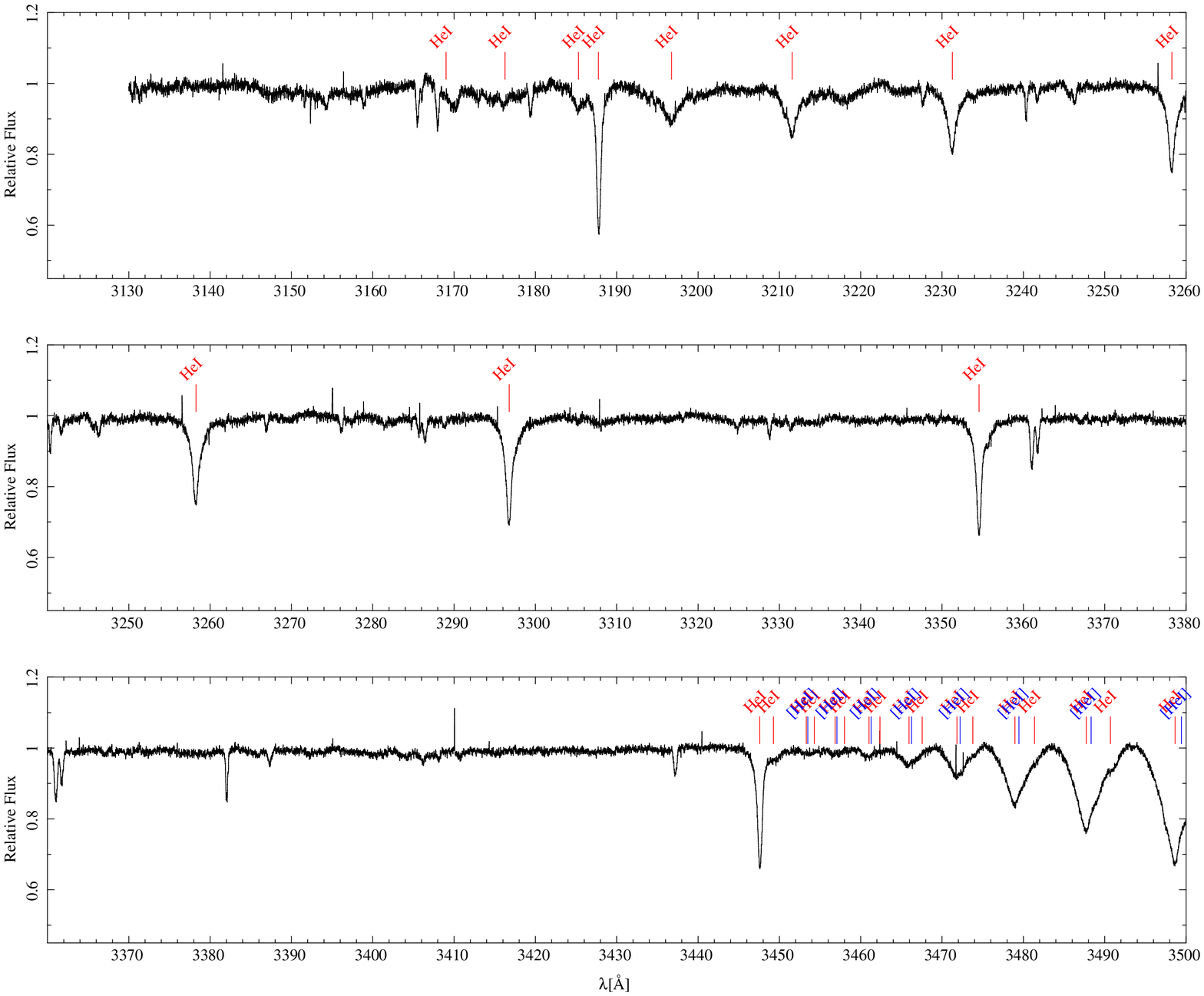}
\includegraphics[width=.83\textwidth]{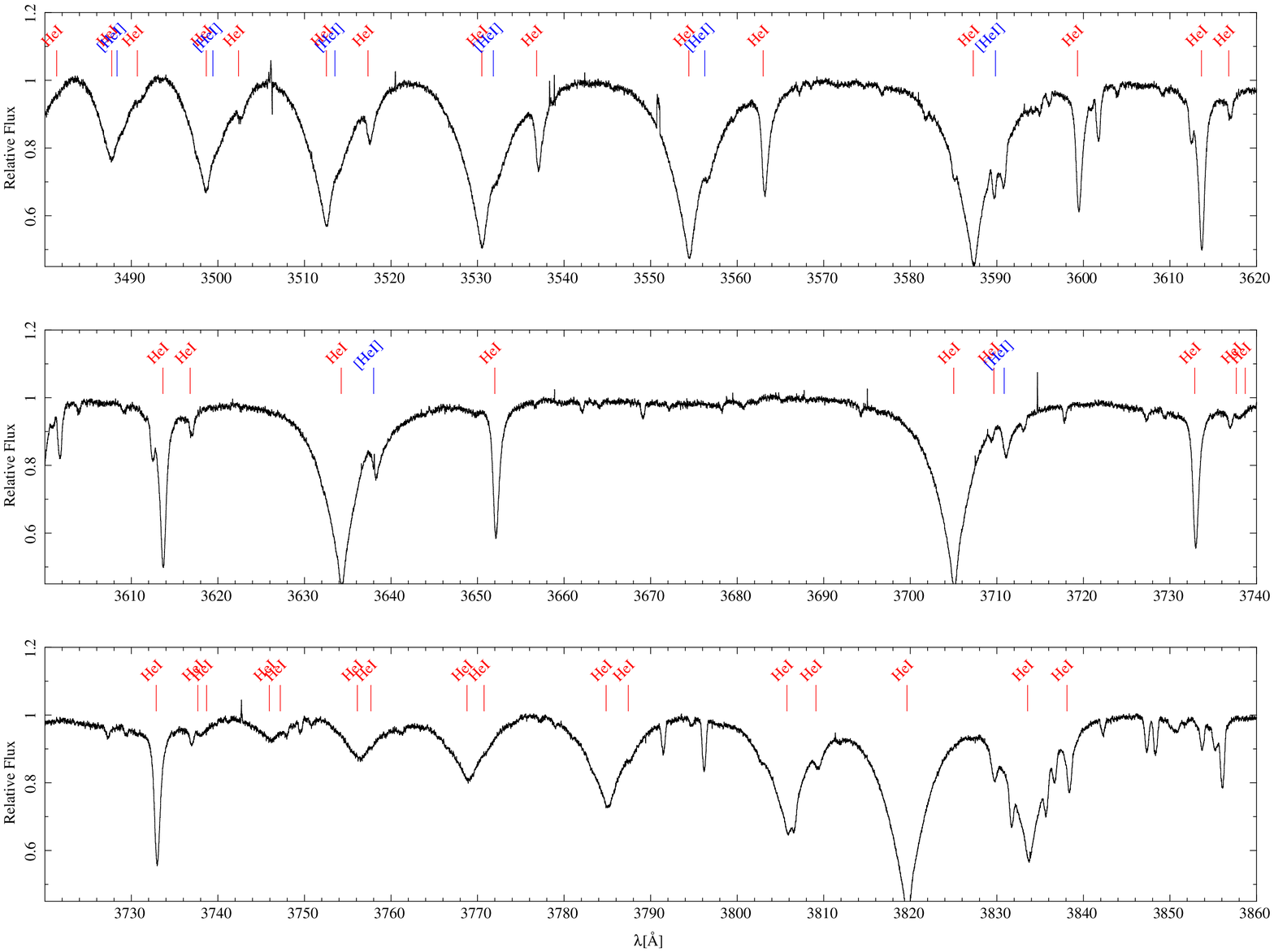}   
\caption{\ion{He}{i} lines in the optical-UV as traced by our UVES spectrum. Blue marked are the new detected series of forbidden \ion{He}{i} given in Table\,\ref{tab:new_forbidden}.}
\label{fig:heuv}
\end{figure*}

\begin{figure*}
\centering
\includegraphics[width=.85\textwidth]{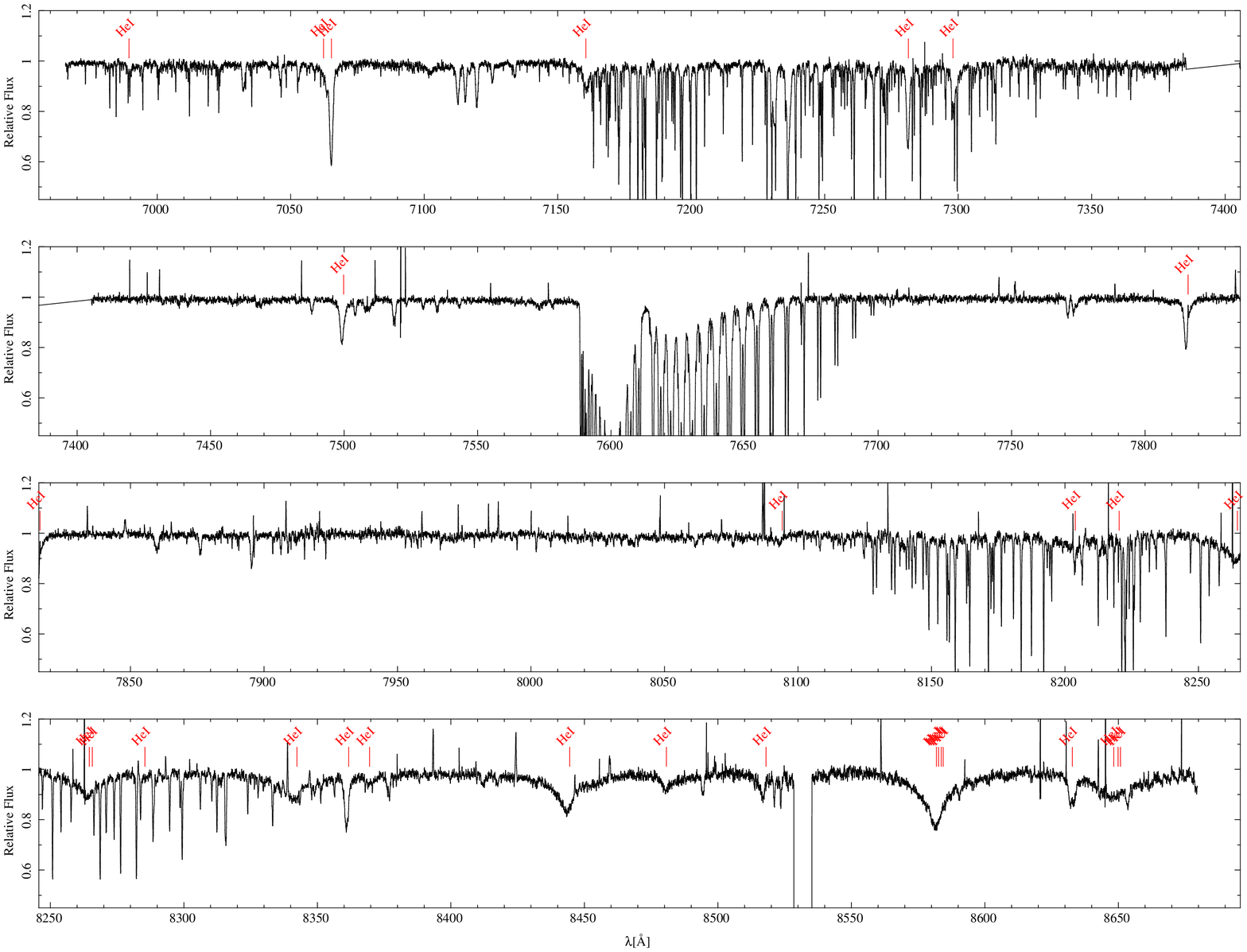}
\includegraphics[width=.85\textwidth]{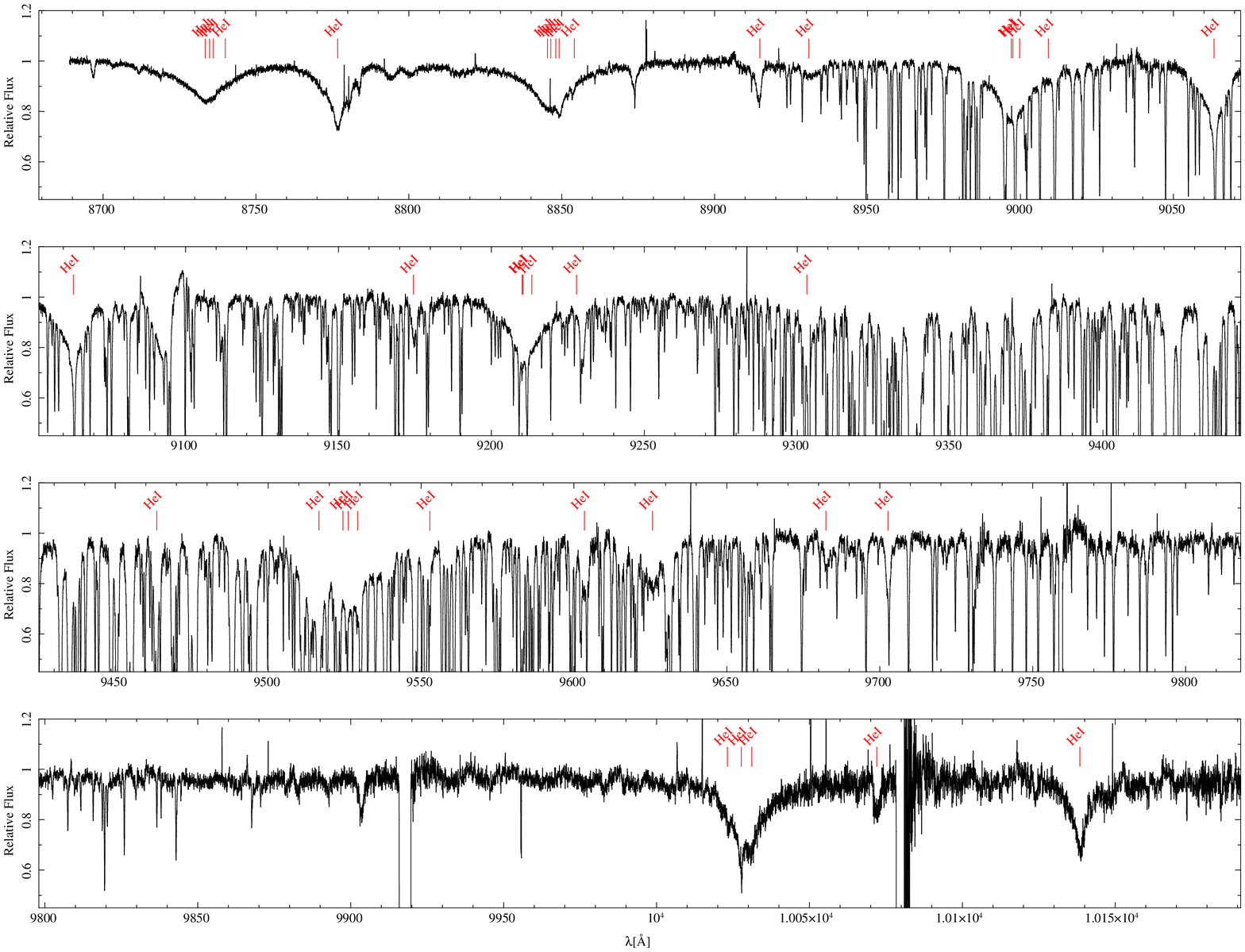}
\caption{\ion{He}{i} lines in the near-IR $IzY$ bands as traced by our UVES and FEROS spectra.}
\label{fig:heir}
\end{figure*}

\begin{figure*}
\begin{center}
\includegraphics[width=.328\textwidth]{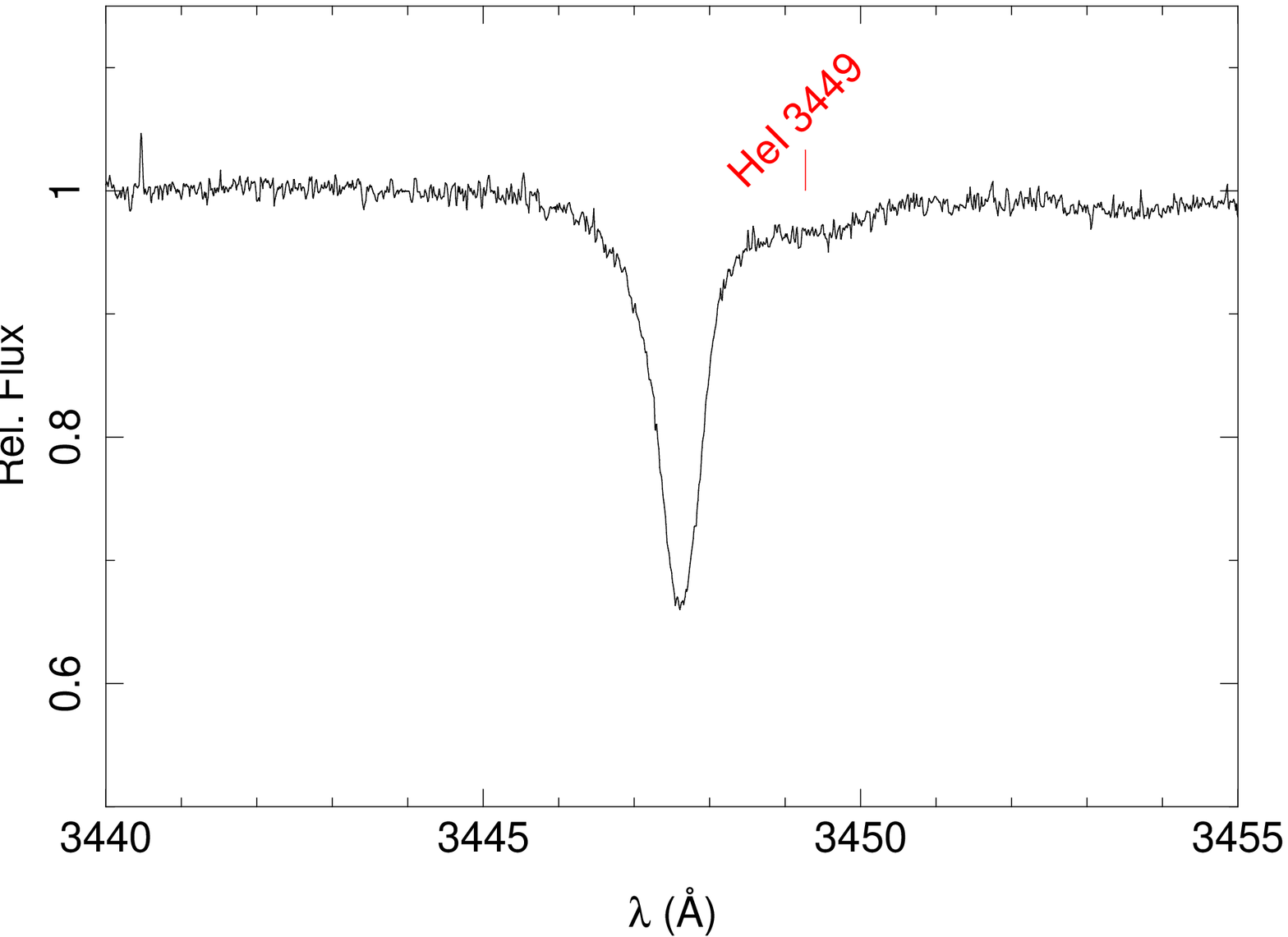}
\hspace{0.1cm}
\includegraphics[width=.31\textwidth]{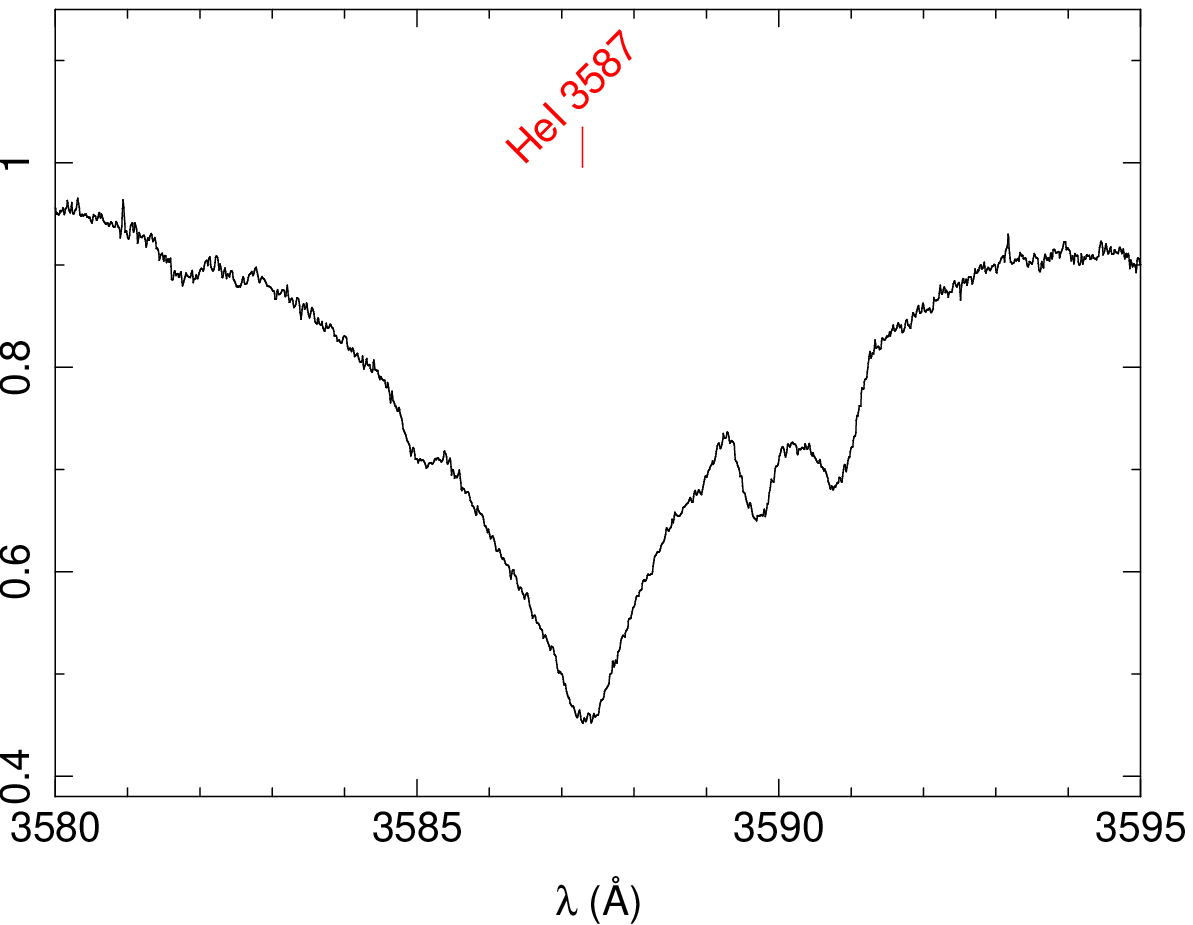}
\includegraphics[width=.30\textwidth]{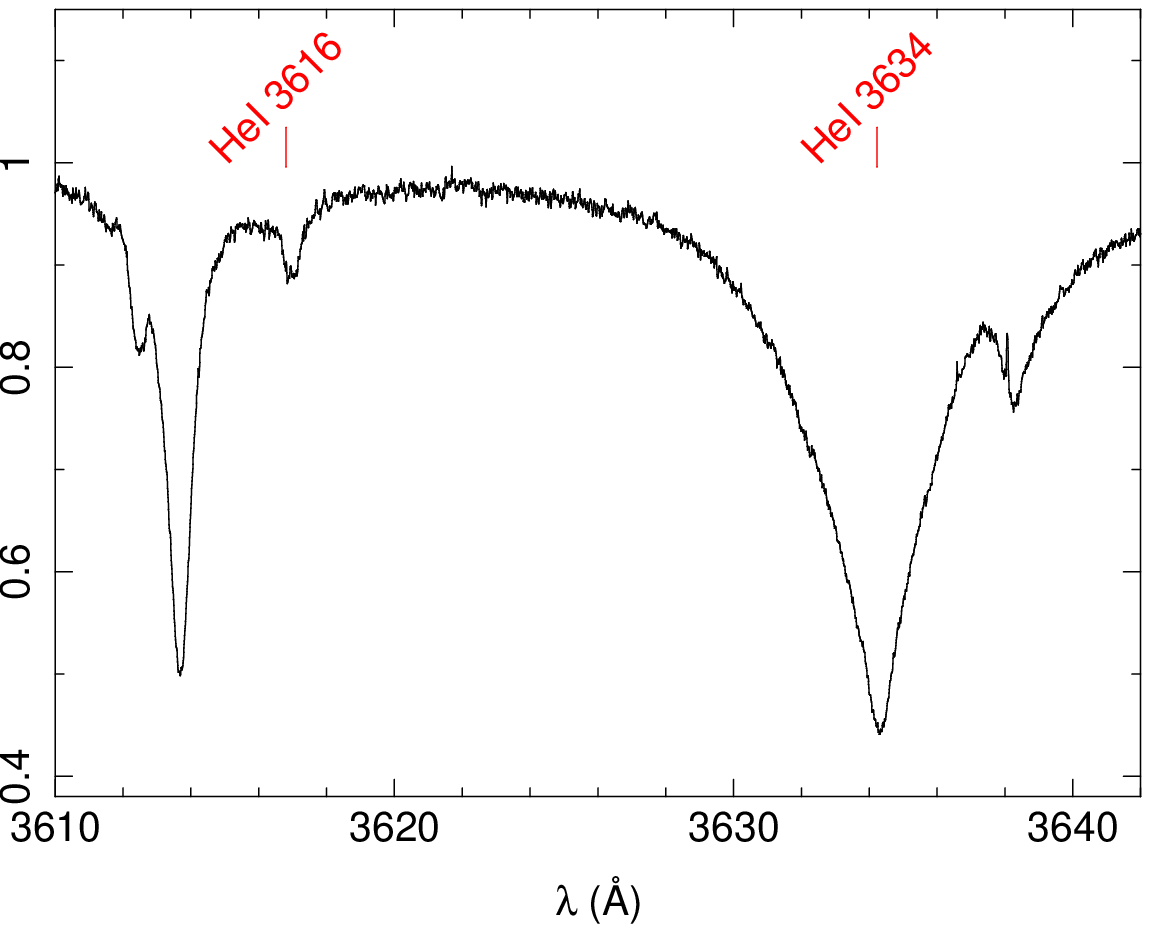}
\includegraphics[width=.325\textwidth]{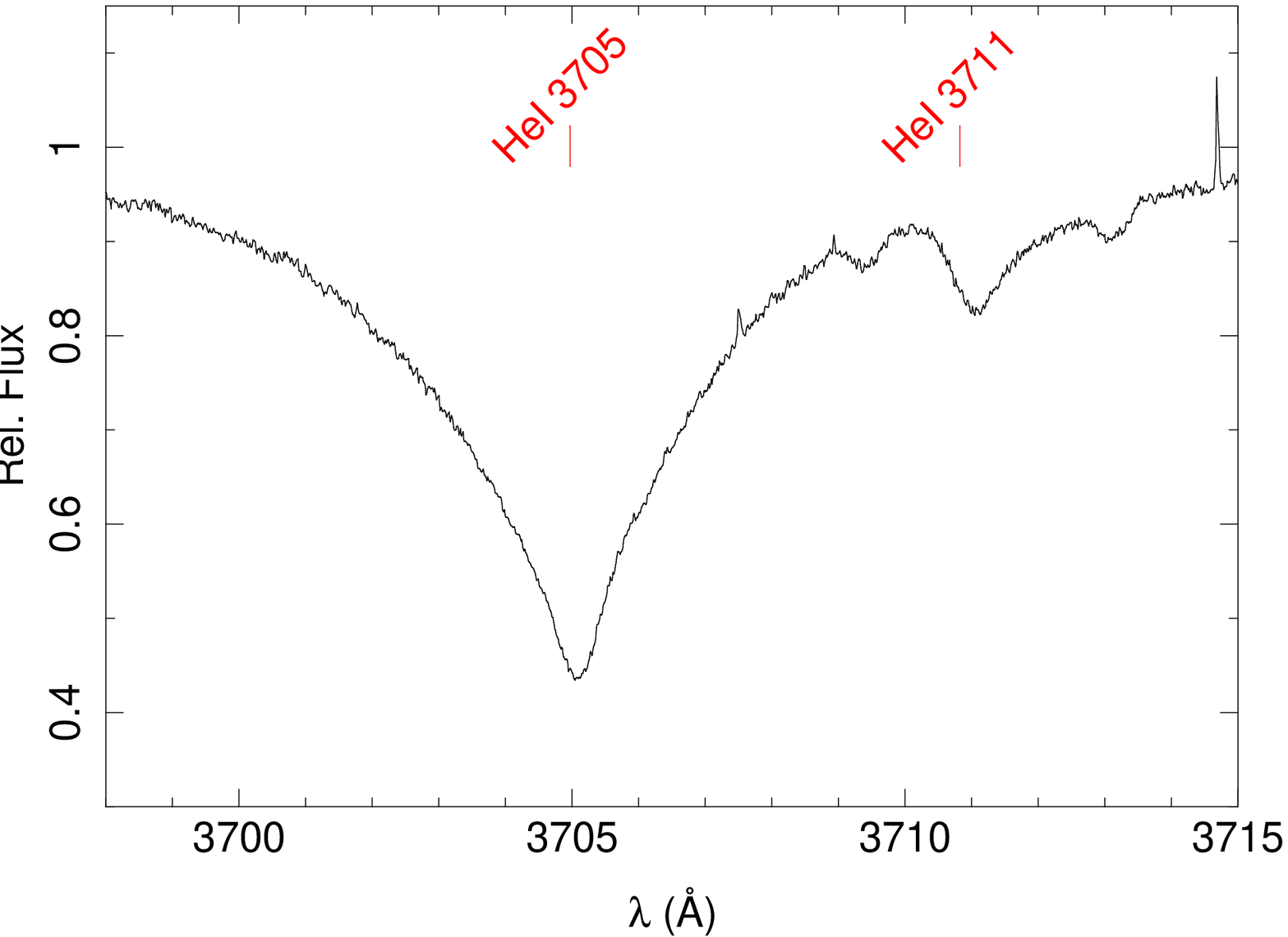}
\hspace{0.1cm}
\includegraphics[width=.303\textwidth]{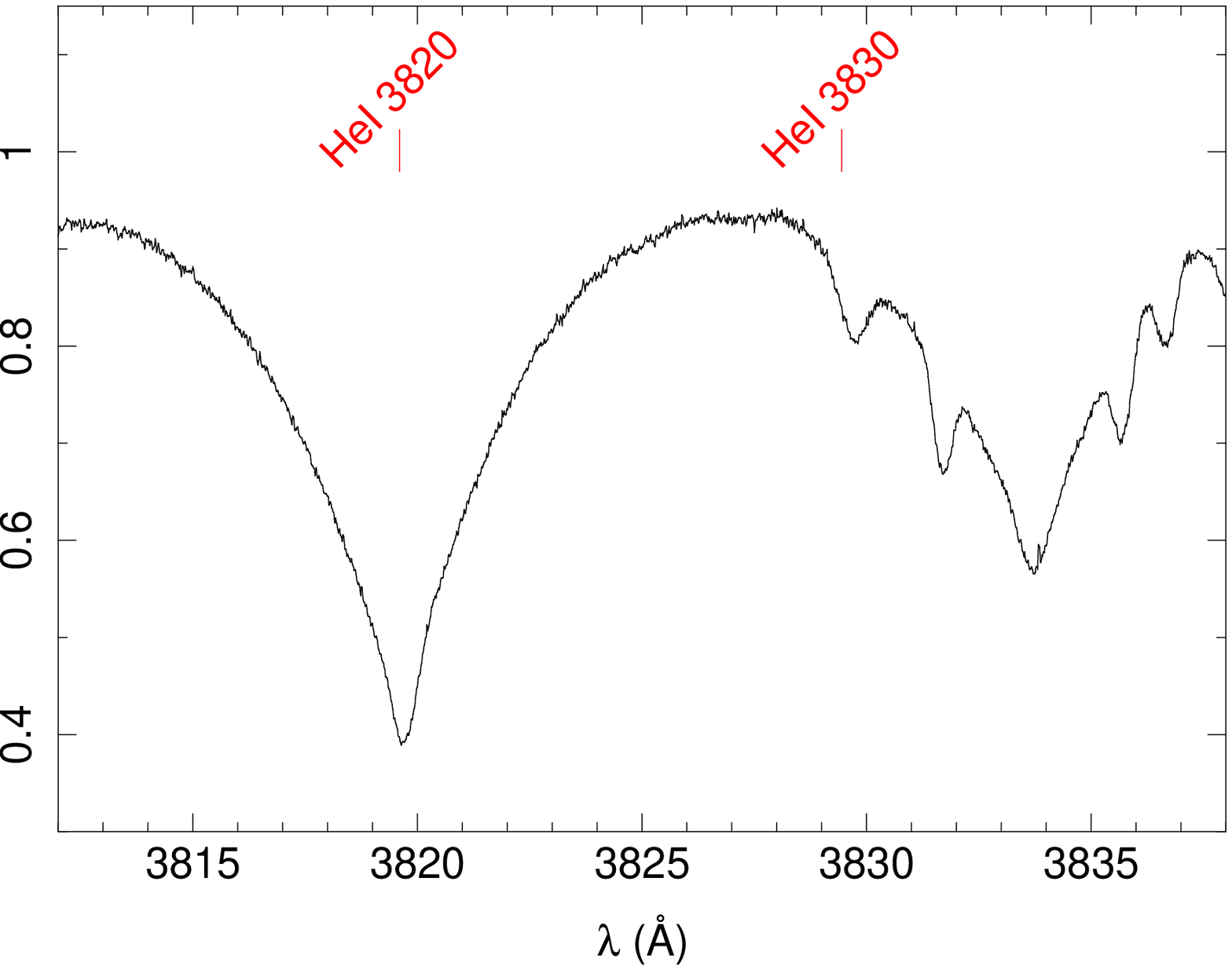}
\hspace{0.1cm}
\includegraphics[width=.315\textwidth]{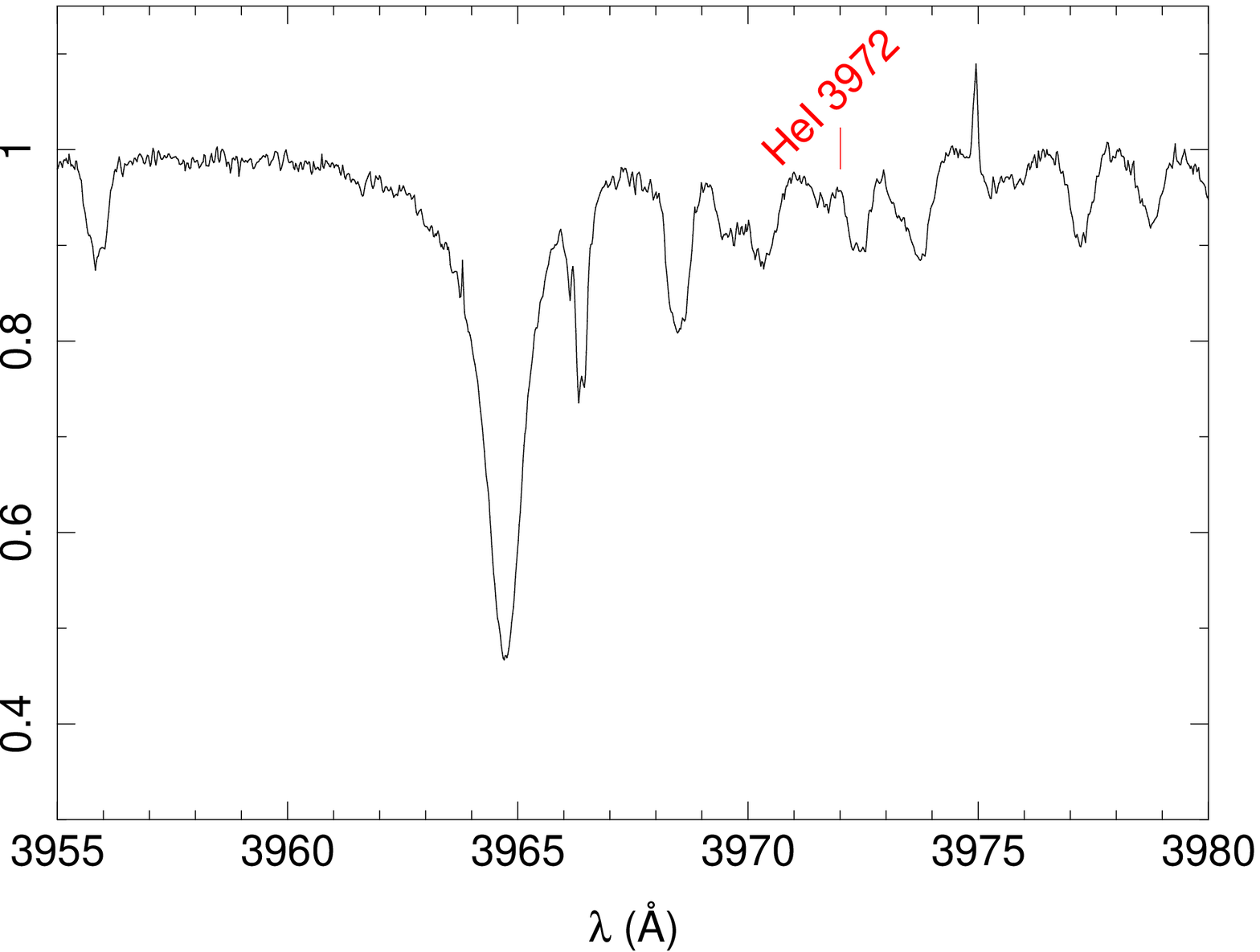}
\includegraphics[width=.32\textwidth]{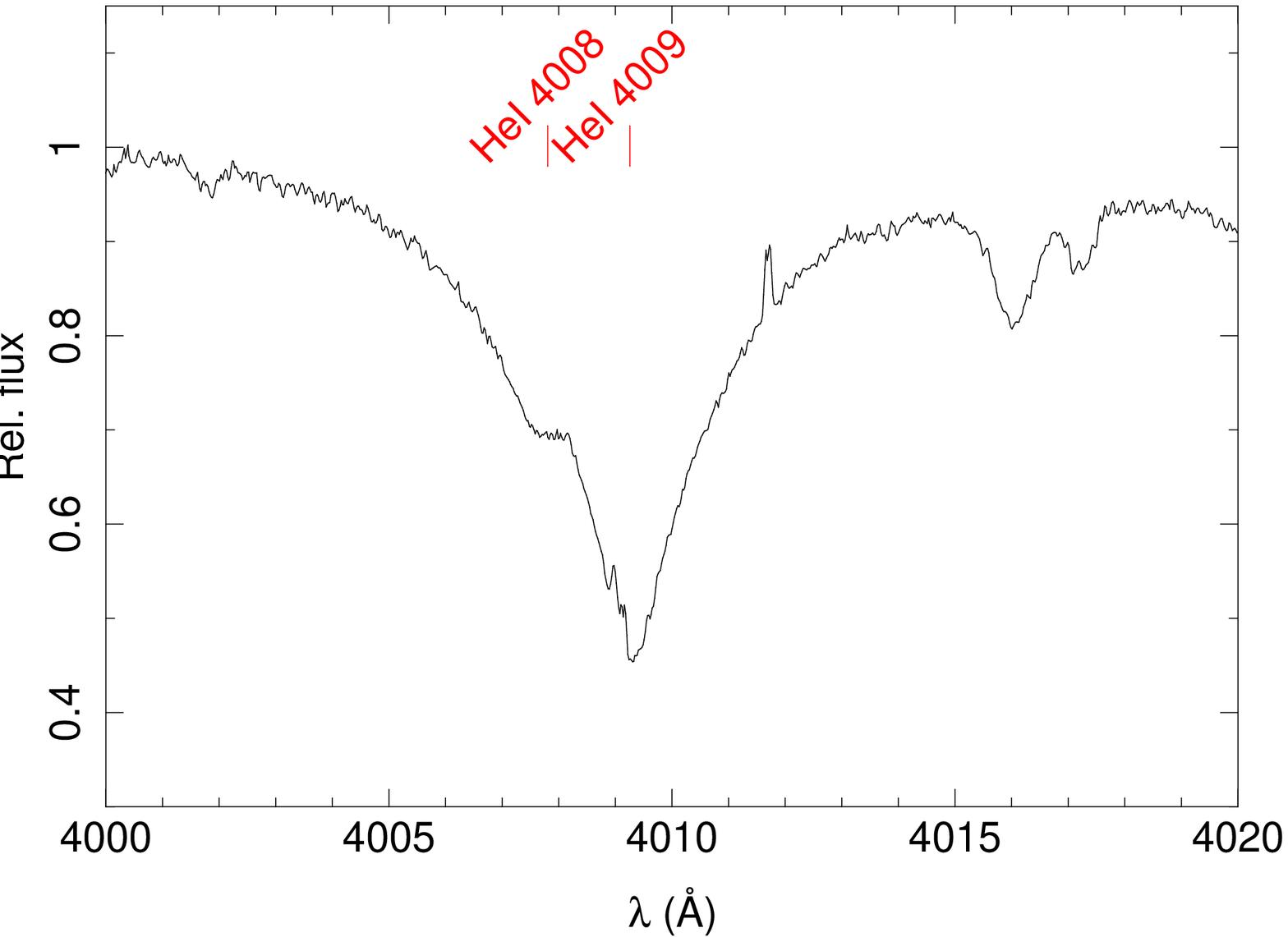}
\hspace{0.1cm}
\includegraphics[width=.30\textwidth]{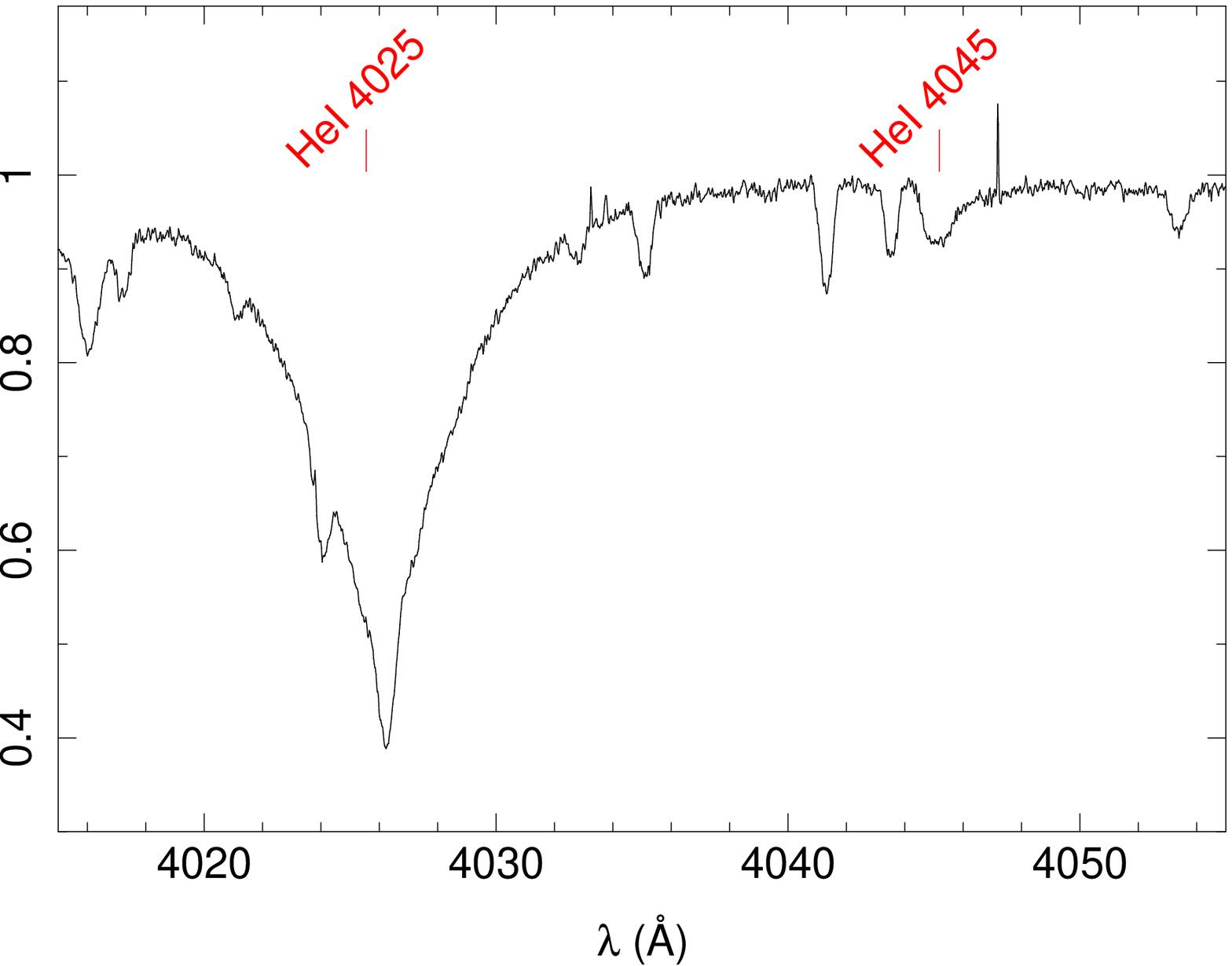}
\hspace{0.1cm}
\includegraphics[width=.30\textwidth]{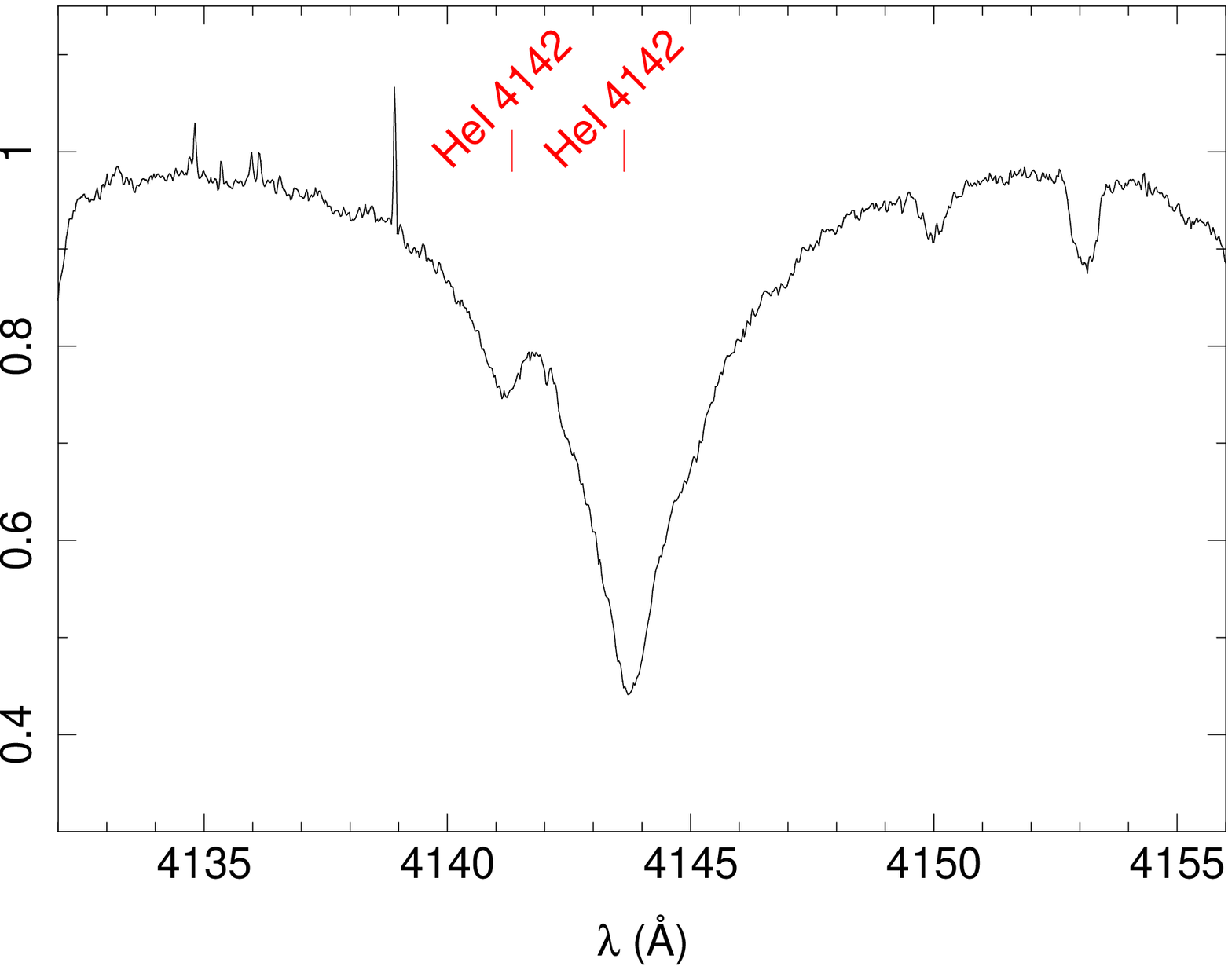}
\includegraphics[width=.317\textwidth]{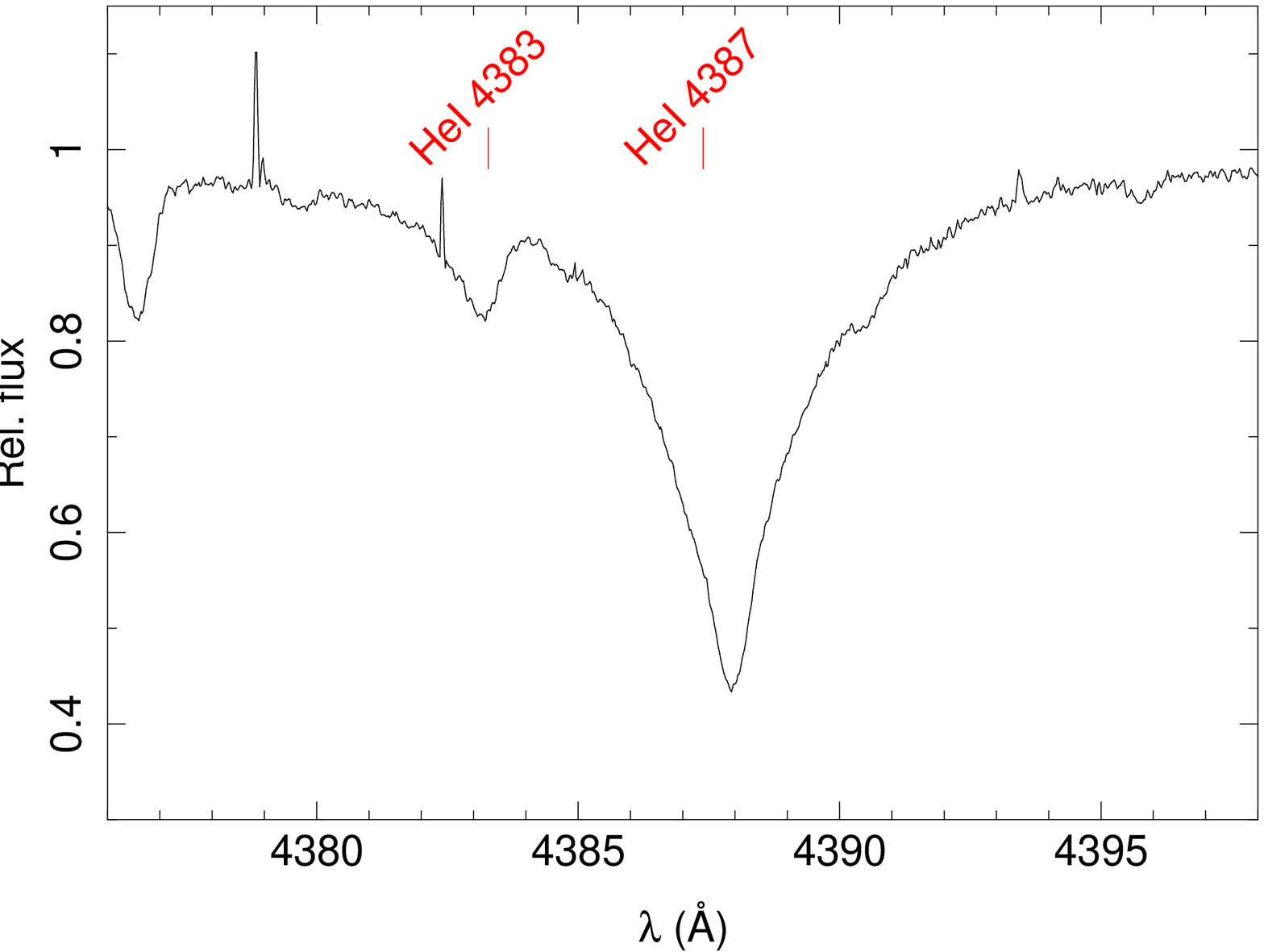}
\hspace{0.1cm}
\includegraphics[width=.307\textwidth]{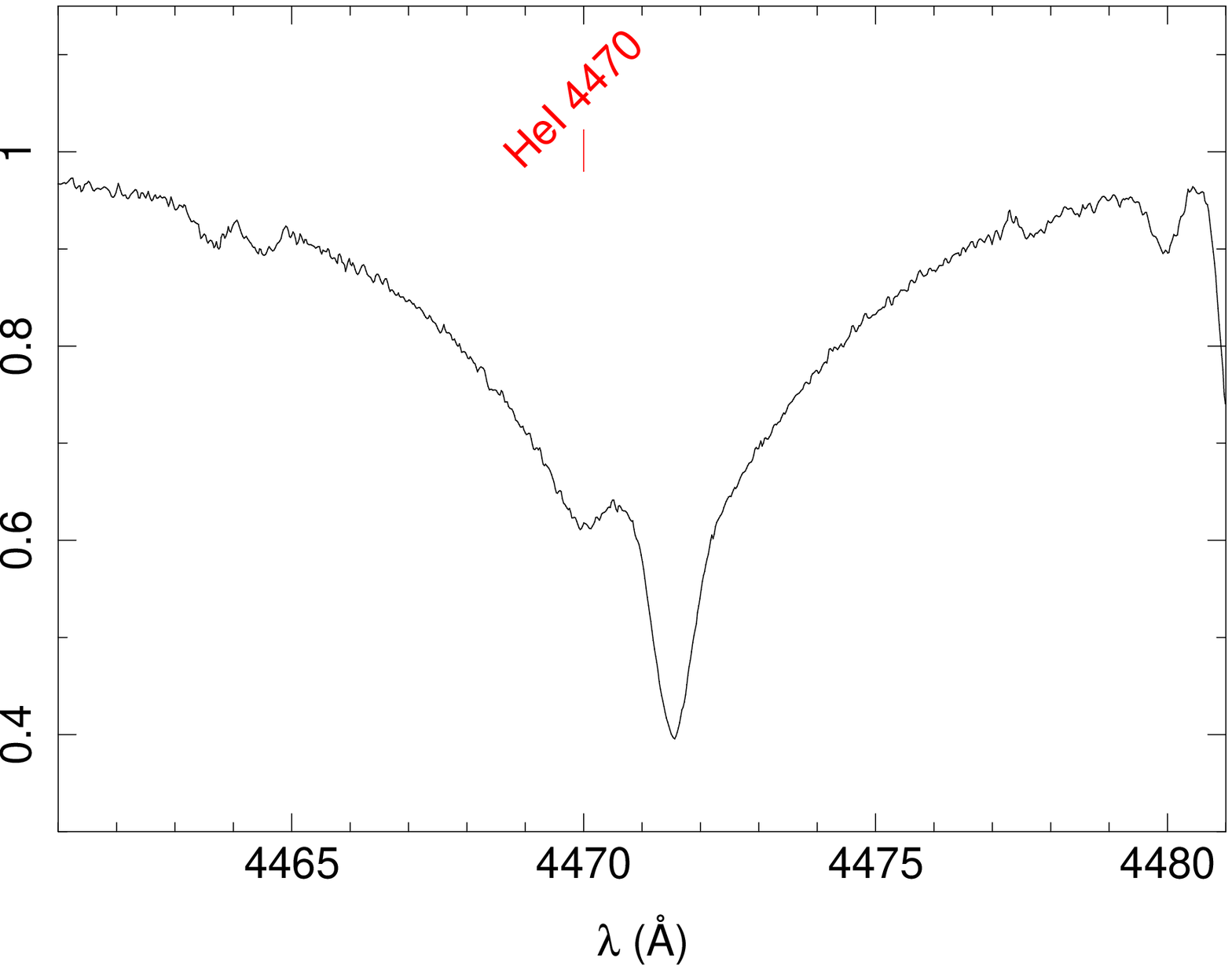}
\hspace{0.1cm}
\includegraphics[width=.317\textwidth]{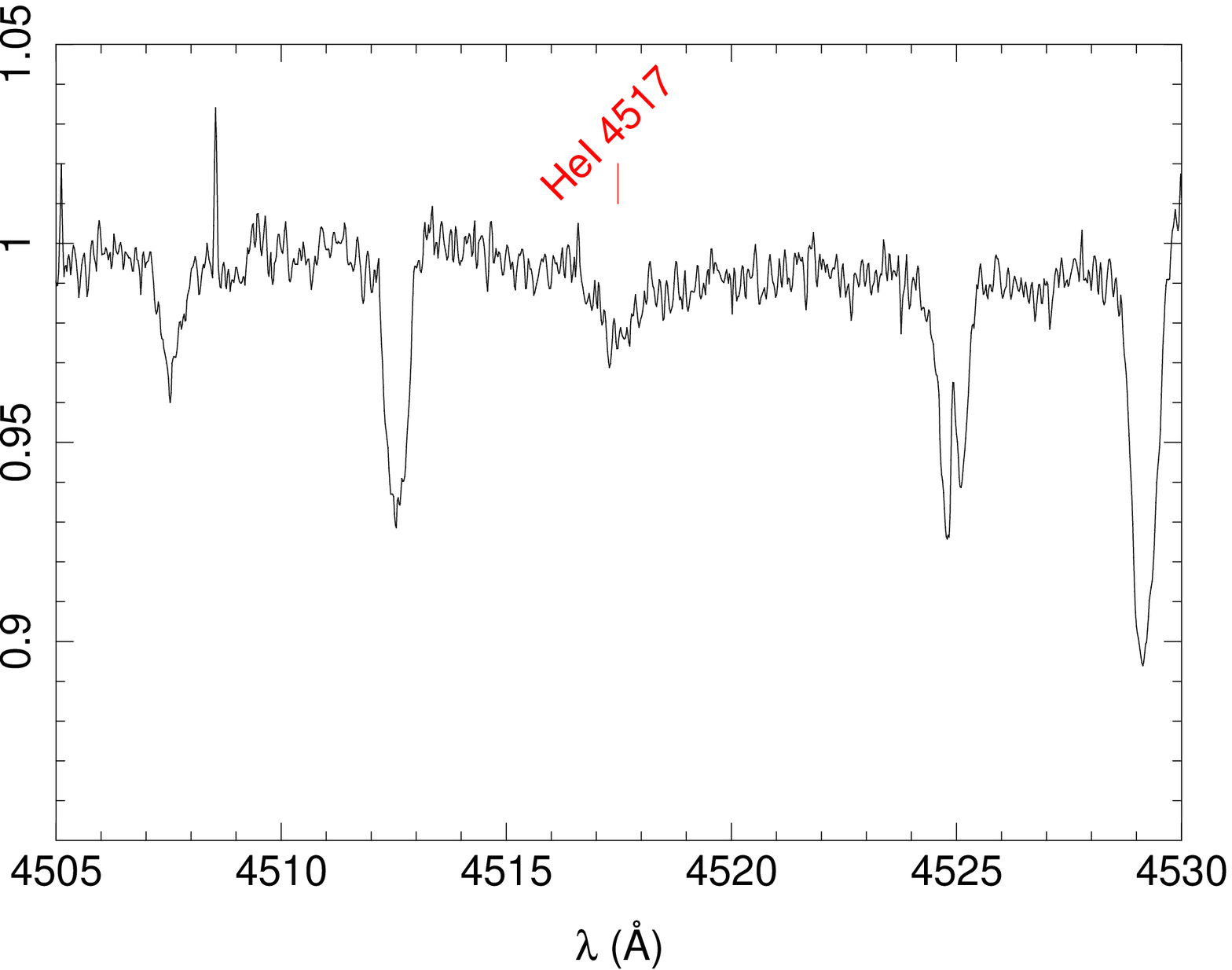}
\includegraphics[width=.32\textwidth]{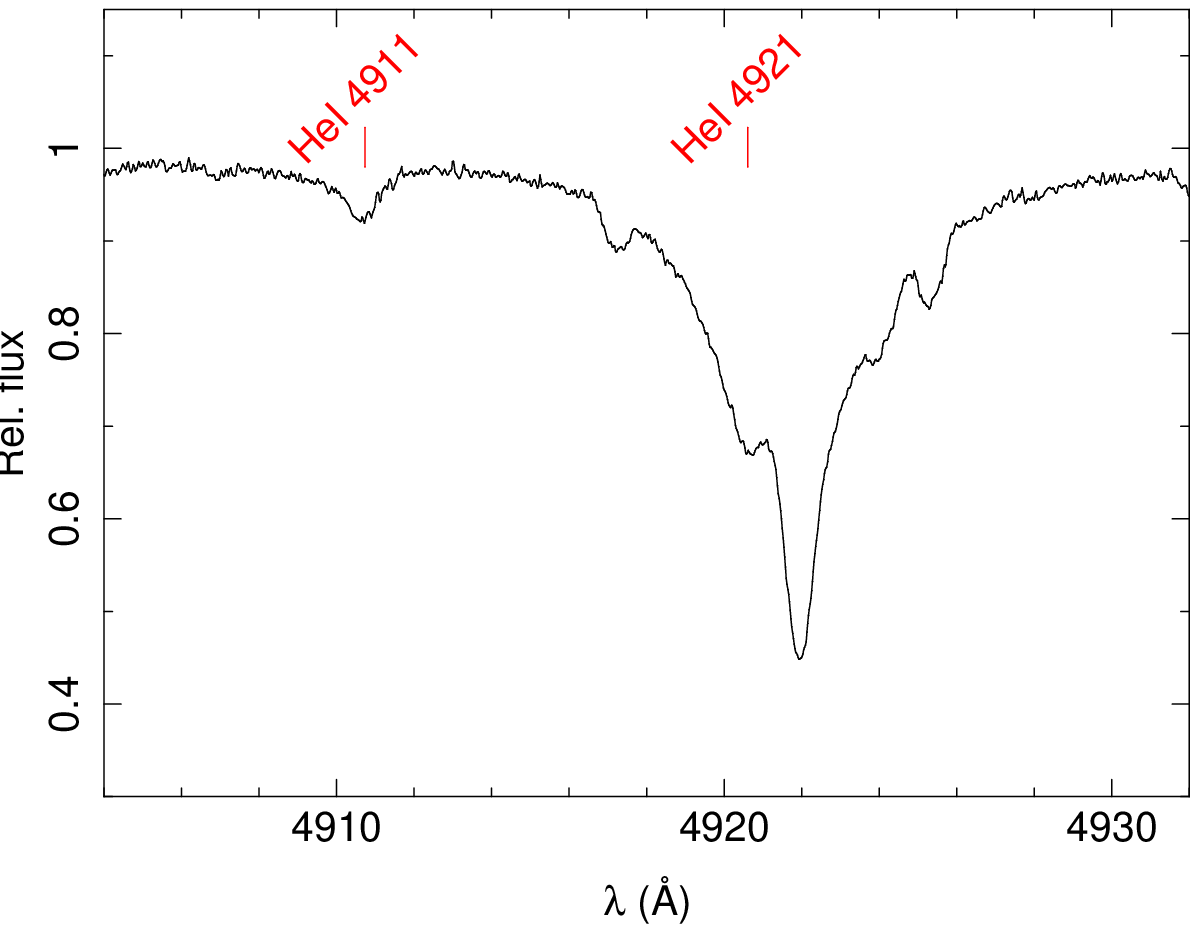}
\hspace{0.1cm}
\includegraphics[width=.32\textwidth]{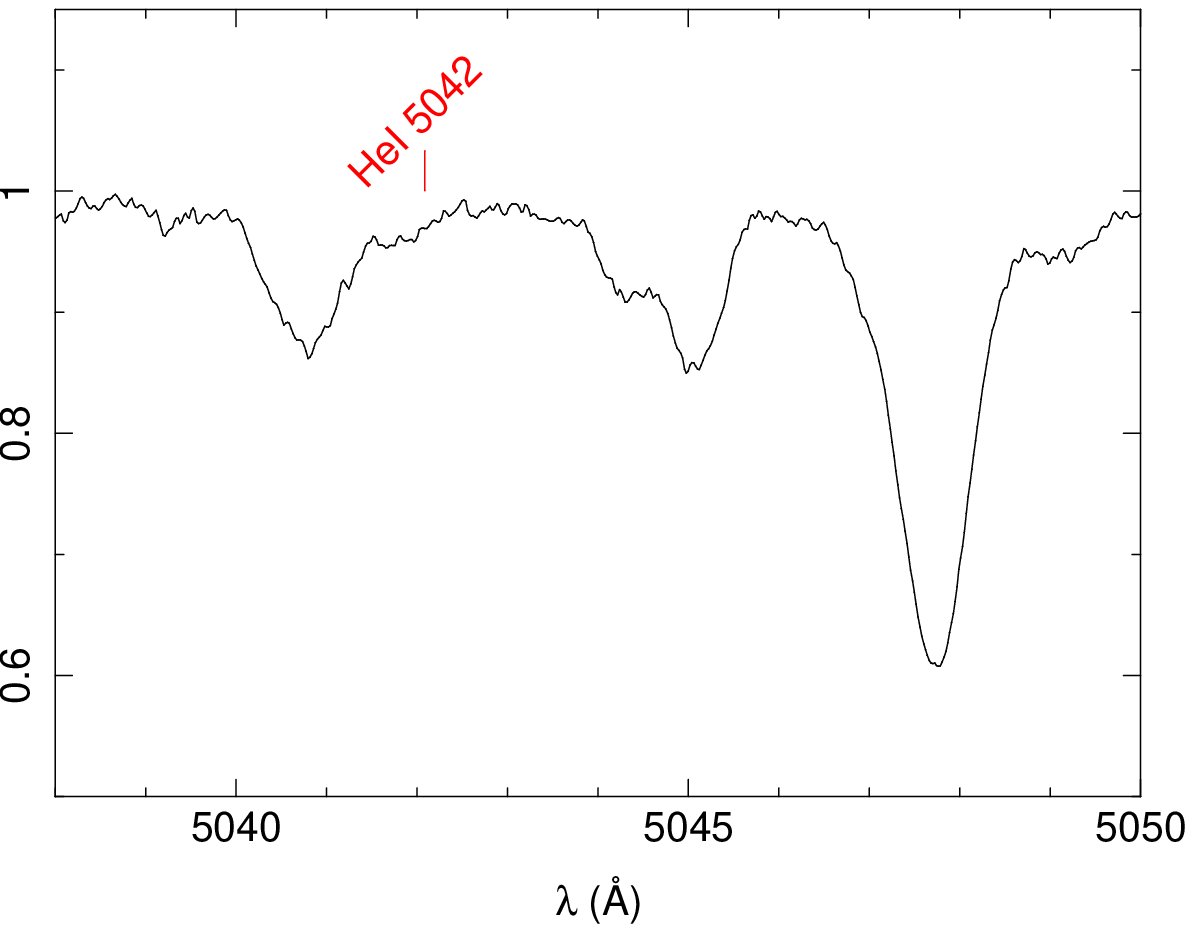}
\caption{Isolated forbidden \ion{He}{i} components in the optical spectrum.}
\label{fig:heforb}
\end{center}
\end{figure*}

The inspection of the optical-UV spectrum with overlaid identifiers of the known He\,{\sc i} lines
shows a conspicuous series of isolated or blended features between the lines of the higher 
$2p$\,$^3$P$^\circ$ to $nd$\,$^3$D and $ns$\,$^3$S series, see the 4th and 5th panels of Fig.~\ref{fig:heuv}. This so far undocumented series is produced by E2 transitions from $2p$\,$^3$P$^\circ$ to $np$\,$^3$P$^\circ$ with $n$\,$\ge$\,8, i.e. transitions to higher levels than described in any previous work. We characterise this series in Table~\ref{tab:newforb}, providing information on the transition, wavelengths in air $\lambda_\mathrm{air}$ and vacuum $\lambda_\mathrm{vac}$, level energies, statistical weights and transition type. The data are given from $n$\,=\,7 to 20, despite we can infer the presence of this component only up to $n$\,=\,16 or 17 because of line asymmetries. Higher series members may be detectable in EHe stars with lower gravities than BD+10$^\circ$2179.

\begin{table*}
\caption{New detections and extension of the $2p$\,$^3$P$^\circ$\,--\,$np$\,$^3$P$^\circ$   forbidden line spectrum of He\,{\sc i}}\label{tab:newforb}
\begin{tabular}{lllccc}
\hline\hline
Transition & $\lambda_\mathrm{air}$\,({\AA}) & $\lambda_\mathrm{vac}$\,({\AA}) & $E_i$--$E_k$\,(cm$^{-1}$) & $g_i$--$g_k$ & Type\\
\hline
$2p$\,$^3$P$^\circ$\,--\,$\,~7p$\,$^3$P$^\circ$ & 3710.85 & 3711.90 & 169086.9085\,--\,196027.316 & 9--9 & E2\\
$2p$\,$^3$P$^\circ$\,--\,$\,~8p$\,$^3$P$^\circ$ & 3638.00 & 3639.04 & 169086.9085\,--\,196566.712 & 9--9 & E2\\
$2p$\,$^3$P$^\circ$\,--\,$\,~9p$\,$^3$P$^\circ$ & 3589.85 & 3590.87 & 169086.9085\,--\,196935.331 & 9--9 & E2\\
$2p$\,$^3$P$^\circ$\,--\,$10p$\,$^3$P$^\circ$ & 3556.26 & 3557.27 & 169086.9085\,--\,197198.332 & 9--9 & E2\\
$2p$\,$^3$P$^\circ$\,--\,$11p$\,$^3$P$^\circ$ & 3531.83 & 3532.84 & 169086.9085\,--\,197392.720 & 9--9 & E2\\
$2p$\,$^3$P$^\circ$\,--\,$12p$\,$^3$P$^\circ$ & 3513.53 & 3514.53 & 169086.9085\,--\,197540.190 & 9--9 & E2\\
$2p$\,$^3$P$^\circ$\,--\,$13p$\,$^3$P$^\circ$ & 3499.43 & 3500.43 & 169086.9085\,--\,197654.830 & 9--9 & E2\\
$2p$\,$^3$P$^\circ$\,--\,$14p$\,$^3$P$^\circ$ & 3488.34 & 3489.34 & 169086.9085\,--\,197745.650 & 9--9 & E2\\
$2p$\,$^3$P$^\circ$\,--\,$15p$\,$^3$P$^\circ$ & 3479.45 & 3480.45 & 169086.9085\,--\,197818.830 & 9--9 & E2\\
$2p$\,$^3$P$^\circ$\,--\,$16p$\,$^3$P$^\circ$ & 3472.22 & 3473.21 & 169086.9085\,--\,197878.690 & 9--9 & E2\\
$2p$\,$^3$P$^\circ$\,--\,$17p$\,$^3$P$^\circ$ & 3466.25 & 3467.24 & 169086.9085\,--\,197928.260 & 9--9 & E2\\
$2p$\,$^3$P$^\circ$\,--\,$18p$\,$^3$P$^\circ$ & 3461.27 & 3462.26 & 169086.9085\,--\,197969.750 & 9--9 & E2\\
$2p$\,$^3$P$^\circ$\,--\,$19p$\,$^3$P$^\circ$ & 3457.07 & 3458.06 & 169086.9085\,--\,198004.850 & 9--9 & E2\\
$2p$\,$^3$P$^\circ$\,--\,$20p$\,$^3$P$^\circ$ & 3453.49 & 3454.48 & 169086.9085\,--\,198034.800 & 9--9 & E2\\
\hline
\end{tabular}
\label{tab:new_forbidden}
\end{table*}

\clearpage

\end{document}